FLORIDA AGRICULTURAL AND MECHANICAL UNIVERSITY

COLLEGE OF SCIENCE AND TECHNOLOGY

# Developing Hands-on Labs for Source Code Vulnerability Detection with AI

By

Maryam Ramezanzadehmoghadam

A Thesis Submitted to the
Department of Computer and Information Sciences
In partial fulfillment of the
Requirements for the degree of
Master of Science in Computer Science

Tallahassee, FL

Summer Semester, 2021

Major Professor: Hongmei Chi.

The Committee approves the Thesis entitled, *Developing Hands-on Labs for Source Code Vulnerability Detection using AI;* by *Maryam Ramezanzadehmoghadam* defended on 07/23/2021.

Hongmei Chi, Ph.D.
Committee Chair

Samia Tasnim, Ph.D.
Committee Member

Jinwei Liu, Ph.D.
Committee Member

Shonda Bernadin, Ph.D.
Outside Committee Member

Approved by:

Idongesit Mkpong-Ruffin, Ph.D., Chairperson, Department of Computer and Information Sciences

Richard A. Alo, Ph.D., Dean, College of Science and Technology

Reginald K. Ellis, Ph.D., Interim Dean of the School of Graduate Studies & Research

## ABSTRACT

As the role of information and communication technologies gradually increases in our lives, source code security becomes a significant issue to protect against malicious attempts. Furthermore, with the advent of data-driven techniques, there is now a growing interest in leveraging machine learning and natural language processing (NLP) as a source code assurance method to build trustworthy systems. Therefore, training our future software developers to write secure source code is in high demand. In this thesis, we propose a framework including learning modules and hands-on labs to guide future IT professionals towards developing secure programming habits and mitigating source code vulnerabilities at the early stages of the software development lifecycle. In this thesis, our goal is to design learning modules with a set of hands-on labs that will introduce students to secure programming practices using source code and log file analysis tools to predict/identify vulnerabilities. In a Secure Coding Education framework called (**SeCodEd**) we will (1) improve students' skills and awareness on source code vulnerabilities, detection tools, and mitigation techniques; (2) integrate concepts of source code vulnerabilities from Function, API, and library level to bad programming habits and practices; (3) leverage deep learning, NLP and static analysis tools for log file analysis to introduce the root cause of source code vulnerabilities.

## ACKNOWLEDGMENT

Writing this thesis during Covid-19 was a challenging experience to me. During this hard time, many people were instrumental directly or indirectly in shaping up my academic career. It was hardly possible for me to thrive in this work without the precious support of these personalities. I would like to send my eternal gratitude to my boyfriend, Peyman Taeb, for his enduring support. He gave quite encouragement and positive belief in my success that kept me going regardless of the challenges that I faced. I am forever indebted to my parents, who from way far away, never failed to say that they were proud of me for doing this work, no matter how worried they were about me, and specially to my mother whose life has led me to be who I am today. I am deeply grateful to my brother, Amir, for his willingness to help. He always believed in my ability to achieve my dreams. A very special word of thanks goes for my Friend, Mahsa, who have been great over the years and always showed how proud she is of me. All of these people selflessly encouraged me to explore new directions in life and seek my own destiny. This journey would not have been possible if not for them, and I dedicate this milestone to them.

It is a genuine pleasure to express my honest sense of thanks and gratitude to my mentor, advisor, and guide Dr. Hongmei Chi. The completion of this study could not have been possible without her expertise. A debt of gratitude is also owed to Dr. Jinwei Liu for his continuous support, patience and guidance to finish this study. Recognition must also be given to my committee, Dr. Shonda Bernadin and Dr. Samia Tasnim and Associate Chairperson, Dr. Edward Jones, for their prompt inspirations and timely suggestions with kindness. I would like to thank Dean of the College of Science & Technology, Dr. Alo, FAMUs' Convergent Data Science Research Center, and the Department of Education Title III: Enhancing Graduate Science Programs for their financial support as I wouldn't be able to finish this program if it wasn't for them.

# DEDICATION

I would like to dedicate my work to all female scientists -past, present, and future. May your inquiring minds lead you forward with confidence, bravery, and grace into the great unknown.

## Table of Contents

# LIST OF TABLES

# LIST OF FIGURES

## LIST OF EQUATIONS

# Chapter 1 Introduction

In recent years, IT companies such as Microsoft, GitHub, Facebook, have faced a substantial amount of scrutiny for exposure of customer data. To be more precise, they have been victims of many different attacks such as zero-day attacks, denial of service, command injection, and many more critical vulnerabilities. The software has become necessary for various societal industries, including technology, health care, public safety, education, energy, and transportation. Therefore, there must be no errors made by the software architects engineers to create secure, reliable, and secure software. A minor flaw in source code can span over the complete software and cause a severe vulnerability to make the system an easy target for attackers. For example, two more lines of code could have easily avoided Mac OS XNU memory leak (CVE-2017-13868). Cyber-attacks are increasing threats to government, businesses, and consumers; however, there have been only few learning modules for students to understand the basic concepts of a software vulnerability. Students need to be educated on techniques used to recognize, mitigate, and avoid vulnerabilities in the early stages of the software development lifecycle. Therefore, there is a considerable learning curve for students to overcome and designing a set of hands-on labs can address these challenges one by one.

## 1.1 Background

Cyber risk is one of the most important priorities for any organization whose services are connected to the Internet, which an attacker can leverage to disrupt the organization's economy. Organizations realize the importance of security corporations in every product, software, framework, or service they offer their customers. Cybersecurity and software vulnerability detection continually extend with more aspects and components to secure software development and mitigation techniques. Such software vulnerabilities and cyberattacks can eventually significantly impact customers. Attackers can use any data bridge to access personal user information, which is a significant loss to any organization. The software industry is divided into four main categories, including system services, programming services, open-source, and software as a service (SaaS). The programming services sector has traditionally been the largest sector, including pioneers like Microsoft Corporation, Automatic Data Processing (ADP), Oracle Corporation, and SDC Technologies (Zippia, 2021).

Today's leading companies require skilled IT talent capable of understanding and analyzing any cyber threat condition. Future IT professionals should help the company stay competitive by mitigating any flaws that a cyber attacker can leverage. However, Intel's 2020 study demonstrated that there exists a lack of cybersecurity skills between organizations. The information security workforce has predicted that in the next three years, high demand for personnel who has relevant security will be increasing by 13% each year. Diagram in Figure 1 indicates that 93.40% of the CVE vulnerabilities are rising from software developments.

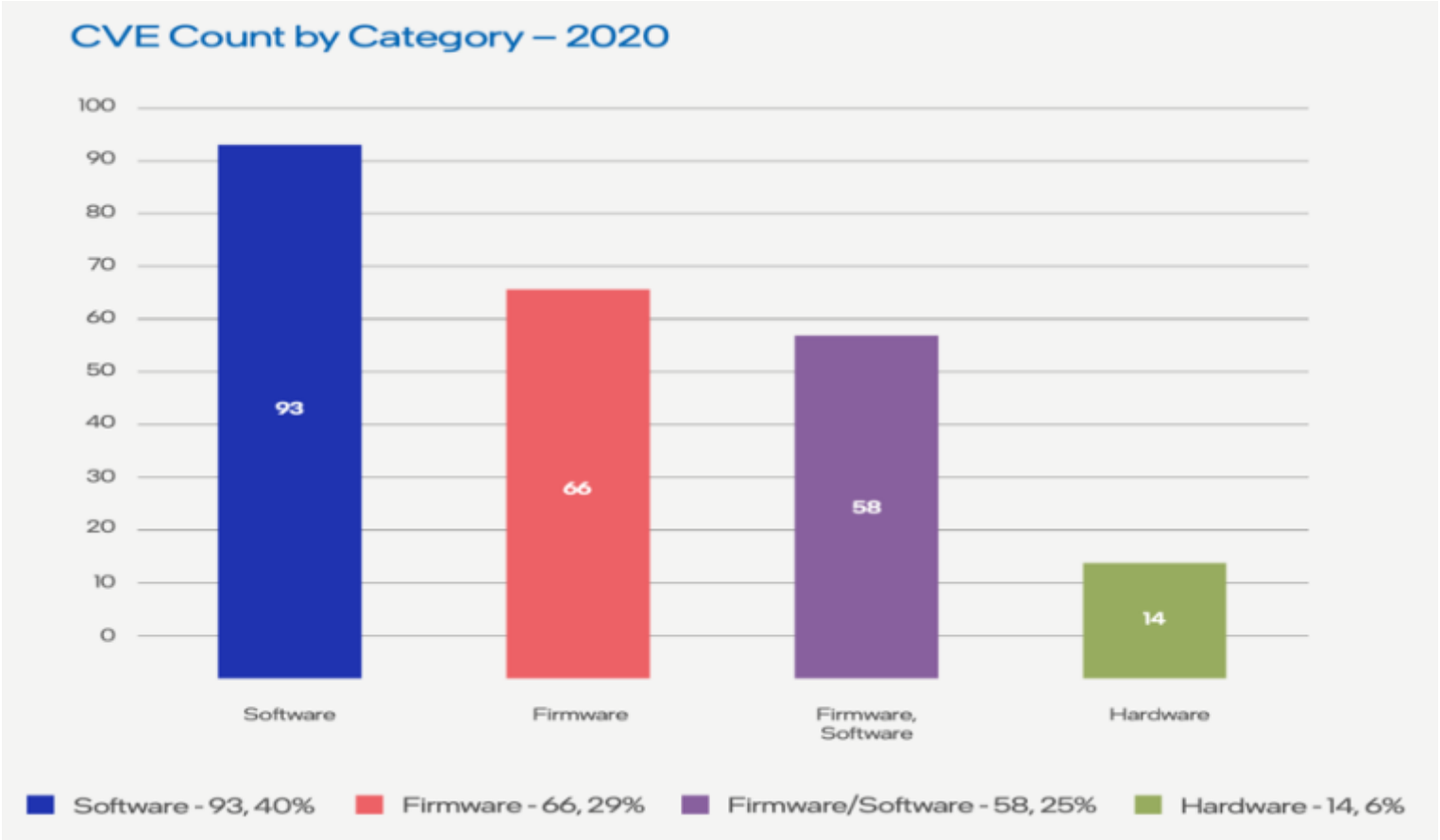

Figure 1 CVE Count by Category (Intel, 2020)

Rowe and Ekstrom (2011) in their work demonstrated the need for security programs in the curriculum. They have suggested that IT curriculums in higher education should include the latest cybersecurity standards report and analysis techniques that help students for real cyber world vulnerabilities.

### 1.2 Identify the need

Some other aspects of cybersecurity, such as cryptography, web security, and the basic software development security metrics, are provided in the computer science undergraduate curriculum. Students will learn the basic concepts and origins of malware and cyber-attack techniques throughout those courses. However, these courses do not demonstrate the importance of source code vulnerability analysis, one of the earliest vital solid foundations of the cyber world. The existing courses do not provide students with practical exposure and sufficient tools/techniques during the teaching process. Therefore, once IT students graduated, they do not have enough skills to relate the core topics of the courses to an actual experience and perform static analysis to avoid vulnerabilities (White and Nordstrom, 1996). Irvine and Chin (1998) in their board suggest that there must be an integration of security into existing programming classes for computer science students to help them understand security implications in software/program development.

Most research projects and related works like Chen and Lin (2007) are focused on educational approaches for real-world problems by reading and understanding the underlying concepts without experimenting with any interactive tools. There are also some suggestions on gaming approaches to spread essential cybersecurity awareness, but that also cannot prepare students for higher-level risk in computer systems and services (Alotaibi et al., 2016). Even though all the mentioned approaches include a theoretical training aspect to them but the essential part of developing training which incorporates practical and tactical skills along with critical thinking, problem-solving, as well as using existing tools to combat industry-level threats and mitigate software vulnerabilities, is being missed. Othmane et al. (2013), Chen and Lin (2007), and Boleng and Schweitzer (2010) support that experiential learning techniques such as virtual labs outside classroom learning activities and certifications improve the interactive learning for students to gain necessary knowledge and experience. Our business lives are more accessible with software applications as we often use software applications like managing databases, accounts, billing, payroll, and more. We use the software in healthcare, banking, general communication, and more.

Figure 2 displays expected exponential growth in the software development industry. Valued at over $533 billion, the software industry is projected to reach a market value of $773 billion by 2025. Therefore, due to this high demand and intense competition of cybersecurity talent, it is of high importance that educational institutions include core security programs for software vulnerability analysis in their curriculum.

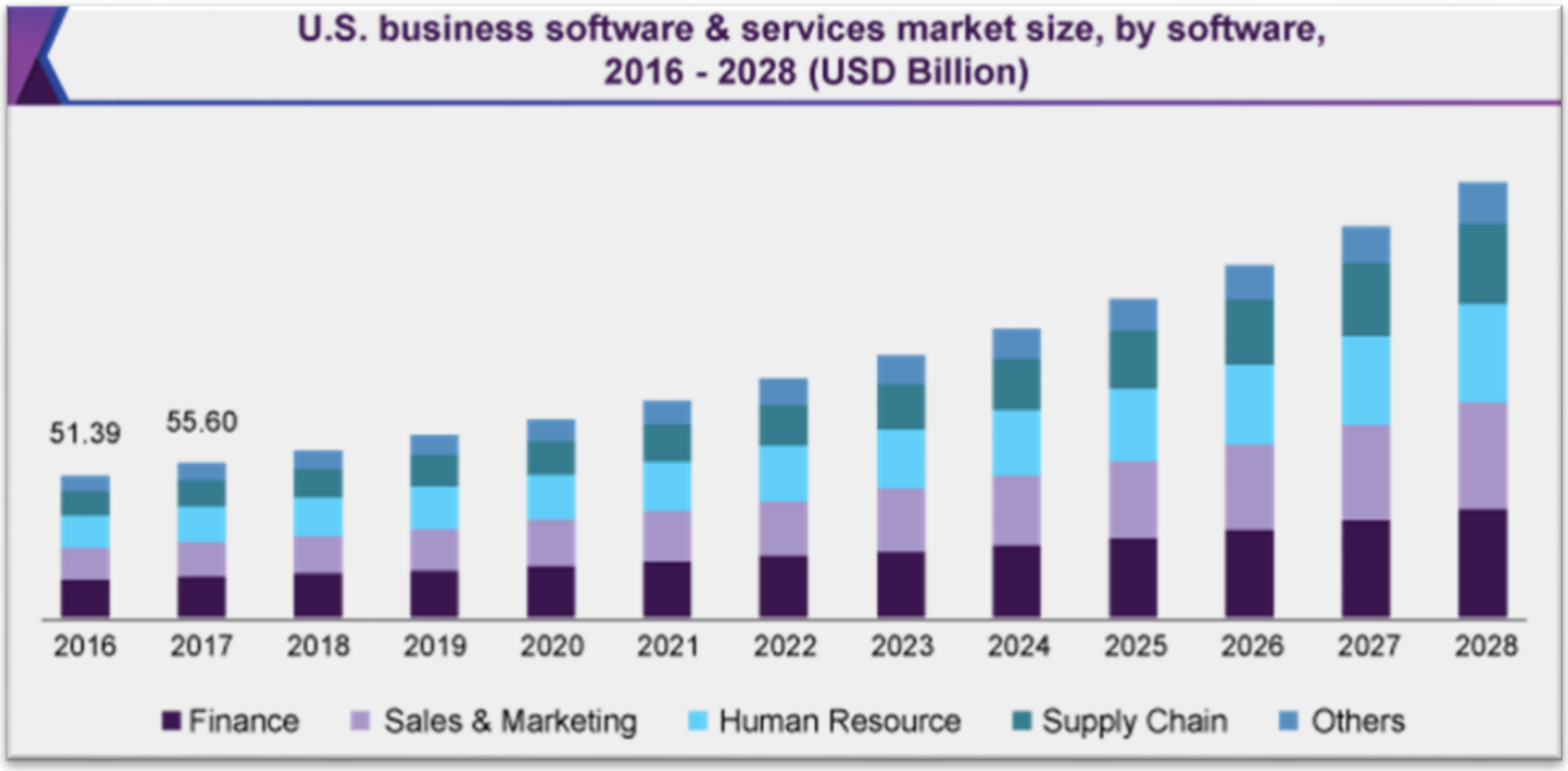

Figure 2 Software Development Growth Rate (Ming, 2020)

## 1.3 NLP Application in Vulnerability Detection

The ideal situation for source code vulnerability analysis and detection is an automated way that is precise and significantly fast, and accurate for code inspection. Such approach has to recognize vulnerability patterns without identifying subjective features manually to adapt to new challenges automatically.

By reducing human expert requirements, such tool can eliminate the need and resources for vulnerability detection and promote a more accurate prediction in vulnerability detection for software developers. With the ever-growing rise of cyber threats and bad actors, it's more important than ever for companies to stay compliant. Artificial intelligence (AI) can address this problem, yet even this approach also has its caveats. Even with many AI-powered cybersecurity solutions, they require human intelligence; and are not automated at their core. NLP is categorized as a subset of Machine Learning (ML) and has excellent applications for cybersecurity professionals seeking to improve their compliance processes continuously (Peacock, 2021). Figure 3 demonstrates the architecture of NLP application for source code analysis.

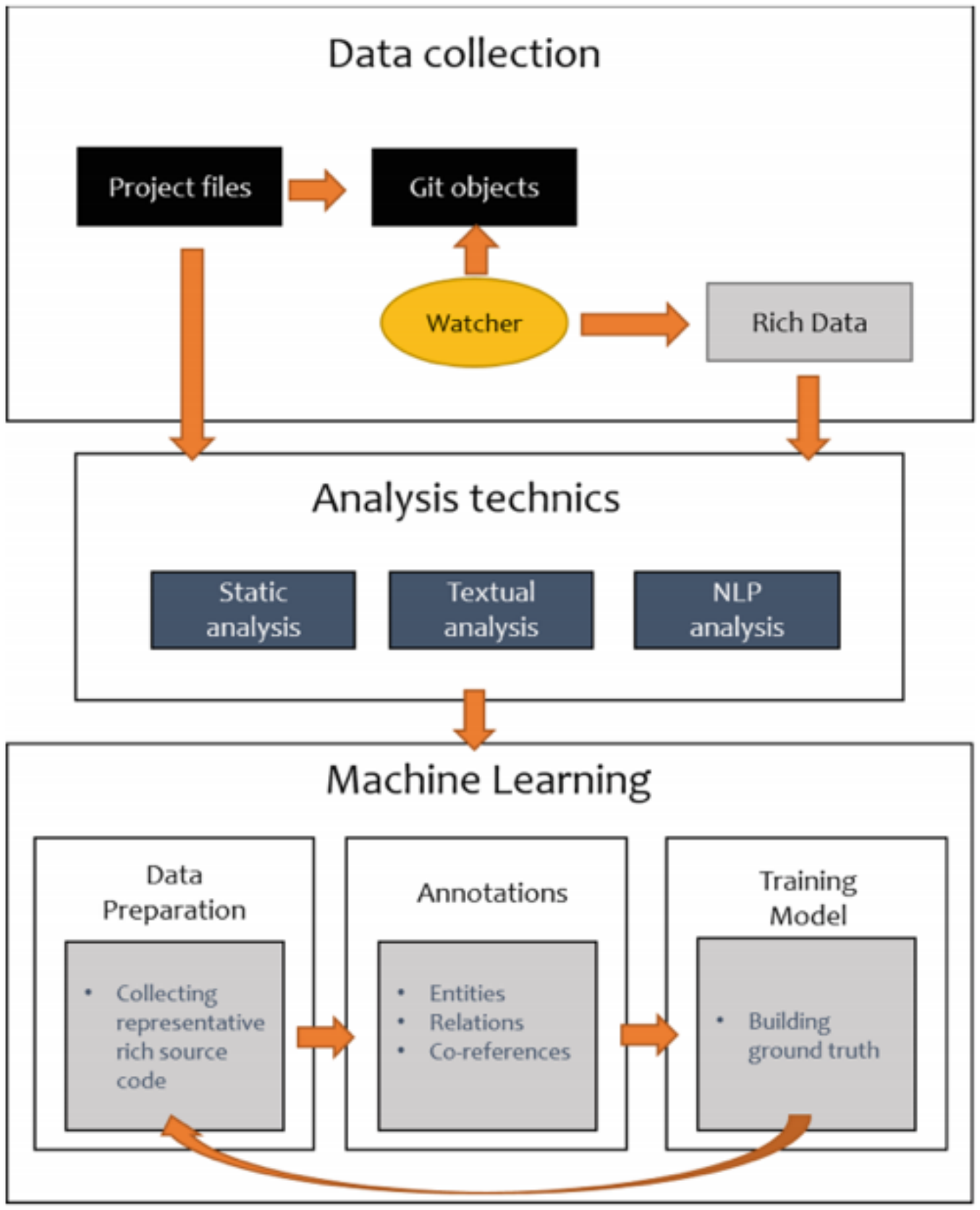

Figure 3 Identify Learner Attitude from Source Code (Itahriouan, 2021)

As the branch of AI-based deep learning that deals with the interaction between humans and computers using natural everyday language, NLP offers a wealth of capabilities to augment human ability. Given those interactions between computers and human (natural) languages are based on language processing, it became increasingly essential to upskill computers in processing and analyzing large amounts of language data. Thus, the role of Natural Language Processing in the humans-machines interaction, and consequently also in the cybersecurity world, starts to gain traction (Masernet, 2021). For example, NLP in risk and compliance can identify overlaps in standards and frameworks and data from an organization's tech stack and threat feeds to identify vulnerabilities in your security infrastructure. NLP's ultimate objective is to "read," decipher, and understand language that's valuable to the end-user (Peacock, 2020). Figure 4 demonstrates applications of NLP in Cybersecurity domains.

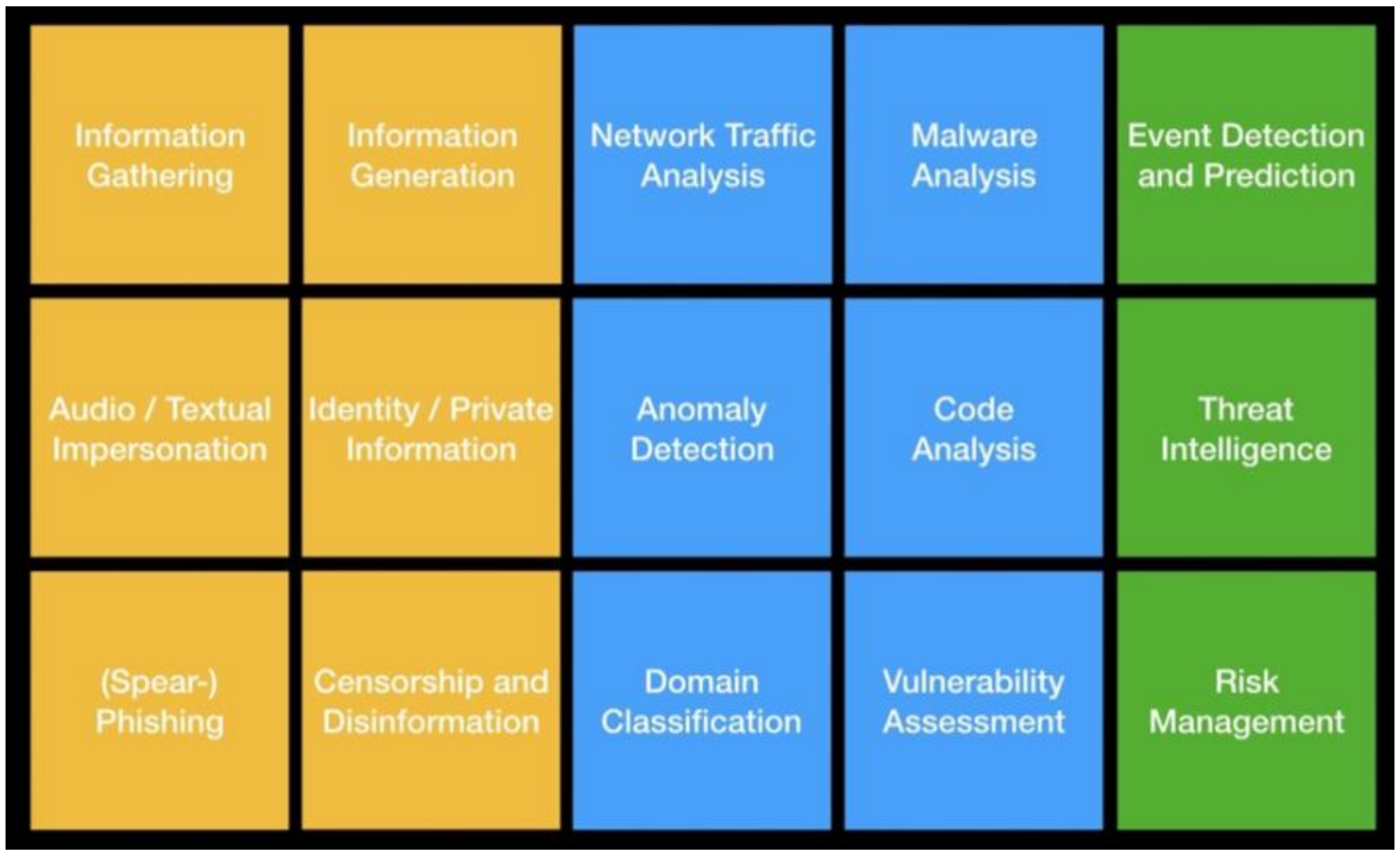

Figure 4 Role and applications of NLP in Cybersecurity (Ursachi, 2019)

The domains that are demonstrated in yellow color represent those applications that use automated information gathering and generation of NLP to mitigate attacks or propose defense strategies in human and machine interactions. The domains demonstrated in blue color represent the application of NLP, which leverages thread actors and attacks against a machine to strengthen their defensive posture. The fields shown in green represent the applications supporting the automation of course security operations and compliance activities. Village (2021) demonstrated that natural language processing could be applied for network traffic analysis and protection of web applications. Reutov and Sakharov (2018) in their work have provided a Natural Language Processing based approach to detect malicious HTTP requests, which is an extension to seq2seq-web-attack-detection.

Different techniques have been applied for representing code and creating a model for static analysis using pattern recognition techniques of NLP. Still, some of them, like the bag of words, word2vec, and Glove combined with simple classification algorithms, cannot capture the sequential nature and semantic structure of source code, which is the traditional method used in static vulnerability analysis tools. There have been new approaches such as Richardson (2019) to improve such analysis using new contextual language representation models. Bert for source code/log files (CyBERT) proves the applicability and advantages of such approaches. This thesis examines the related work of each of these techniques and how developing a novel hands-on laboratory exploration in academia can significantly enhance undergraduate student development.

### 1.4 Approach

The goal of this project is to develop a series of hands-on labs that address the main techniques of source code static analysis using NLP and provide practical tools to educate IT professionals and equip them to address vulnerabilities in source code. In addition, this approach can help students systematically learn and comprehend the fundamental concepts of source code vulnerability analysis. Figure 5 demonstrates the activities that are involved with Active and Passive Methods of Learning.

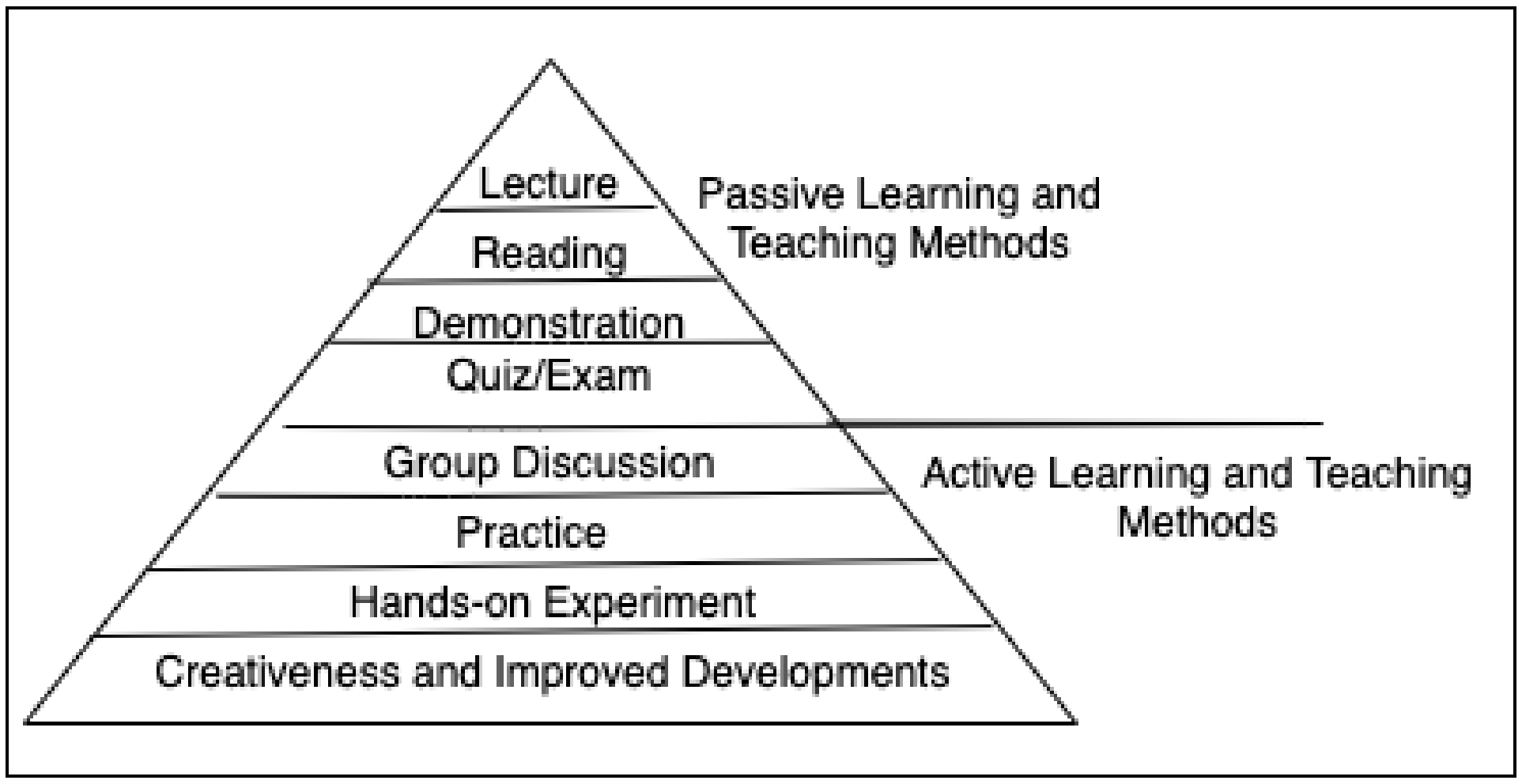

Figure 5 Activities Involved in Active & Passive Methods of Learning

An active learning model starts from a hands-on educational approach which involves group discussion practice and hands-on experiments to make students use the game knowledge in coursework and help them towards creativity to have a more secure and improved development. This project presents the design for a specific hands-on lab application to help students understand the fundamental concepts of source code vulnerability. This project will further explain the usage of natural language processing techniques to recognize vulnerability patterns in source code. This will allow students to perform the most effective static analysis that mitigates any threats due to bad programming habits in the early stages of development. The labs are built based on different static analysis techniques which leverage machine learning and natural language processing techniques with:

1. Sequence (Distributed Representations) program understanding model: Converts the source code into a sequence in a particular order, including character, token, and API → It retains native information.
2. Structure-based program understanding model: Includes Abstract Syntax Tree (AST), Control Flow Graph (CFG), Program Dependence Graph (PDG), and Code Property Graphs (CPG)

To enhance the students' ability to understand and mitigate source code vulnerabilities. This integrated approach exposes the students to the cost of the risk involved in each application cycle. Figure 6 presents the Secure Coding  Education (SeCodEd) framework consisting of the following elements.

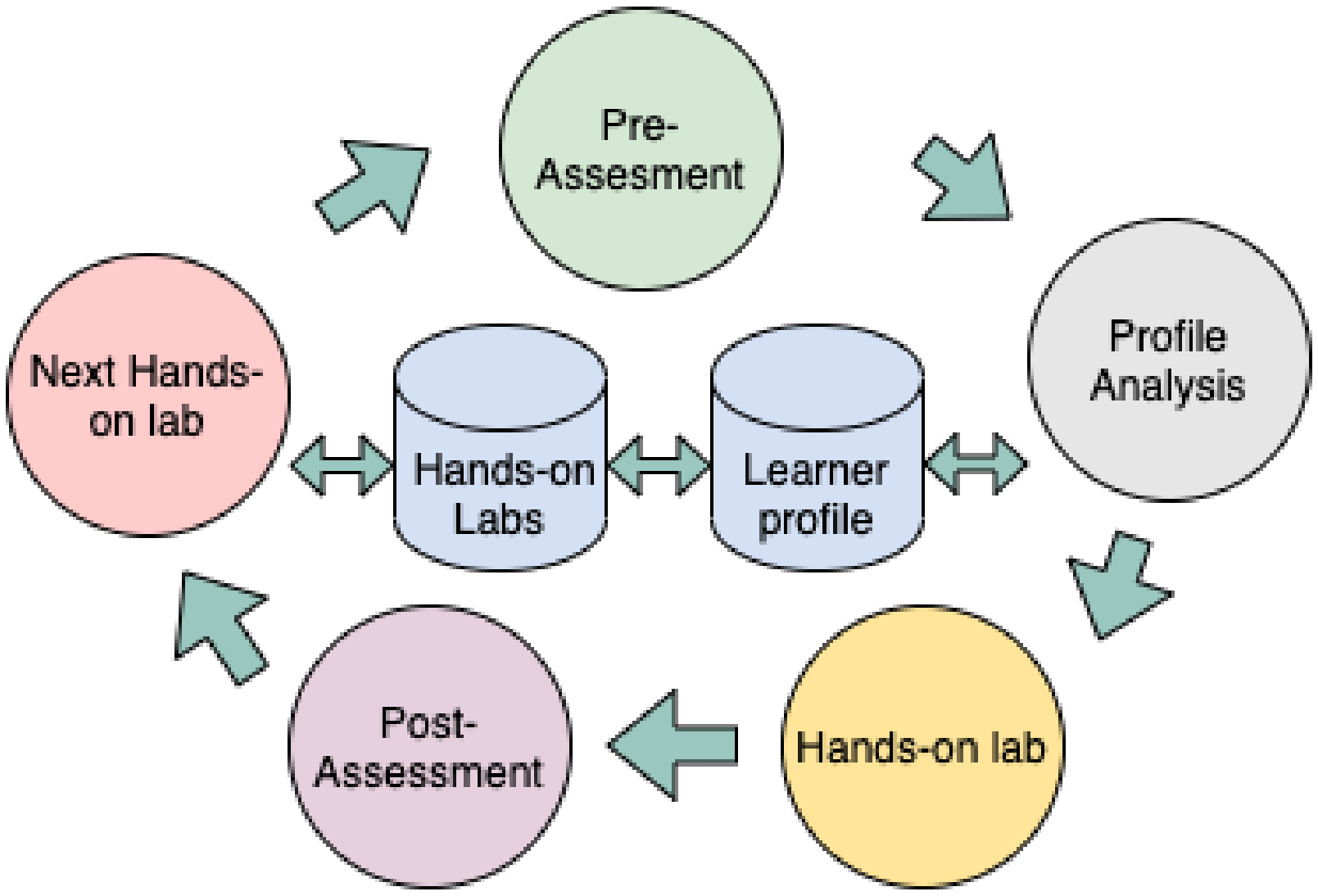

Figure 4 Source Code Vulnerability education framework
(SeCodEd)

- The hands-on labs will be shared via Piazza and Code Ocean. Code Ocean has been chosen because it allows users to have a Jupiter notebook up and running on the cloud to run the code for each technique used and described in the lab.
- Each hands-on module will have a scientific paper with implementations and techniques along with the code and the dataset used for each task.
- A data analytics component includes an analysis of the preassessment to determine which module students receive based on the following criteria: academics, age, and skill set to provide the most practical module.
- Following the preassessment, students receive a hands-on module to complete.
- Next, students complete a post-assessment to determine whether they have met the core objectives.

- A data analytics component analyzes the student's post-assessment to determine the next module
- Last, students continue to the next hands-on module.

### 1.5 Contribution

This research covers the concepts of source code vulnerability detection. As the number of open-source communities grows and companies use the source code to implement any software development, the importance of source code vulnerability analysis and recognizing vulnerability patterns rise too. Using new technologies such as machine learning, deep learning, and natural language processing, many future IT professionals are unaware of how lousy programming habits and not having a structured syntax for source code can lead to vulnerabilities. future IT professionals are unaware of the higher-level damages that such exposures can cause to any company or individual. This research aims to develop an innovative application tool that can introduce future IT professionals to the nuances of static analysis tools and source code vulnerability detection using NLP. Figure 7 demonstrates the importance and effects of such an educational model.

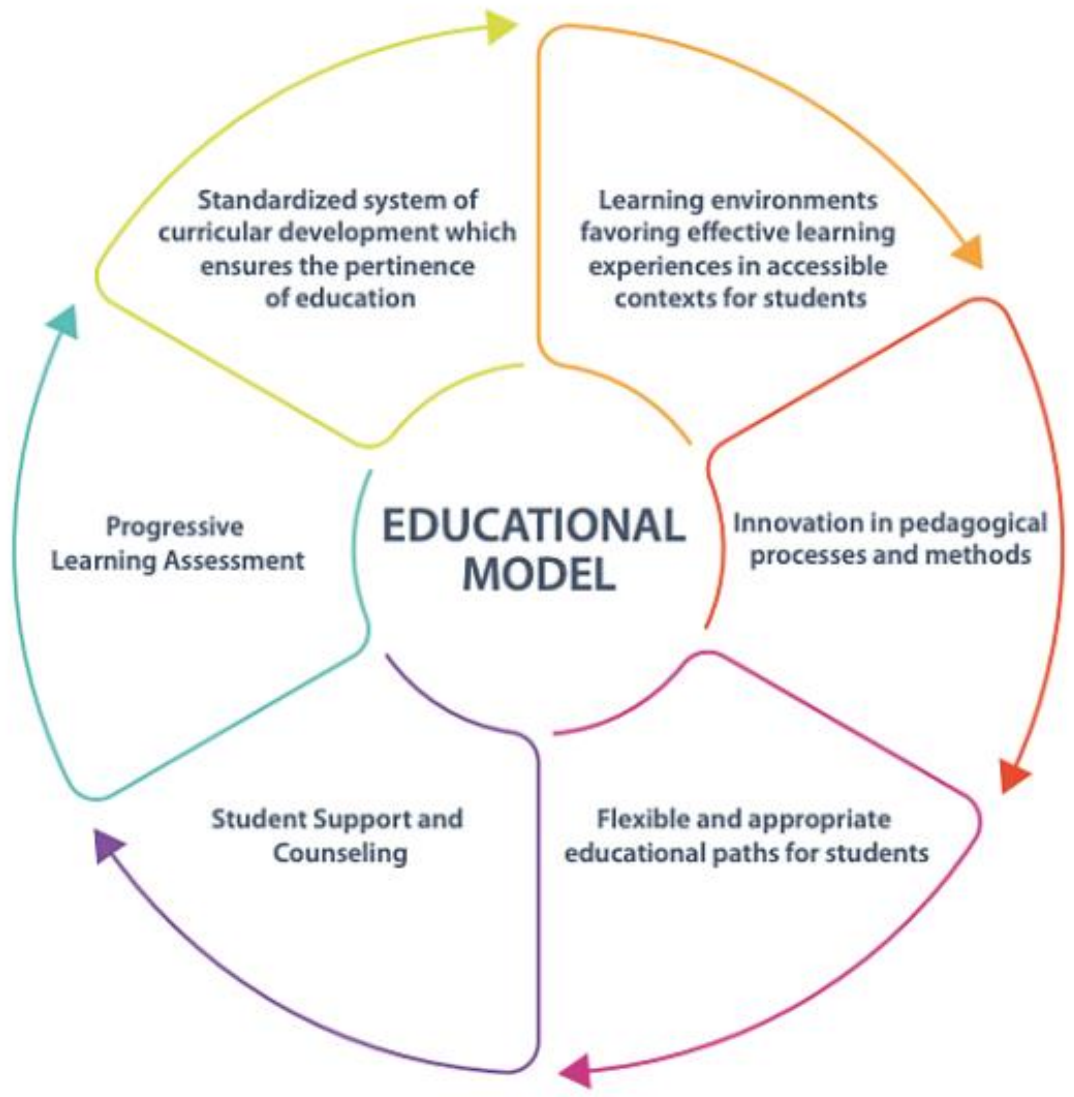

Figure 5 Educational Model (IPLACEX, 2021)

The following list provides detailed information on the contributions of this work:

1. Improve future IT professionals' skills and awareness of source code vulnerabilities by designing a set of hands-on labs that leverages deep learning and static analysis tools to further expand the usage of them for log file parsing.
2. Integrating concepts of source code vulnerabilities by learning modules:
    a. Function, API, and Library level
    b. Bad programming habits and practices
3. Identify Learner Attitude from unsecured Source Code styles to develop secure programming habits.
4. Integrating deep learning static analysis tools into secure hands-on labs by expanding the usage of such tools and techniques for log file parsing.
5. Create a set of hands-on labs by adopting different language representation models (Non-Contextual & Contextual) combined with different ML techniques to perform static analysis.

### 1.6 Thesis Organization

The thesis consists of 6 chapters. Chapter 1 provides a high-level overview of the proposed research, establishing the importance and application of the project in the world today, identifying the motive for the research project, proposed approach, and the contributions made towards the project.

Chapter 2 discusses the concepts used in this thesis, including Natural language processing, Static analysis tools, discussing the architecture behind each static analysis approach for source code vulnerability detection. The architecture behind performing a machine learning approach vs. a Lexicon-based approach was briefly addressed, highlighting each stage found in both architectures. The difference between the traditional ways of solving problems and the challenges faced during the machine learning/deep learning evaluation was also discussed.

Chapter 3 provides in-depth background knowledge and literature review on the concepts used in this thesis, including Natural language processing, Static Analysis tools, Source code vulnerability detection, and its different techniques. It also includes the study of the different contexts where to apply different algorithms. It compares different approaches and algorithms that have been used to perform this task by other researchers and discusses their advantages and disadvantages and provides the reasoning behind the chosen method of this thesis. Furthermore, this chapter provides a literature review on previous and related efforts on developing learning modules for source code vulnerability detection and secure programming habits.

Chapter 4 illustrates details about the design of the proposed framework and a comprehensive understanding of the application. This chapter also discusses the details on various types of vulnerabilities that have been covered in the hands-on labs and tools that have been leveraged to equip students with the proper set of skills.

Chapter 5 describes the implementation of the entire application in detail. In addition, it provides the results of this study, student Feedback on surveys, Success, measure rates, and the limitations and challenges faced during the development of this framework.

Chapter 6 discusses the study's conclusion by summarizing the contributions made to the project and avenues for further improvement of the project.

# Chapter 2 Background

Many techniques are available to help developers find bugs in their code, but none are perfect: an adversary needs only one to cause problems (CMU, 2020). This chapter will discuss how a branch of artificial intelligence called natural language processing, or NLP, is applied to computer code and cybersecurity. NLP is how machines extract information from naturally occurring language, such as written prose or transcribed speech. Using NLP, we can gain insight into the code we generate and find bugs that aren't visible to existing techniques.

## 2.1 Software vulnerability Analysis

Vulnerability's definition by NIST is a *weakness in an information system, system security procedures, internal controls, or implementation that could be exploited or triggered by a threat source* (NIST, 2021); NIST also provides another description of it as *specific flaws or oversights in a piece of software that allows attackers to do something malicious: expose or alter sensitive information, disrupt or destroy a system, or take control of a computer system or program' in a standard handbook on software security assessment* (Dowd et al., 2006). Thus, even though they occur less frequently, vulnerabilities could be considered a subset of defects. Shin and Williams' (2013) research demonstrated that roughly 21% of files in Mozilla Firefox have flaws, but only 3% have a security vulnerability.

One of the main concepts for discovering vulnerabilities is identifying *features* that describe vulnerable code and can distinguish it from non-vulnerable code. The three main approaches for analyzing source code vulnerability are **Static analysis**, which processes code without executing it, **Dynamic analysis**, in which code behavior is interpreted at runtime, and **Hybrid approaches** combining the two previous methods.

## 2.2 Dynamic Analysis

Dynamic code analysis techniques systematically analyze and monitor programs' execution and their traces at runtime (Shar et al., 2013). Therefore, dynamic analysis can easily miss problems related to executing and exploring all the possible inputs. Another practical problem with Dynamic analysis is that there needs to be a lot of computational time and resources dedicated to this type of analysis which will be difficult to provide when analyzing more significant software collections.

The hybrid analysis complements the dynamic approach with a static analysis tool that combines both approaches' advantages and disadvantages. Kim et al.'s (2016) work proposed a method for a hybrid approach which verifies the vulnerability in the hybrid analysis by implanting a fault based on the information received from static code analysis.

## 2.3 Static Analysis

Static analysis analyzes the form, structure, content, or documentation of a program without executing it by a generalization and abstract rules (Liu et al., 2012). Therefore, it is claimed that the cause for a vulnerability can be named by static approaches (Gupta et al., 2014). They can discover threats stemming from access control, information flow, and incorrect usage of APIs (*e.g.*, cryptographic libraries) with a wide range of different algorithms (Pistoia et al., 2007). Static analyses can also identify potential defects such as misused APIs, performance issues, deadlocks, and good/bad programming practices (Venkatasubramanyam et al., 2014).

There is a wide range of open-source tools available which perform Static code analysis to identify potential issues in source code. Such tools allow developers to fix any mistakes and avoid any vulnerabilities before the program is run for the first time and improved the code quality (Liu, 2018) (Gupta et al., 2014).

The performance of static analysis tools depends on the quality of the underlying patterns, abstractions, or rules used to identify problems. Meaning it is harder to define practices of secure software in a more complex. This brings a more critical issue: if a static tool is not familiarized with a particular vulnerability, it will be blind to this problem and never recognize it. Furthermore, creating such patterns *automatically* is a challenging task (Rolim et al., 2018). Also, building them manually is tedious and significantly time-consuming since technology and computer systems advance quickly. Therefore, keeping track of all possible kinds of vulnerabilities is complex. Therefore, there has to be an automatic system to perform such tasks (Ma et al., 2017).

### 2.4 Machine Learning (ML)

The booming of the open-source software community has made vast amounts of software code available, allowing machine learning and data mining techniques to exploit abundant patterns within software code. Remarkably, the recent breakthrough application of deep learning to speech recognition and machine translation has demonstrated the great potential of neural models' capability of understanding natural languages. This has motivated researchers in the software engineering and cybersecurity communities to apply deep learning for learning and understanding vulnerable code patterns and semantics indicative of the characteristics of vulnerable code (Lin et al., 2020). Machine Learning describes computational algorithms that allow computer systems to solve problems without explicit programming. It enables them to learn from experience and form concepts from examples by extracting patterns from the raw data. Machine Learning and Data Mining techniques have been used successfully in security. For example, Black et al. (2018) exploited deep learning methods to review source code based on code property graphs. They implemented their approach on public datasets Software Assurance Reference Dataset (SARD) of C/C++ command injection compared with current popular methods and proved a significant improvement on precision.

The performance of this model depends on the features that are going to be defined and remember the features there the vectors that we obtained previously after tokenization by combining different methods after the organization that will be further discussed in the literature review with varying methods of vectorization and also applying various machine learning algorithms and approaches, we can obtain better and more improved accuracy.

More details about the approach taken during this study will be provided in the literature review chapter.

**2.5 Natural Language Processing (NLP)**

The whole scope of this project is in the area of Natural language processing and machine learning. Natural Language Processing is a theoretically motivated range of computational techniques for analyzing and representing naturally occurring texts at one or more levels of linguistic analysis to achieve human-like language processing for a range of tasks or applications (Liddy, 2001). Part of speech is a category of words with similar grammatical properties that play similar roles within the grammatical structure of sentences (Wikipedia, 2020). Humans classify, interpret and relate each word in a sentence with others based on their position and meaning with a part of speech tagging and concludes an overall meaning of the sentence. Whereas a machine only understands numerical values. Figure 3 is demonstrating the interpretation of human vs. machine from text data. The machine has to be taught to assign a unique numerical value to each character in the text, understand the sequential structure of the words in the text, and relate the words based on their position. In other words, machine should understand part of speech tagging and think like a human. This task could be done either by developing my own NLP framework or using the preexisting frameworks, further discussed in the next chapter (Literature Review).

### 2.5.1 Tokenization

Before processing text data, which is essentially understanding natural language, we need to identify the words that create a sequence or string of characters. Tokenization is the first and most crucial step towards processing text data since it could quickly help the machine interpret the meaning behind the text by analyzing the words present in the document. For the computer to understand each character in a sentence or the entire document, the phrases, sentences, paragraphs, and essentially full document must be split up into smaller units with some sort of mechanism.

In this process, the available character sequence within a document unit needs to be chopped up into pieces such as individual words or terms. These pieces are then called tokens. The next thing that has to be taken care of during this process is the removal of punctuations since they don't have any meaning or weight in the polarity level of the text data. Several available organization tools will be further discussed in the literature review chapter.

### 2.5.2 Vectorization

As discussed in the NLP section of this chapter, the machine only understands numerical values. Hence, the available text document, after being tokenized, must be converted into numerical values. Feature extraction word embedding, or word vectorization is the methodology in an LP that enables the words or phrases that are available in a document or vocabulary to be mapped to a corresponding vector of- numbers which will be later used to find the semantics or the similarities between the words to give them a polarity value. This step converts the text data into matrixes that facilitate text classification document clustering and computation of similar words. There are many available techniques for word vectorization that will be discussed in detail in the literature review.

## 2.6 Representing Code

When using machine learning to find vulnerabilities in a source code, there are different approaches to preprocess the code and represent it. Sections below will present some of the frequently used techniques in this field.

### 2.6.1 Abstract Syntax Trees

Abstract syntax trees (AST) are a hierarchical representation of any code or any changes to a code used to reduce the tokens. There have been several approaches in creating AST's, such as considering only API nodes and functions names and ignoring everything else (Russell et al., 2018). In addition, working on plain text to mine patterns, although very challenging, can be an efficient approach to create AST’s.

### 2.6.2 Code as Natural text

A source code can be viewed as a natural text, meaning it includes a sequence of words with natural language characteristics. Therefore, features of the source code text could be automatically extracted using natural language processing. Furthermore, because Source code is written by humans, their code, shares characteristics of natural language text such as representing specific structures, common patterns, long-term dependencies, and repetition.  Therefore, natural language processing models can be used on source code vulnerability analysis to generalize and uncover patterns of vulnerabilities. NLP has demonstrated that in cases of facing source codes that contain exposures of a similar pattern that occur very often, for instance, missing a check before a function call, it can have good recognition and prediction of a vulnerability.

### 2.7 Datasets

Data is the most crucial part of all data analytics, machine learning, artificial intelligence. Without data, no model can be trained, and all modern research and automation will not exist. Big companies are spending a tremendous amount of money on gathering as much specific data as possible so they would be able to improve their services. Machine learning helps to find patterns in data; these patterns will later be used to make predictions about new data points. To have an accurate prediction, the in-hand dataset has to be constructed, and the data has to be transformed correctly. The training data is used to make sure the machine recognizes patterns in the data, the cross-validation data is used to ensure better accuracy and efficiency of the algorithm used to train the machine, and the test data is used to see how well the machine can predict new answers based on its training. Figure 8 demonstrates the application of training and testing data sets.

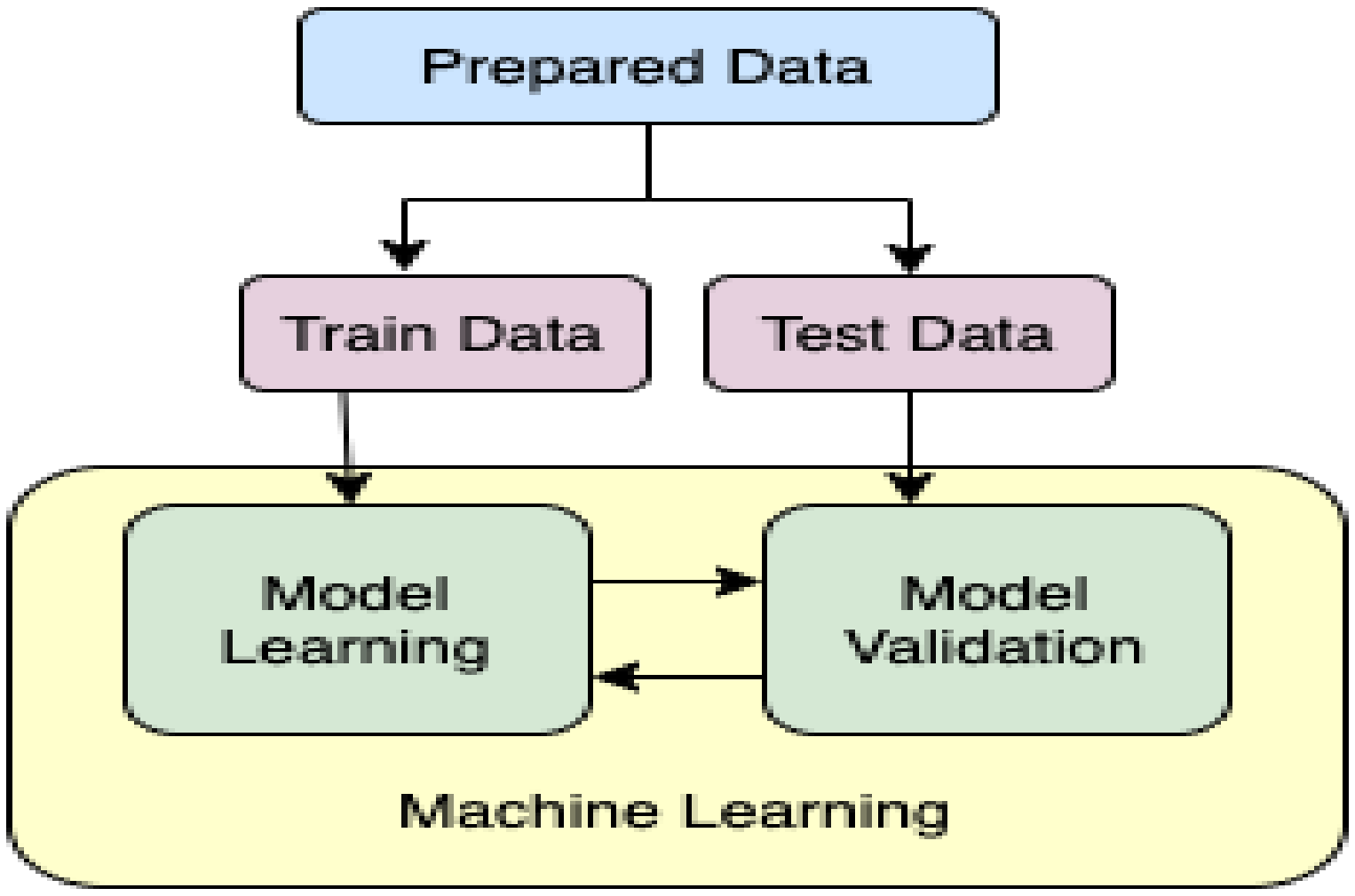

Figure 6 Training & Testing Dataset in Machine Learning

### 2.7.1 Training Data

The source code itself contains much valuable information. Fortunately, due to the emergence of giant online repositories and open-source software, it is comparatively easy to access large amounts of code that can then be analyzed somehow. In this case, the interest data is all information relating to code, bugs, or vulnerabilities. Several official bug databases, such as the Bugzilla bug database by Serrano and Cirodia (2020) or the Common Vulnerabilities and Exposures (CVE) database by Corporation (2021). More about the process of extracting the data from software repositories can be found in Chapter 4.

### 2.7.2 Testing Data

GitHub is one of the most significant resources for hosting source code, including a tremendous amount of proper source code to perform vulnerability analysis. most of the source codes shared on GitHub are developed by communities that work together and communicate throughout the source code to implement a project; therefore, GitHub can be a public resource for this work to use source codes that include natural language for writing a program. Zhou and Asankhaya (2017) demonstrated that it's a good practice to use source codes and commits from GitHub to detect vulnerabilities.

# Chapter 3 Literature Review

The emergence of computer technology and social networking services promoted open communication between all the users and provided a collaborative platform for discussions, feedbacks, and software development. Open source by providing publicly available source code enabled developers to modify and tweak OSS applications to enhance their capabilities. However, the availability after source code doesn't mean that such applications are immune from vulnerabilities. If a security hole emerges in an open-source product, the damage will be widely felt throughout all uses and reuse after source code.  Monitoring all these contents could be very beneficial yet very difficult. This chapter will Investigate the previous work in vulnerability detection/classification, different approaches and algorithms that have been used, and their advantages and disadvantages.

## 3.1 Natural Language Processing

This thesis is focused on natural language processing combined with artificial intelligence technologies. This section will cover different feature processing methods and evaluate other available classical machine learning models and their performances in previous research.

### 3.1.1 Definition and Distinction

Artificial intelligence is developing a machine or a computer system that can perform tasks usually done by humans and require intelligence and decision-making such as speech recognition translation between different languages and understanding the definition of a word or a term. According to Hurwitz and Kirsch (2018) there are four main subsets to AI. As shown in Figure 9, AI, by using both machine learning, natural language processing, reasoning, and planning, aims to simulate human intelligence in machines.

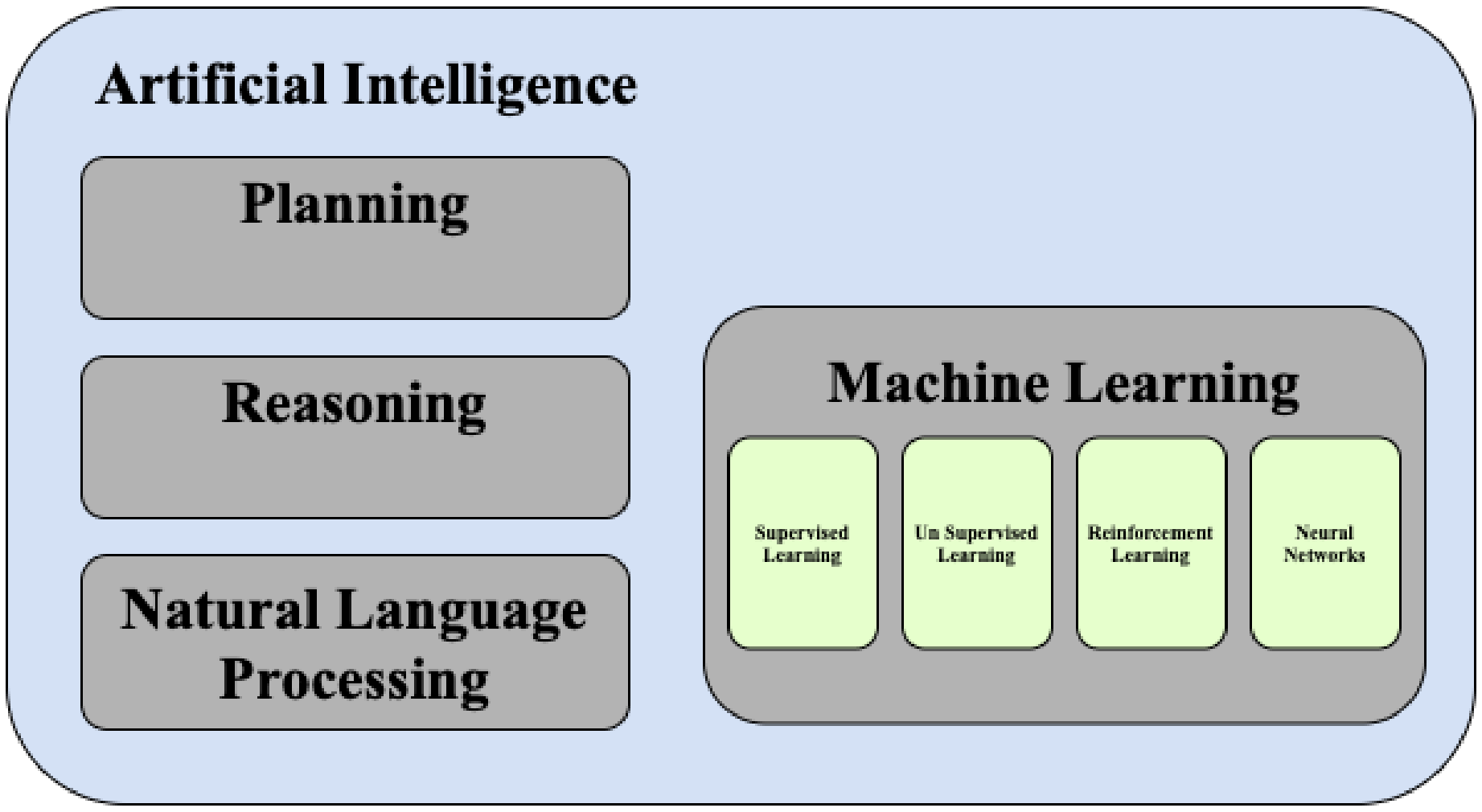

Figure 7 Subset Of AI

Machine learning's goal is to learn and adapt a model based on the training data that it will receive. Machine reasoning, however, makes inferences helps to fill in the gaps in incomplete data and connect them to make an inference based on them. Natural language processing is a combination of computational techniques to analyze and represent naturally occurring texts and human speech at one or more levels of linguistic analysis to interpret text and human spoken language (Liddy, 2001) and (Hurwitz and Kirsch, 2018). Machine planning is the intelligence system's ability to take a sequence of actions autonomously, which will help the system to adapt to its surrounding context and accomplish the goal it's trying to achieve. There are various applications to AI which indeed Refer to humans on structured natural language. Some natural language processing tasks are text classification and information retrieval speech recognition named entity recognition machine translation and dialogue systems (Maglogiannis et al., 2007).

### 3.1.2 General Design

By cross-referencing all the literature that has been reviewed for this study, it was observed that most of the machine learning studies fall into three main sections in terms of algorithm and design: Data Collection, Feature Extraction, and Classification. Data Collection includes the source that the data for both training and testing will be collected from, the process that will be performed to have clean and formatted data, splitting and scaling on training and testing data sets, and also how the transfer learning will be achieved to improve the intelligence of the model.

Feature Extraction involves extracting features of the text and formatting them so that the classifier can read them as input.

Classification is the process in which, using a probabilistic statistical algorithm, we will enable the machine to learn from the training data and then use the distribution probability to classify the polarity labels of the testing dataset.

### 3.2 Feature Extraction

A feature in machine learning is a measurable property, a piece of information that can be observed as a characteristic when solving a problem. Selecting relative, quantitative and qualitative features is a crucial step in model construction since it can simplify the model, shortening the training time, having a practical algorithm, and enhanced results. Ghaffarian and Shahriari (2017) in their review of Software Vulnerability Analysis and Discovery Using Machine Learning and Data Mining Techniques, they identified three major areas of features classification methods (Figure 10): vulnerability detection based on **software metrics**, **anomaly detection,** and **vulnerable code pattern recognition**. Some tools may solely use each of these techniques or combine all three of them like SCALe by Seacord et al. (2012) that includes all three of the methods applied to perform static analysis.

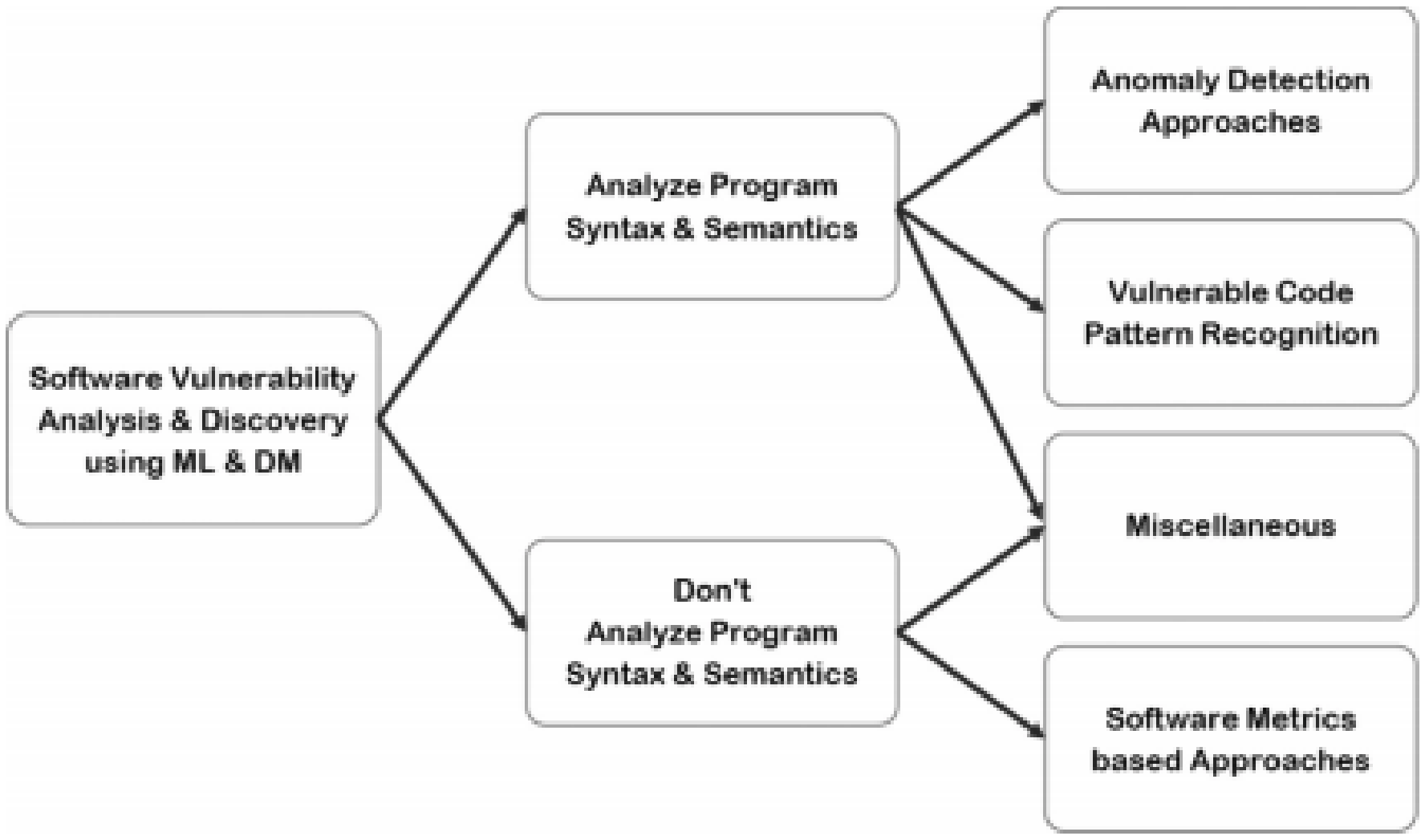

Figure 8 Feature Classification Methods in Vulnerability detection (Ghaffarian and Shahriari, 2017)

### 3.2.1 Software Metrics

Fault prediction models are computational models trained based on historical data gathered from software projects and provide a list of software artifacts that are more likely to contain faults to prioritize software testing efforts (Ghaffarian and Shahriari, 2017). Previously, the most commonly used features in fault prediction models were identified by standard software engineering metrics (software quality and reliability assurance) such as the size of the code, cyclomatic complexity, code, number of dependencies, and legacy metrics which are not the best features for a vulnerable source code (Patrick Morrison, 2015).

As an example, Nagappan et al. (2008) suggested their metric scheme, which includes metrics like number of ex-engineers, number of engineers, depth of master ownership, edit frequency, percentage of an organization contributing to development, overall organization ownership, level of organizational code ownership, and organization intersection factor in measuring organizational complexity. They state that in comparison to the traditional metrics like code churn, code coverage, code complexity, code dependencies, and pre-release defect measures, the organizational metrics perform better in predicting failure-proneness.

Using traditional metrics, two pieces of code can have the same metrics and, therefore, the same likelihood of vulnerabilities, but that is not necessarily true because they can have completely different behavior, leading to a different probability of vulnerabilities. Shin and Williams (2013) in their investigation of vulnerability prediction based on complexity and code churn metrics, using 18 complexity metrics and classification techniques received an 83% recall and 11% precision rate in a vulnerability prediction task. Yu et al. (2019) take many different possible features into account, including software metrics such as the number of subclasses or number of methods in a file, as well as crash features and code tokens with their tf-idf scores. Their approach is, therefore, a mix of many different angles. They predict vulnerabilities on the level of whole files and achieve very satisfying results in narrowing down the volume of code that human experts must inspect to find a vulnerability.

Other researchers have been able to make predictions just with commit messages. Sharma et al. (2017) leverage a K-fold stacking algorithm to analyze commit messages to predict whether a commit contains vulnerabilities, reportedly with great success. In contrast, Russell et al. (2018) found that both humans and Machine Learning algorithms performed poorly at predicting build failures or bugs just from commit messages. The proposed approach, VUDENC, does not take external code metrics into account but learns features directly from the source code itself.

Williams et al. (2013) claims that the traditional method does not capture the semantics of the actual source code, the program behavior, or the data flow. Therefore, the conventional approaches can provide insights into the software security metrics but are not ideal predictors for security vulnerabilities in source code.

### 3.2.2 Anomaly Detection

Anomaly detection refers to the study of finding unnormal and unexpected behavior or patterns, often referred to as anomalies or outliers (Chandola et al., 2007). Anomaly detection in software quality assurance aims to find:

1. Segments of source code that do not conform to the usual or expected code patterns for application programming interfaces
2. Neglected conditions or missing checks.

This approach will automatically extract specifications, rules, and patterns used as the basics of detecting deviant behaviors. Fabian et al. (2015) proposes a system named Chucky to automatically detect missing checks in source code to assist manual code audit. Chucky combines machine-learning techniques with static program analysis to determine missing checks. The authors discriminate two types of security checks in source code: (1) Checks implementing security logic (*e.g.,* Access Control); and (2) Checks to ensure secure API usage (*e.g.*, checking buffer size). Anomaly detection approaches are only effectively applicable for mature software systems. Most of the previous approaches had a high false-positive rate which indicates that such systems are not reliable. There needs to be further progress for this approach in the field of vulnerability discovery. Some notable works suffering from high false-positive rates include (Zhou et al., 2015) and (Andrzej et al., 2007)

### 3.2.3 Pattern Recognition

The pattern recognition approach for vulnerability detection utilizes data mining and machine learning techniques to extract features and patterns of vulnerable code segments automatically via pattern matching techniques. Similarly, instead of extracting models and rules of normality compared to the previous approach, the pattern recognition technique aims to extract models and vulnerable code patterns. One of the best applications of this approach is when we have identical or nearly identical code clones which the code fragments and their inherent structure is very similar, which is the case that happens in the open-source community through code sharing (Li et al., 2018). This pattern analysis process involves gathering a large set of data processing to extract feature vectors and apply machine learning algorithms to classify vulnerable from non-vulnerable source codes (Ghaffarian and Shahriari, 2017). In many cases, those approaches also rely on a very coarse granularity, classifying whole programs in Grieco et al. (2016) work, files in Shin et al. (2010), components in Neuhaus et al., (2007), or functions as in Yamaguchi et al. (2012), which makes it impossible to pin down the exact location of a vulnerability. Some, like Zhou and Yuanyuan (2005) and Russell et al. (2018), use a more fine-grained representation of the code. Furthermore, the approaches differ in many aspects: the language used, the source of the data (real-life projects or synthetic databases) and the size of the dataset, the creation process for labels, the granularity level of the analysis (whole files down to code tokens), the machine learning model that was used, the examined types of vulnerabilities, and whether the model is usable in cross-project predictions or just on the project it was trained on.

### 3.3 Program Understanding Model

Be basic of automatic vulnerability detection is the program understanding model. There are two main approaches which are **sequence-based** and **structure-based** models.

#### 3.3.1 Sequence-Based Model

The sequential understanding model retains native information meaning, it converts source code into a sequence with a specific order, including any character talking or API. The structured program understanding model includes different techniques such as:

- Abstract Syntax Tree (AST): A method that represents the syntax structure of a source code in a format of a tree.
- Control Flow Graph (CFG): A technique that considers code with a sequential relationship as a basic black and concatenates the blocks into an ordered graph based on their control dependency
- Program Dependency Graph (PDG): A technique that adds data dependency and control dependency on the nodes of CFG, which it's estimates statements that affect sensitive operations can be accurately and efficiently identified
- Cold Property Graph (CPG): A technique that combines the previous methods to provide a piece of more detailed information that is accurate but compromises the prediction efficiency due to a high load of data analysis

### 3.3.2 Structure-Based Model

A source code can be viewed as a natural text, meaning it includes a sequence of words with natural language characteristics. Because Humans write source code, their code shares characteristics of natural language text such as representing specific structures, common patterns, long-term dependencies, and repetition. Therefore, features of the source code text could be automatically extracted using natural language processing.

Therefore, natural language processing models can be used on source code vulnerability analysis to generalize and uncover patterns of vulnerabilities. NLP has demonstrated that in cases of facing source codes that contain vulnerabilities of a similar pattern that occur very often, for instance, missing a check before a function call, it can have good recognition and prediction of a vulnerability.

### 3.4 Representation Methods

Source codes can be vectorized and represented to the machine using different natural language processing and vectorization techniques. The most common techniques used to express a natural language are categorized into context-free and contextual groups. The most recent classification technique is converting each of the words of text data into word embeddings, a numerical representation of terms. These vector representations capture the underlying words in relation to the other words present in a sentence. By this method, words with similar meanings will be grouped closer together in a hyperplane that distinguishes them from the position further in the hyperplane. This step could be done by creating a corpus-specific word embedding based and trained on the features introduced from the training data set, or it could be done using a pre-trained word embedding.  The problem with the embedding layer is that its features will be relevant for only this specific problem. There won't be enough words introduced to map the encoded categorical variables to vectors of floating-point numbers. There are a few pre-trained words embeddings available, and the top three popular of them are Google's Word2Vec, Stanford's GloVe, and Google's BERT, each of which will be explained below.

#### 3.4.1 Bag of Words (Bag of N-grams)

**Tokenization** For processing written natural language data, each text message has to be split into smaller or basic units [Tokens], which need not be decomposed in subsequent processing [Tokenization] (Webster et al., 1992). This process facilitates computers to distinguish single text entities that usually represent simple words (the minor independent units of natural language). Although there are a few downsides to this approach, due to the lack of identifying words that semantically belong together, a simple word tokenizer can be realized in many languages by splitting the text at the occurrences of space symbols (Webster et al., 1992). For example, by using tokens, n-grams could be created, which indicate a token set with the length of “n”. Figure 11 demonstrates an example of a tokenized sentence and stop words indicated.

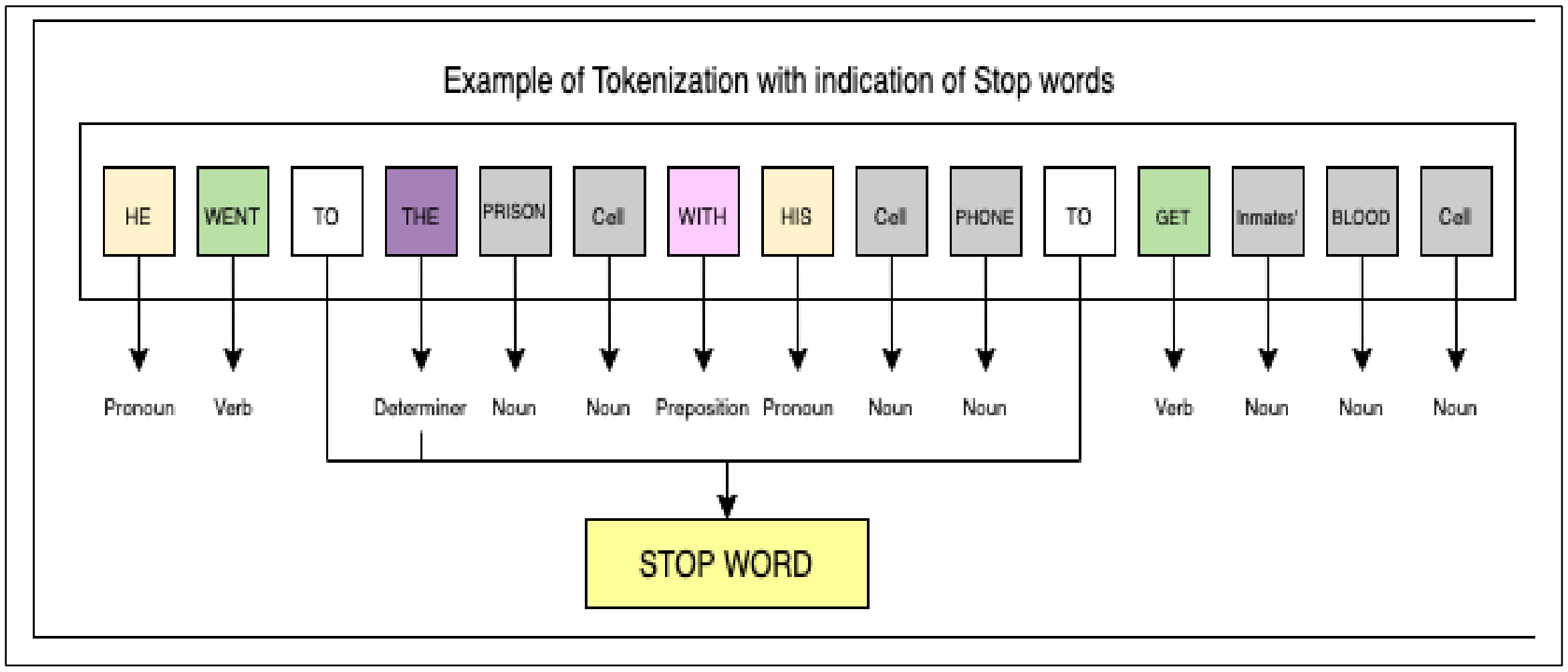

Figure 9 Tokenized sentence with Stop words indicated

**N-gram** means a sequence of N words. Each token is an example of a unigram. This analysis could be performed at both character and word levels. A unigram is of size 1 and consists of one token or word. These tokens, however, could be used to create a bigram or trigram. Using these n-grams and the probabilities of the occurrences of certain words in specific sequences could improve the predictions of the overall weight and polarity of a sentence (Brown et al., 1992). Figure 12 is an example of a punishment tokenized using the bigram and trigram method.

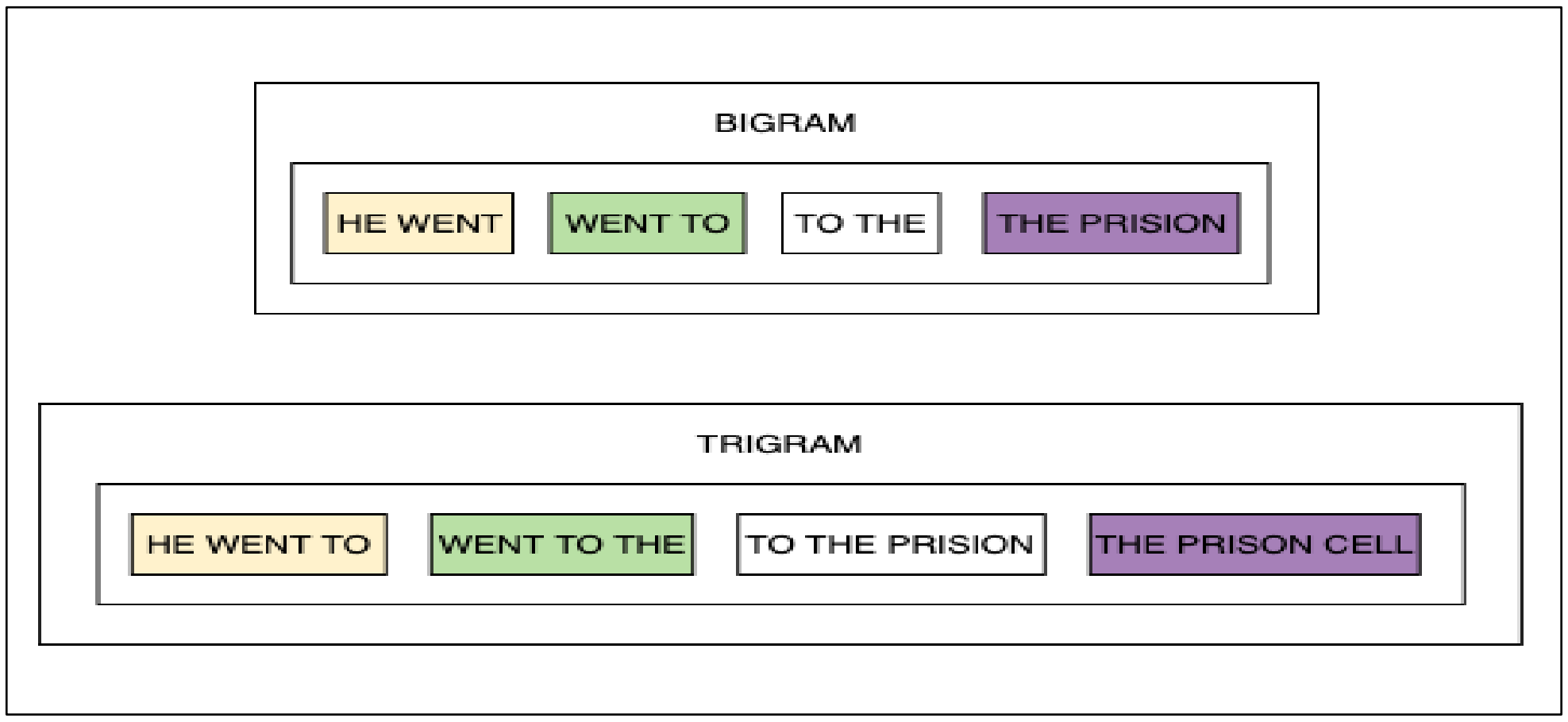

Figure 10 Example of Bigram and Trigram Method

Rout et al. (2019) in their research of performing a sentiment and emotion analysis of unstructured social media text, the accuracy obtained using existing Bag of words libraries varies and is limited by their domain since they contain many words and the number of ties each word has appeared. Still, the order of the tokens is not taken into account. N-grams help discovers sequential patterns in code. However, they suffer from problems in their performance when dealing with high dimensionality, and therefore the context is mostly only very few code tokens (Dam et al. (2016).

### 3.4.2 Word2Vec

Mikolov et al. (2013) introduced the word2vec technique, which obtains word vectors by training text corpus. Word2Vec is a two-layer neural net that processes text by "vectorizing" words using a neural probabilistic language model (Bengio et al., 2003). The idea originated from the distributed representation of words, its input is a text corpus, and its output is a set of feature vectors representing words in that corpus. By feeding enough information, Word2vec can make highly accurate guesses about a word's meaning based on its past appearances and establish an association for each word with others to cluster them by topic in a spatial position. Therefore, as shown in Figure 13 mathematically, the cosine of the angle between such vectors (Equation 1) should be close to 1, *i.e.*, an angle close to 0.

$$Similarity\ of\ (A.B) = Cos(\theta) = \frac{A.B}{||A||B||}$$

Equation 1 Cosine Similarity

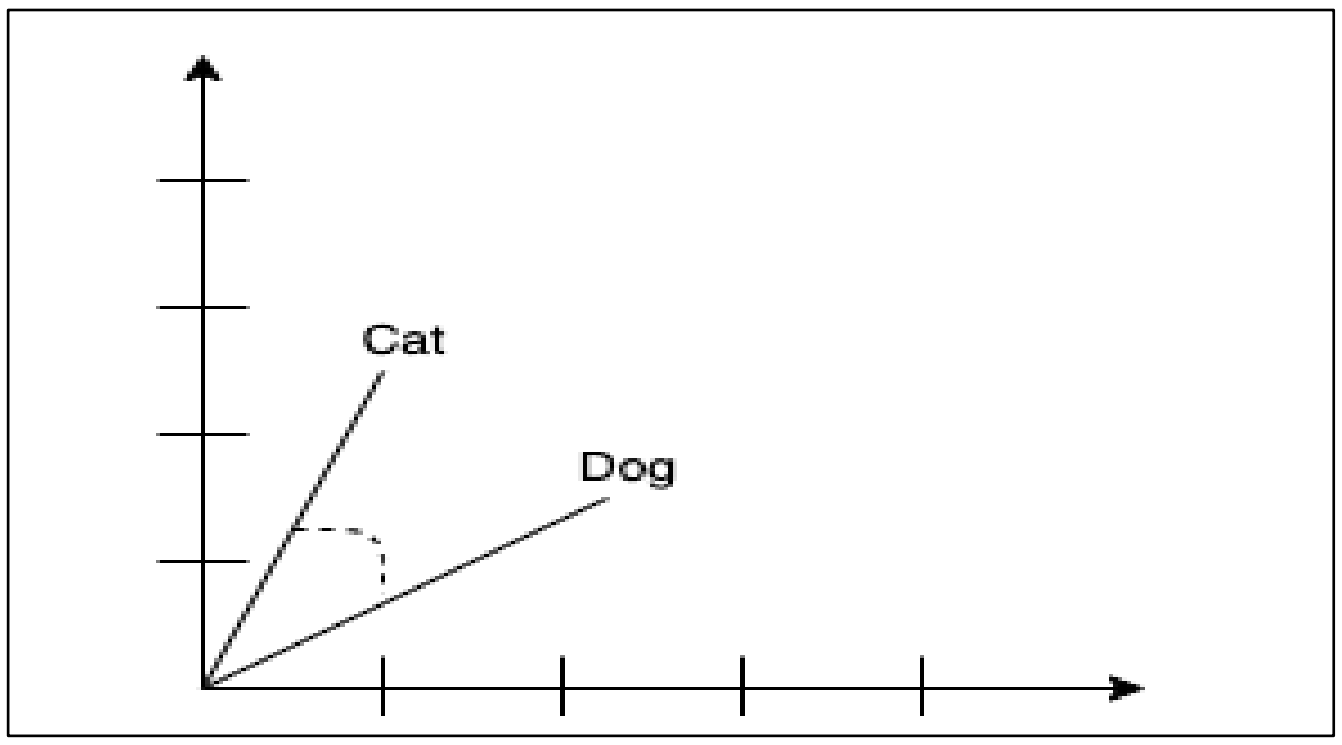

Figure 11 Cosine Similarity

The output of the Word2vec neural net is a vocabulary in which each item has a vector attached to it, which are trained by optimizing loss functions with gradient descent and could be fed into a deep-learning net or simply queried to detect relationships between words. The loss function in this method is calculated by measuring how well a particular word can predict its surroundings words. Those clusters can form the basis of search, vulnerability analysis, and recommendations in many diverse fields. The effectiveness of word embedding representations in vulnerability analysis tasks has been proven by many researchers (Tang et al., 2014). There are two main word2vec models: Continuous Bag of Words (CBOW) and Skip-Gram (Meyer, 2016). In the CBOW model, a word will be predicted given a context (like a sentence), whereas, in Skip-Gram, the context will be predicted given an input word. Figure 14 is demonstrating the difference between CBOW and Skip-gram models. The only downside to the word2vec is that it only takes local contexts into account and do not look at global count statistics

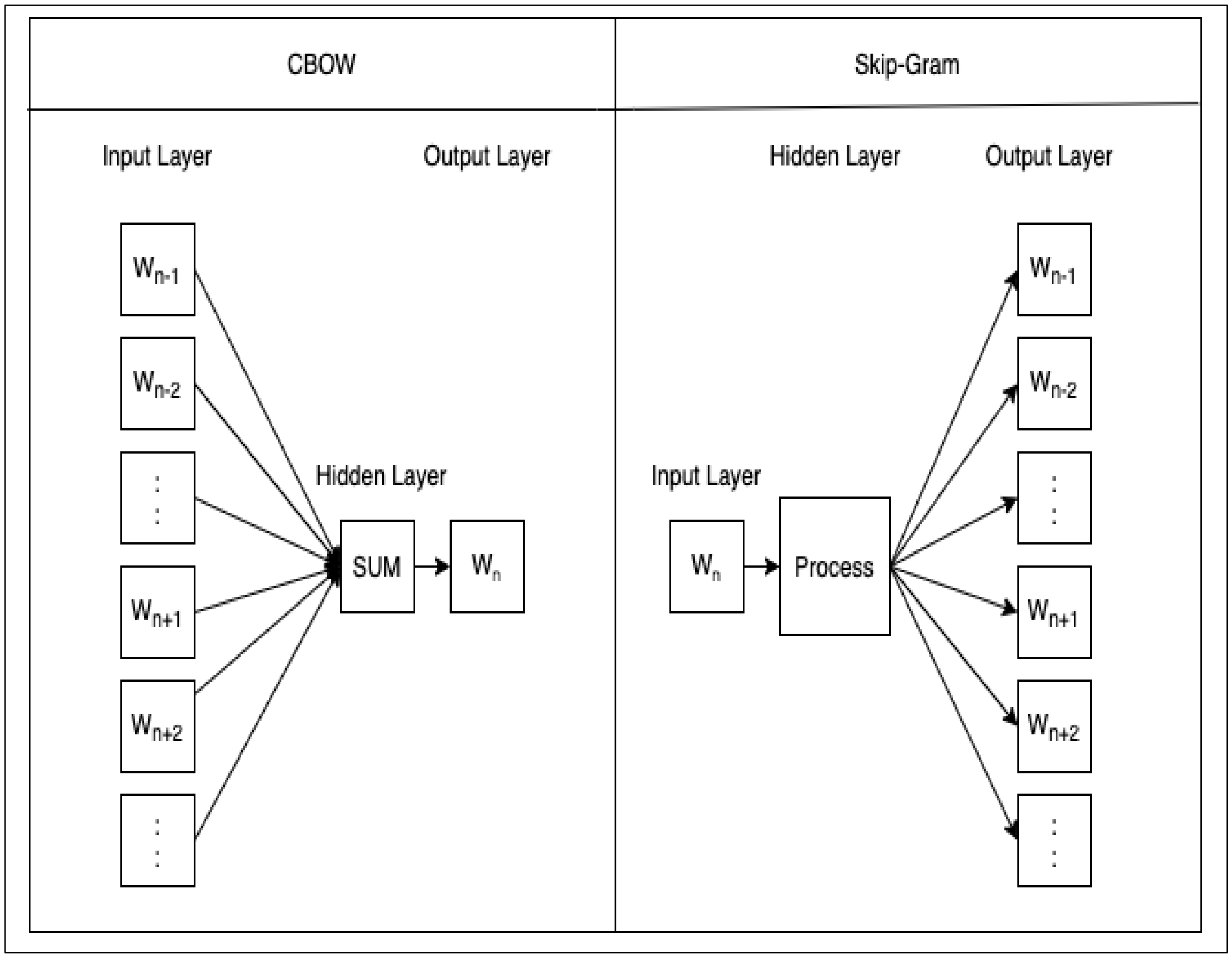

Figure 12 CBOW vs Skip-Gram model

### 3.4.3 GloVe

As it was mentioned before, GloVe was developed to solve the local context issue with the Word2Vec model. The GloVe model is trained on the non-zero entries of a global word-word co-occurrence matrix, tabulating how frequently words co-occur with one another in a given corpus. A single pass through the entire corpus needs to happen to collect the statistics and populate this matrix. This first pass can be computationally expensive for large corpora; however, it will be much faster in future training iterations since the number of non-zero matrix entries is typically much smaller than the total number of words in the corpus (Pennington et al., 2014). GloVe, when building the co-occurrence matrix, takes local context into account by computing the co-occurrence matrix using a fixed window size (words are assumed to co-occur when they appear together within a fixed window). The GloVe is essentially a log-bilinear model with a weighted least-squares objective. The relationship of these words can be revealed by finding the co-occurrence ratios between two words in a context. For example, consider the co-occurrence probabilities for target words ice and steam with various probe words from the vocabulary. Figure 15 is demonstrating the probabilities of the word k appearing in the context of word w.

| Probability and Ratio | $k = solid$ | $k = gas$ | $k = water$ | $k = fashion$ |
|---|---|---|---|---|
| $P(k\|ice)$ | $1.9 \times 10^{-4}$ | $6.6 \times 10^{-5}$ | $3.0 \times 10^{-3}$ | $1.7 \times 10^{-5}$ |
| $P(k\|steam)$ | $2.2 \times 10^{-5}$ | $7.8 \times 10^{-4}$ | $2.2 \times 10^{-3}$ | $1.8 \times 10^{-5}$ |
| $P(k\|ice)/P(k\|steam)$ | 8.9 | $8.5 \times 10^{-2}$ | 1.36 | 0.96 |

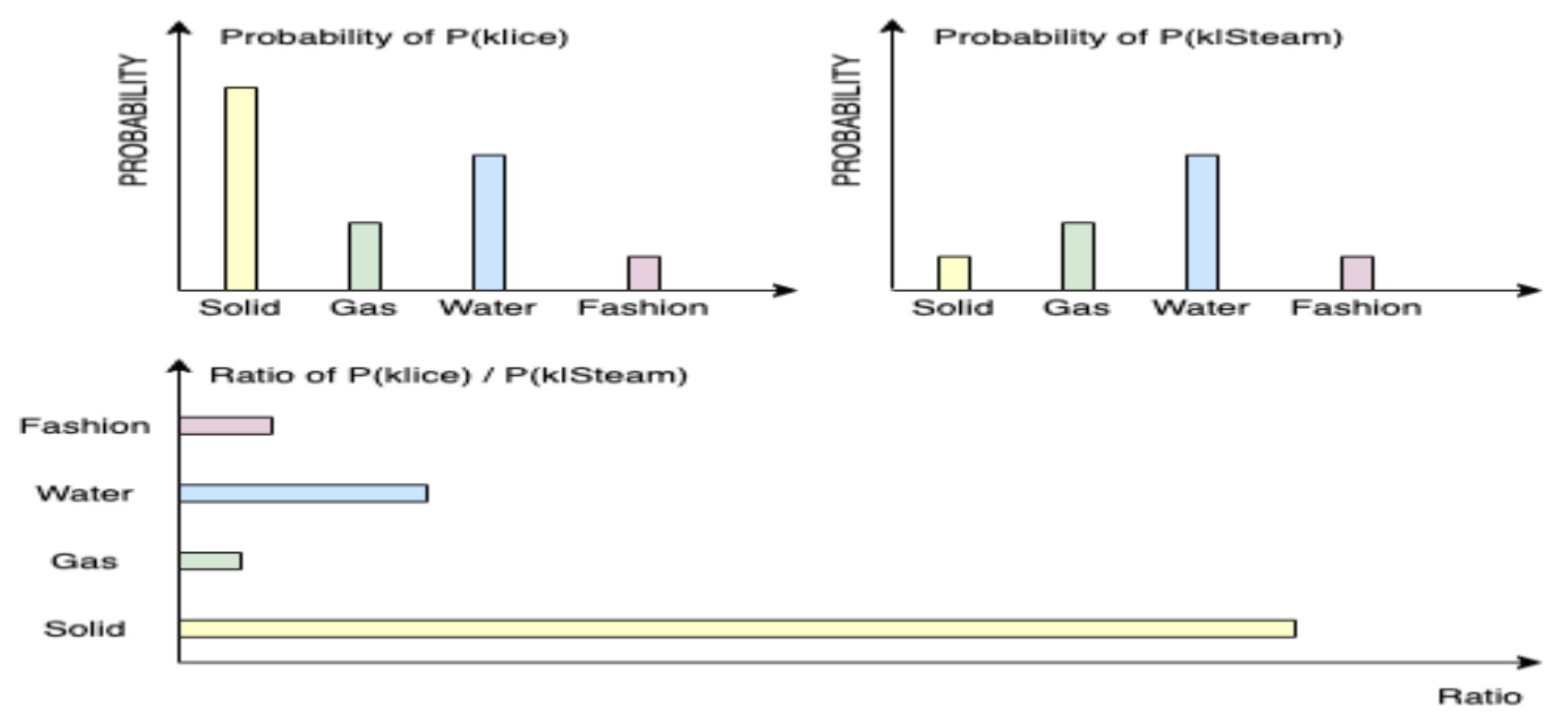

Figure 13 Probabilities of word k appearance in a context

As shown in figure 15, ice co-occurs more frequently with solid than it does with gas, whereas steam co-occurs more frequently with gas than it does with solid. Both words co-occur with their shared property water frequently, and both co-occur with the unrelated word fashion infrequently. Only in the ratio of probabilities does noise from non-discriminative words like water and fashion cancel out so that large values (more significant than 1) correlate well with properties specific to ice, and small values (much less than 1) correlate well with properties specific of steam. In this way, the ratio of probabilities encodes some crude form of meaning associated with the abstract concept of the thermodynamic phase.

The training objective of GloVe is to learn word vectors such that their dot product equals the logarithm of the words' probability of co-occurrence. Because the logarithm of a ratio equals the difference of logarithms, these objective associates (the logarithm of) ratios of co-occurrence probabilities with vector differences in the word vector space. Because these ratios can encode some form of meaning, this information gets encoded as vector differences. GloVe predicts surrounding words by maximizing the probability of a context word occurring given a center word by performing a dynamic logistic regression. As shown in Figure 16, you can see GloVe associates more than a single number to the word pair.

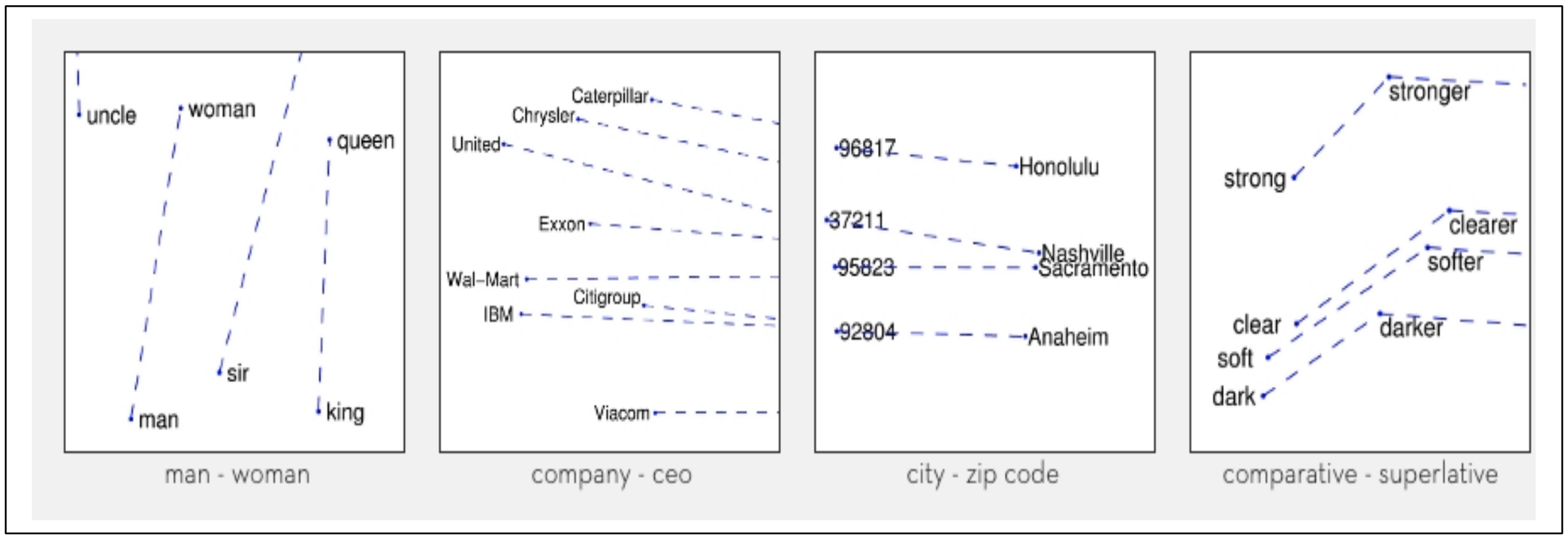

Figure 14 Visualization of the distinguishing and associating words by GloVe

The model utilizes the main benefit of count data, the ability to capture global statistics while simultaneously capturing the meaningful linear substructures prevalent in recent log-bilinear prediction-based methods like word2vec. As a result, GloVe becomes a global log bilinear regression model for the unsupervised learning of word representations that outperforms other models on word analogy, word similarity, and named entity recognition tasks.

### 3.4.4 BERT

BERT stands for Bidirectional Encoder Representations from Transformers. It is designed to pre-train deep bidirectional representations from the unlabeled text by jointly conditioning both left and proper context. As a result, the pre-trained BERT model can be fine-tuned with just one additional output layer to create state-of-the-art models for a wide range of NLP tasks (Devlin et al., 2018).

**Recurrent neural networks (RNNs)** are dynamic systems with the internal state at each step of the classification because of the circular connections between higher- and lower-layer neurons and optional self-feedback connections. These feedback connections enable RNNs to propagate data from earlier events to current processing steps. Therefore, RNNs build a memory of time series events (Staudemeyer and Morris, 2019). The problem with RNNs is that keeping the context from a word far away from the current word being processed decreases exponentially with the distance from it. Therefore, when dealing with long series of text data, the model often forgets the content of distant positions in the sequence. The other problem with RNN models is that a sentence has to be processed word by word, making it hard to parallelize the work for processing the sentences. To be able to deal with the linear distance between positions, and inhabitation of parallelization in sequential computations, which makes the RNN be able to only attend over the output of another RNN and focus only on different positions in the other RNN at every time step, new techniques like attention and Transform architecture was introduced.

**Transformer architecture** advances the recurrent neural network model that does not use recurrent connections and uses attention over the sequence instead. The Transformer is composed of multiple attention blocks which are essentially trying to parallelize the computation. Each block transforms the input using linear layers and applies attention to the sequence. The transformer consists of encoders and decoders. All encoders have the same architecture, and all the decoders have the same architecture as well. Each encoder consists of two layers: Self-attention and a Feed Forward Neural Network. The encoder's inputs first flow through a self-attention layer which helps the encoder access other words in the input sentence as it encodes a specific word. The decoder has both those layers, with a difference of having an attention layer between them which helps to focus on relevant parts of the input sentence. It is shown in Figure 17 BERT architecture is trained in several different ways on unsupervised data. In other words, each sample in the data set does not have any labels and the way BERT is trained.

Word2Vec and GloVe word embedding models had trouble capturing the meaning of the combination of words and taking the context into account. They only allocated a single vector for each word formed to represent a wide range of meanings.

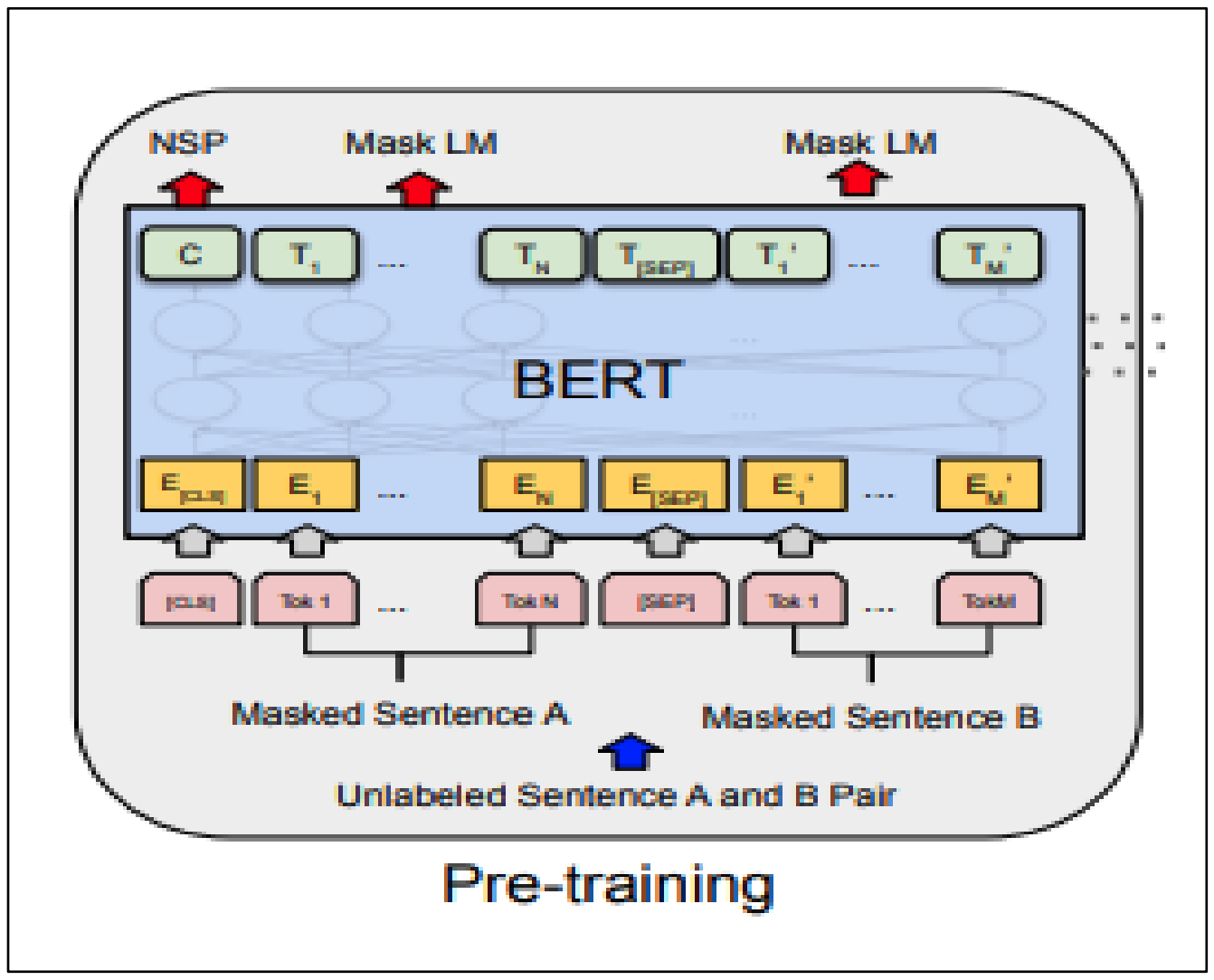

Figure 15 BERT Architecture

Using transfer learning, BERT, instead of just training a model to map a single vector for each word, introduces a complex deep neural network to map a vector to each word based on the entire sentence or context. Furthermore, language models have generally been trained from "left to right." Given a sequence of words, they would predict what word would come next, but BERT randomly masks words in the sentence, predicts them, and forces the model to learn how to use information from in the entire sentence to deduce what words are missing.

There exist bidirectional Long short-term memory-based language models like Hochreiter and Schmidhuber (1997), where they train a standard left to suitable language model and the reverse. These models have a single LSTM for both the forward and backward language models, which helps them predict previous words from subsequent words. However, BERT uses the information from the entire sentence simultaneously regardless of their position and considers both the prior and subsequent tokens simultaneously. Therefore, BERT will be used as a pretrained embedding layer in this thesis. The three mentioned word embedding are compared in terms of learning level and their sensitivity to the full context in Table 1 below.

Table 1 Word Embeddings Comparison

| Model Name | Context-Sensitive embeddings | Learned representations |
|---|---|---|
| Word2Vec | No | Words |
| GloVe | No | Words |
| BERT | No | Sub words |

### 3.5 Classification Algorithm (Deep Learning)

Machine-learning-based techniques are usually divided into two groups for vulnerability analysis problems: (1) traditional models and (2) deep learning models. Traditional models include the classical machine learning techniques, such as the naïve Bayes classifier in Murphy (2006), maximum entropy classifier in Nigam et al. (1999), or support vector machines (SVM) (Tong and Koller, 2001). The accuracy of these systems depends on which features are chosen (Dang et al., 2019). However, Deep learning models are proven to provide better results than traditional models (Torfi et al., 2020). Different kinds of deep learning models can be used for vulnerability analysis, including convolutional neural networks (CNN), and recurrent neural networks (RNN).

### 3.5.1 Convolutional Neural Networks (CNN)

CNN is a subclass of feed-forward neural networks, and the human visual cortex inspires its architecture. This structure includes convolutional and pooling layers. When utilizing CNNs for NLP tasks, the input data, which could be either sentences or documents that are represented as matrices (each row of the matrix is associated with a language element such as a word or a character) will be passed into a stringing of one or more convolutional layers, which are followed by a pooling layer.

This convolutional-pooling layer structure could be repeated to form various CNN architectures or deep CNNs. A convolutional layer applies a filter pattern (essentially performing the specific mathematical operation) to the input data, which helps the network map the input data to an underlying feature map and find local connections between them. Filters that will be applied map parts of the input to one feature and reduce the number of weights in the network. Furthermore, the semantically similar features in the feature map will merge to minimize the feature map size by the pooling layer that will be applied in the network. Figure 18 is demonstrating the basic architecture of a convolutional neural network.

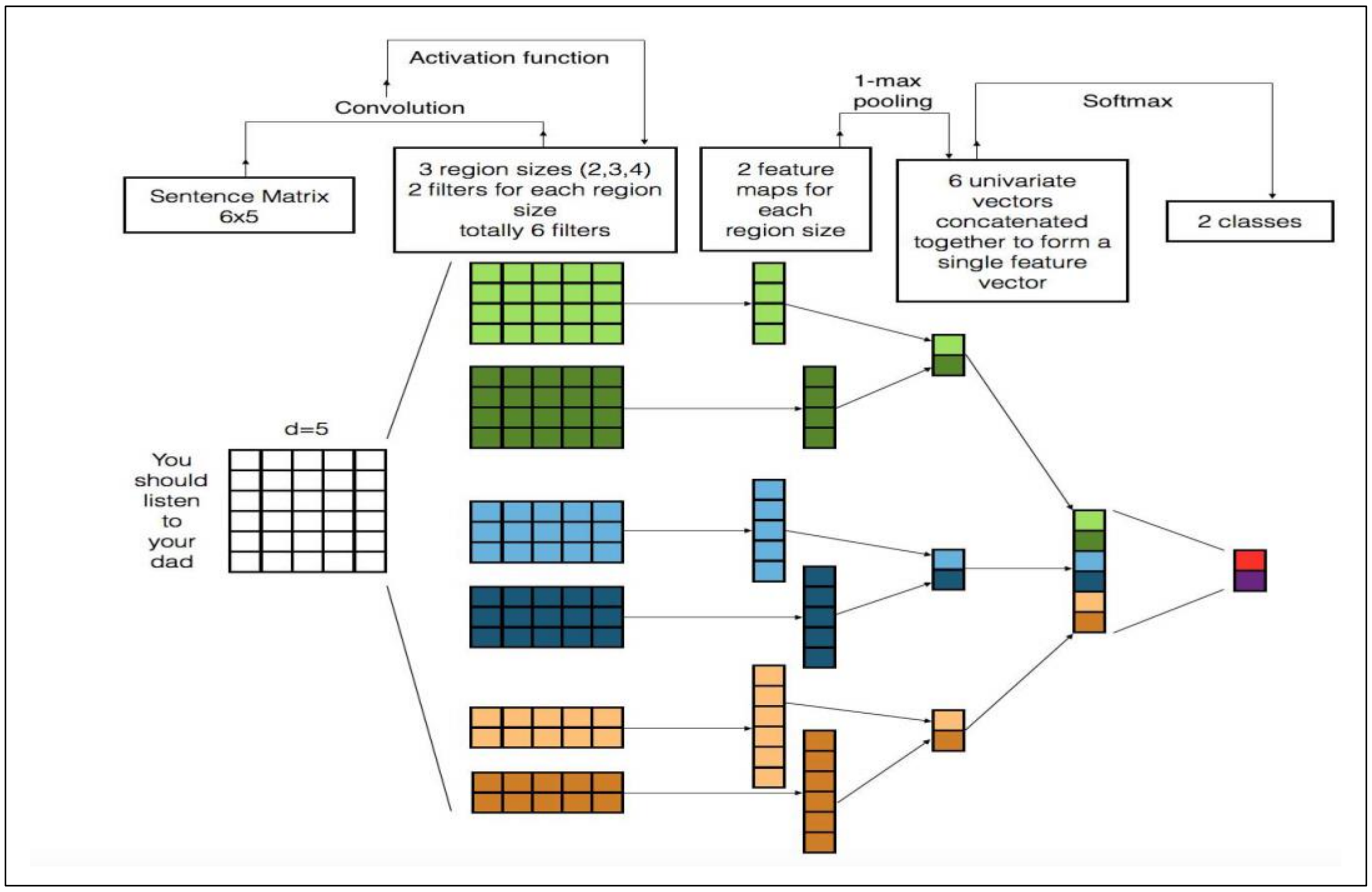

Figure 16 Basic CNN Architecture

Dos Santos (2014) has proposed a deep convolutional neural network that exploits from character-level to sentence-level information to perform vulnerability analysis achieved a positive accuracy of 76.9%.

### 3.5.2 Recurrent Neural Networks (RNN)

RNN is constructed of a sequence of feedforward neural networks. In this architecture, the output of each layer $y = (y_1, . . ., y_t)$ is an input of the next layer, and the final output ($\hat{y}$) depends on the actual input $x_t$ and preceding inputs $(x_1, . . ., x_{t-1})$. In discrete time frames, sequences of input vectors are fed as the input, one vector at a time; therefore, the input of the next layer depends on and will be fed into it after inputting each batch of vectors, conducting some operations, and updating the weights of the current network and making a prediction. Hidden layers in recurrent neural networks usually have a memory and carry information from the past (hidden vector sequence history $h = (h_1, . . ., h_t)$), which helps them to control the process of analyzing an instance after another. Because of this characteristic, they have been useful for applications that deal with a sequence of inputs such as language modeling (words have to be processed step by step to retrieve their context). The hidden state $h_t$ is calculated by

$$h_t = H(W_{xh}x_t + W_{hh}h_{t-1} + b_h)$$

Equation 2 RNN Hidden state

Where H stands for a hidden layer function, which is often the hyperbolic tangent (tanh), W is again the weight matrix between the layers, and b represents the bias vector. The output is calculated by multiplying the weights with the result of the hidden state and adding a bias.

$$y_t = W_{hy}h_t + b_y$$

Equation 3 RNN Hidden state with bias

In this architecture, as shown in Figure 19, the network is initialized, and the initially hidden state $h_0$ is usually set to 0 at the beginning. Then, during the forward pass, the prediction $\hat{y}$ for the input $x_t$ is calculated. An evaluation of the actual outcome y takes place afterward. In the backward pass, the error's derivative is calculated using backpropagation through time (BPTT) with respect to weights (Salehinejad et al., 2017).

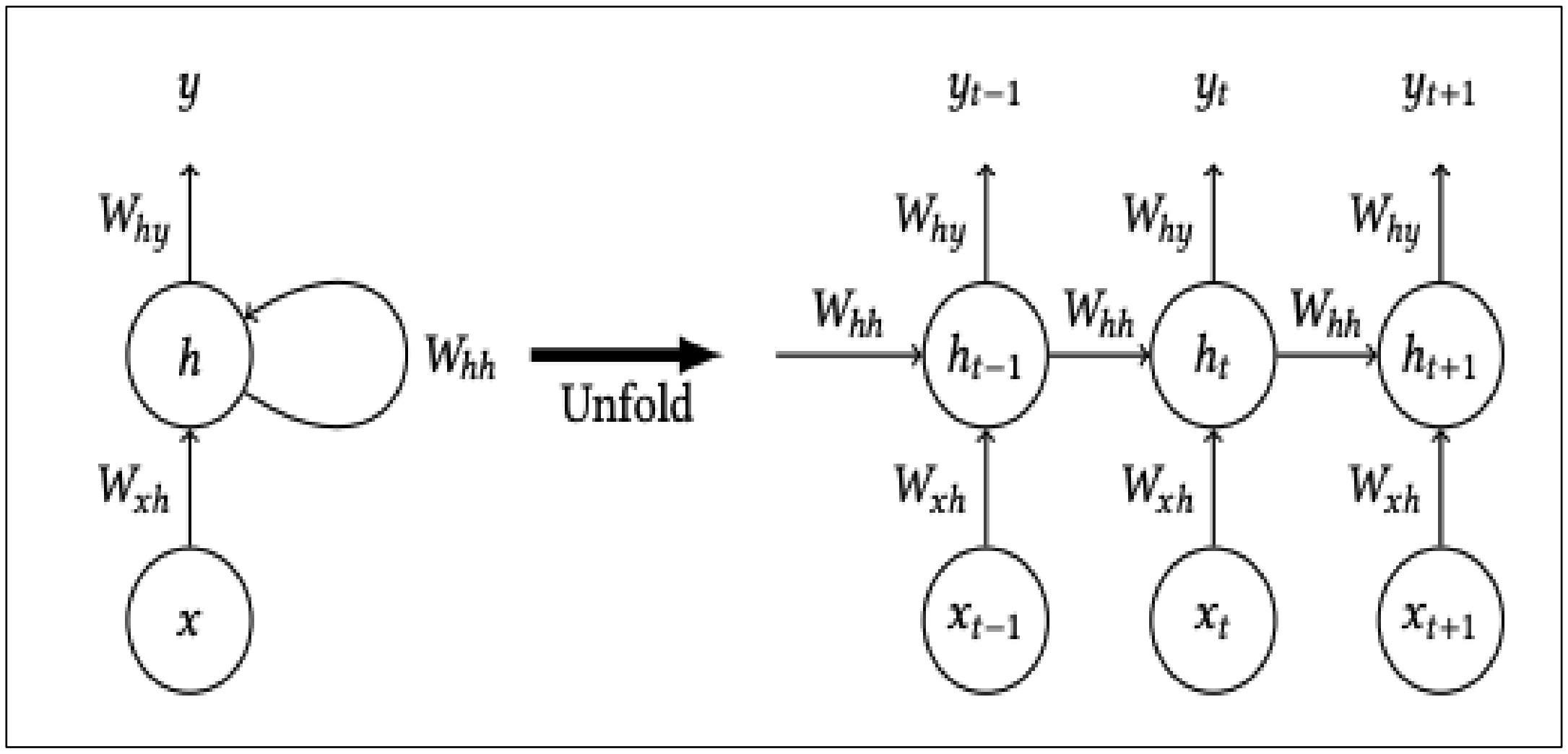

Figure 17 RNN Architecture

**Vanishing and Exploding gradients** are one of the main problems in RNNs. When dealing with a large dataset with long input sequences, it is difficult for the RNNs to keep track of the dependencies for two main reasons.

1. Activation functions such as hyperbolic tangent or sigmoid saturate quickly, and their gradient gets closer to 0.
2. The gradient is also being exponentially reduced by multiplying it with the recurring weight matrices when applying Backpropagation Through Time (BPTT) which eventually leads to the gradient getting closer to 0 in a short time (Salehinejad et al., 2017)

Due to these phenomena, which eventually lead to a high oscillation of the network's weights and increases learning time, the Long Short-Term Memory Network was introduced to avoid network failure.

**Long Short-Term Memory (LSTM)** introduced a forget mechanism (Schmidhuber et al., 1997). LSTM can save the information of the last input and forget earlier processed instances stored in their internal state. Moreover, LSTMs have a cell state on all of their input and output cells, input gate, forget gate, and output gate based on sigmoidal activation functions and regulate the values passed through the LSTM.

**Bidirectional LSTMs** are based on bidirectional RNNs and extend the standard LSTM model (Paliwal et al., 1997). The idea behind this model is to process the sequence $x = x_1, \ldots, x_t$ forward and backward. Their hidden states are connected to the output layer by a Feed Forward Neural Network. The forward part of the network processes $x_1, \ldots, x_t$ and the backward part $x_t$ , .

$$y_t = W_{\rightarrow hy}\, h_y + W_{\leftarrow hy}\, h^{\leftarrow}{}_t + b_y$$

Equation 4 Bidirectional LSTM output gate

. . , $x_1$. Using the last hidden states, the outcome $y_t$ at the output gate is computed similarly to Equation

Arras et al. (2017) in their work has extended the usage of Layer-wise Relevance Propagation (LRP) to recurrent neural networks. They implemented a specific propagation rule applicable to multiplicative connections as they arose in recurrent network architectures such as LSTMs and GRUs and applied their technique to a word-based bi-directional LSTM model on a five-class vulnerability prediction task and evaluated the resulting LRP relevance both qualitatively and quantitatively, obtaining better results than a gradient-based related method. Russell et al. (2018) in their work has jointed CNN and RNN with random forest architecture, taking advantage of the coarse-grained local features generated by CNN and long-distance dependencies learned via RNN for vulnerability detection and achieved an 81% and 82% accuracy accordingly.

### 3.6 Source Code Vulnerability Education

Technology is becoming widely used in all teaching processes. In today's educational system, programming and software engineering assignments are executed in an integrated development environment. Teachers usually analyze source code for lack of errors and relevance to solve the task to perform the grading. However, making sure of the lack of vulnerability in APIs or programming style of the students is currently missed in the educational system.

Itahriouan et al. (2020) in their work suggest that students' personality dimensions can be a factor for teachers to undertake the teaching mission. They propose a method that leverages machine learning and natural language processing to obtain source code understanding and insides from the learner’s attitude.

Escandor-O'Keefe (2020) in Coursera provides a course that will build a foundation of certain programming concepts. Their work covers threat modeling, cryptography applications, exploit attacks, authentication, and resource management. However, this course is designed to cover the concepts only and does not provide any hands-on assignment or experience.

Doshi (2017) in their work present a secure startup which is a novel system that aims to provide startups with a platform to protect their website cost-effectively while educating students about the real-world cyber skills. This system finds potential security problems in startup websites and provides them with practical solutions through a crowd testing framework while teaching students with necessary cyber skills. This system, however very effective, depends on the concept of crowdsourcing which requires a significant number of human resources to introduce students with source code vulnerabilities in real-time.

Turpin (2010) has provided a secure coding practice quick reference guide that includes a checklist of all elements in a source code that has to be analyzed. This guide is an excellent reference for developing hands-on experiences for source code vulnerability education, but it does not offer any learning modules on its own.

Seacord et al. (2012) in their work (SCALe) provide a software that performs both static and dynamic analysis to identify any vulnerabilities for commonly used software development languages and a detailed report of our findings to guide the codes repair. To use this platform, students would have to have some background information on why liability detection, and there are no learning modules or hands-on experiences that students could use to see the application of this platform in real life. In other words, students require some guidance on the usage/application of such tools.

### 3.7 Experiential Personalized Learning

Chen and lin (2007) propose, to meet the current industry demand for qualified security professionals, we need innovative courses that can help students apply information assurance theory into practice. The result of their work demonstrates that students generally agree that they have learned better with hands-on laboratory exercises. Irvine and Chin (1998) approach demonstrations that security topics can contribute to an engineering program by fostering skills required to produce graduates capable of critical thinking. Rowe and Ekstrom (2011) propose hands-on exposure, collaboration, and case studies are the most effective educational methods to familiarize students with high-order skills at a professional level. Werghi and Kamoun (2010) in their work present a decision support system that aims to provide students with an automated program planning and scheduling service that best fits their profiles while meeting academic requirements.

### 3.8 Summary

This chapter provided a detailed understanding of the concepts of artificial intelligence, NLP, and vulnerability analysis. Different algorithms and techniques used for feature extraction and classification have also been explained in detail, and their technical performance has been compared with each other. BERT language representation model combined with Deep Learning techniques Bidirectional LSTM has shown the highest performance on vulnerability analysis tasks than traditional machine learning used for vulnerability classification tasks. Furthermore, it reviewed the previous related work on hands-on lab and learning module developments for further educating students and improving the educational curriculum.

# Chapter 4 Design

With the increasing importance of source code security and vulnerability analysis, it is more critical to address the shortcoming of its learning curves for future IT professionals. Our future IT professionals (current students) are unaware of source code vulnerabilities, the tools/techniques used to recognize and mitigate vulnerability patterns in source code. The design of this work provides practical tools to educate future IT professionals and equip them to address source code vulnerability analysis using various techniques. Lab development must follow a modularized approach to provide future IT professionals with the proper learning material based on their skills for every aspect of source code vulnerability analysis. This hands-on lab series will integrate various most common static analysis tools (software metrics-based and machine learning-based) available for source code vulnerability. This chapter will present an overview of this project's design which leverages experimental learning to fit future IT professionals' needs.

## 4.1 Overview

This work covers several subjects that are related to software vulnerability analysis. The first module will introduce future IT professionals to different components, making a source code vulnerable in the software development life cycle. Future IT professionals will be introduced to source code vulnerabilities and then try to fix them on their own using open-source static analysis tools. After that, future IT professionals will understand the difference between dynamic and static analysis. Furthermore, they will be introduced to different techniques used in the static analysis, both the ones that do not analyze program syntax (*i.e.,* software-based metrics) and the ones that analyze program syntax and semantics, including anomaly detection and pattern recognition approaches.

Once future IT professionals are familiar with the advantages and disadvantages of the previously mentioned approaches, they will use such techniques from scientific works that have been reviewed in the related work section to compare the model's prediction and recognition of vulnerability with the available static tools. This will allow future IT professionals to understand how different data mining approaches and machine learning can improve the model's prediction. Figure 20 presents an overview of the proposed framework consisting of the student applications.

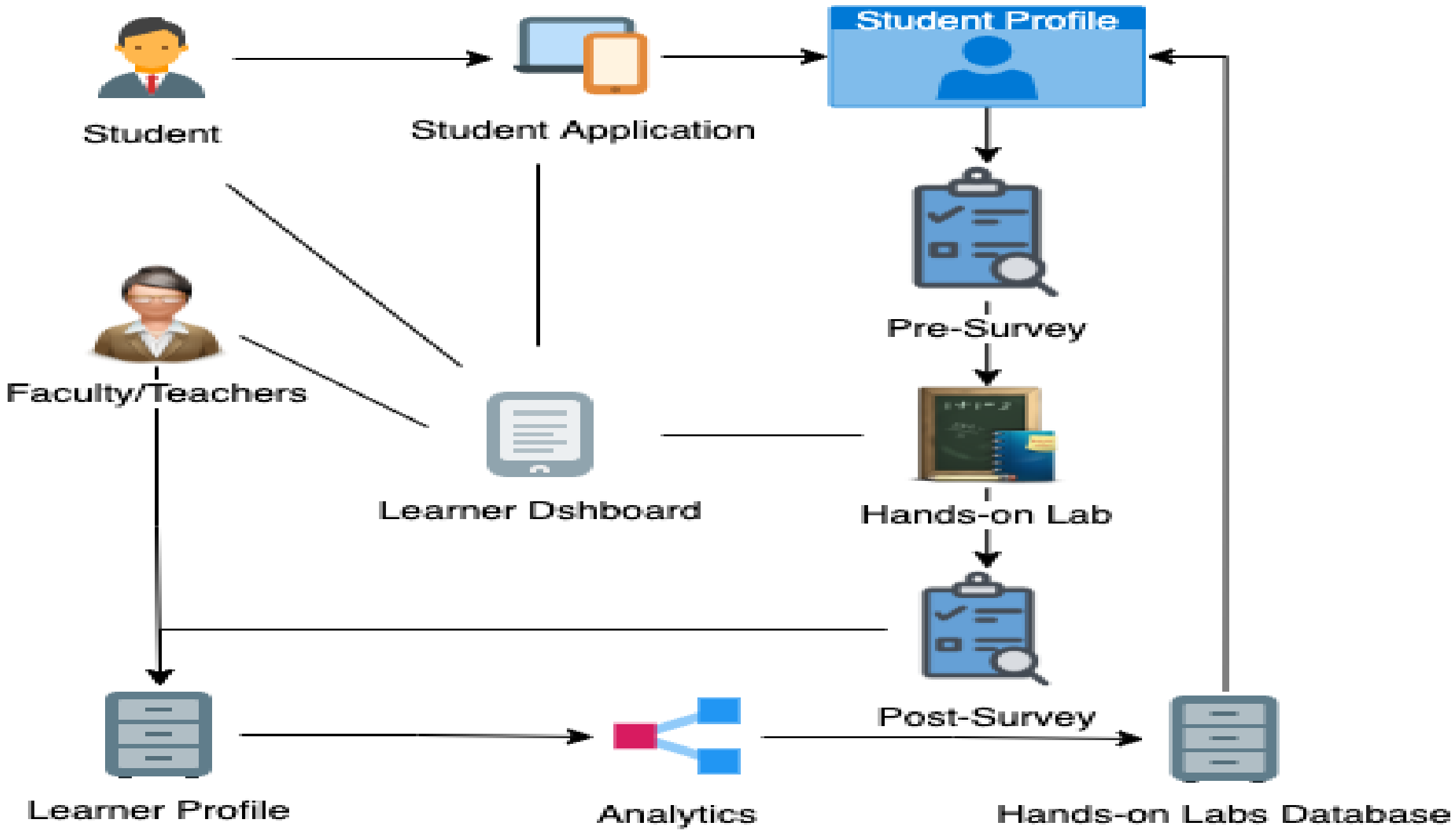

Figure 18 Overview of the SeCodEd Design

The Secure Coding Education (SeCodEd) framework (Student View) homepage serves as a portal for the application on Piazza and is illustrated in the example in Figure 21. Students can view courses and recent publications related to the course material, post any questions or discussion, and make general use of the application. Students also have access to an additional resource for reading about different companies, their job positions, and how to prepare for interviews. There are options for students to create a social profile that will allow them to connect to other students from other colleges.

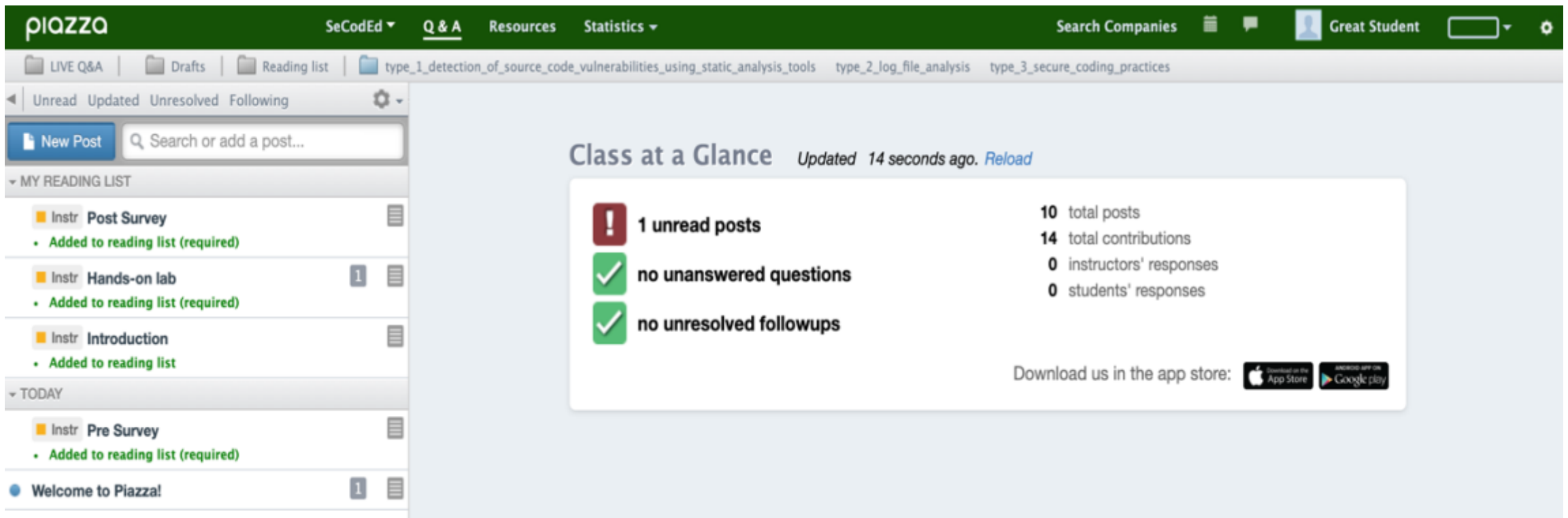

Figure 19 SeCodEd Homepage (Student View)

Figure 22 shows that, when administrators log in to the site, they can view a statistics page containing usage trends per day or month, total posts and contributions, Top student Contribution, and average response time. At the bottom of the administrator, reporting page is a progress overview of the course broken down into day, month, and year to allow administrators to view the completion of the system and download the statistics for further analysis.

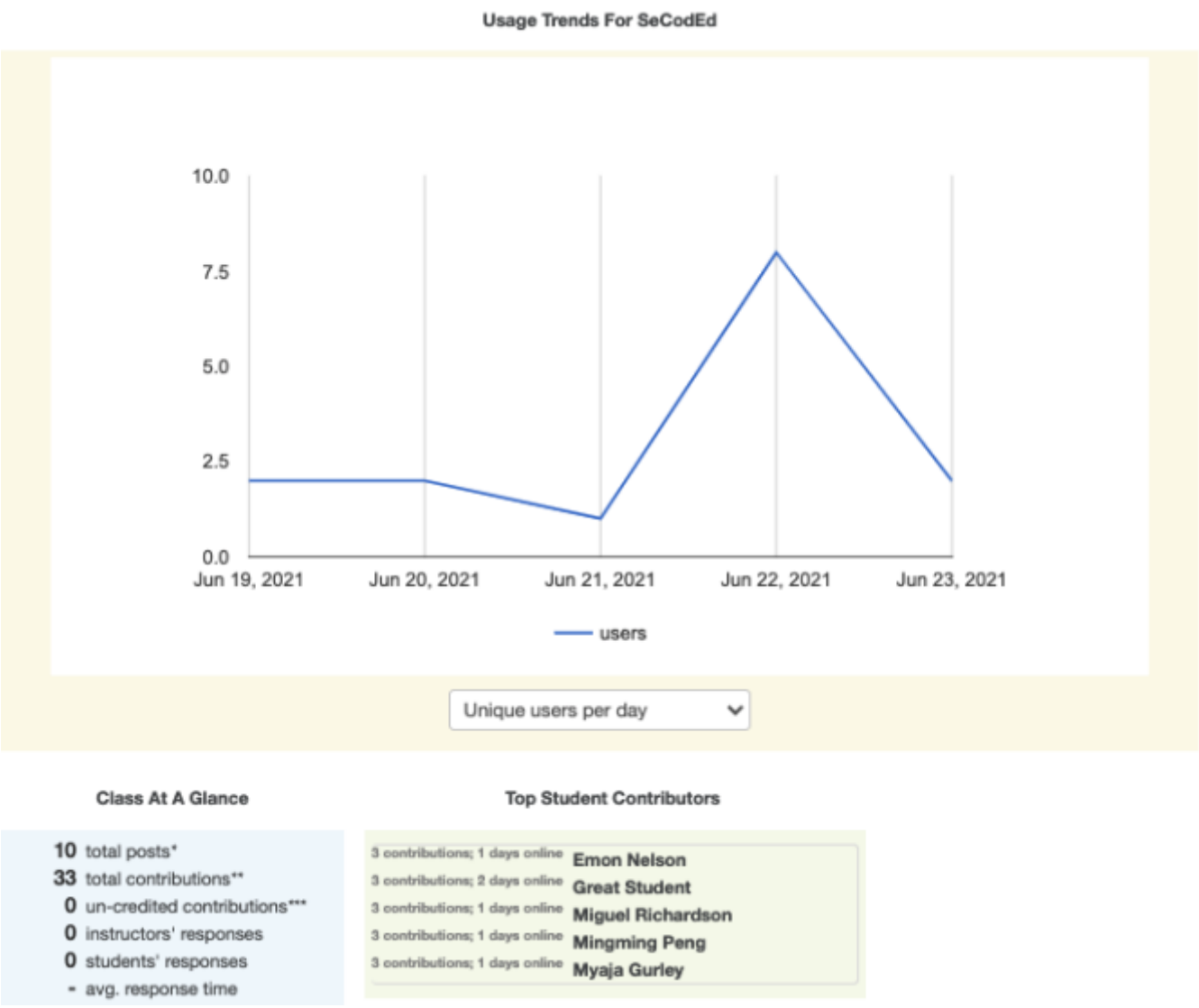

Figure 20 Statistics Page of the Course

On the Administration home page, Professors and teaching assistants can add, remove, or upgrade lecture notes, homework, supplementary materials, or any other format of assignments they would like. As shown in Figure 22, you can see that the “Great Student” has four tasks in their reading list, 1 is the pre-survey, 2 is an introduction task, 3 is the Hands-on Lab, and finally, the Post Survey. Mandatory tasks have a “required” tag next to them.

### 4.2 Case Study

The following scenario is described to demonstrate the use case of the SeCodEd platform. Student Bob creates a profile including his educational background information. He then fills our pre-assessment survey to gather information on his current level of understanding and interest in the concept. He is then provided with the first set of hands-on lab instructions. After finishing the lab and submitting it, He will be filling out the post-assessment survey to test his understanding and improvement on the concept. His submitted work and survey will be analyzed using a Decision Tree. He will be provided with the next hands-on lab or proper learning material based on a Decision Tree of survey responses.

### 4.3 Experiential Learning: Sample Courses and Labs

With the growth of usage of open-source source, vulnerability analysis of source code is becoming an industry standard. However, there are only a few universities that have included source code security in their undergraduate curriculum. Therefore, any hands-on experiment which can improve future IT professionals’ knowledge on source code vulnerabilities and source code security improvement is crucial to the success of computer science education. Three of the most frequently used approaches in Hands-on lab design are: (1) freestyle, (2) dedicated computing environment, and (3) build it from scratch.

SeCodEd, by leveraging and understanding the benefits and drawbacks of each of these approaches, is well designed, with each laboratory having its own goals and tasks.

These series of hands-on labs:

- Do not require any particular computing environment, provides all necessary environment setups and tools which will allow all future IT professionals to be able to finish the lab assignments and conduct security implementation, analysis, testing, and comparison of different tools
- Provides an infrastructure for future IT professionals that will not require them to build everything from scratch and experience making a secure static analysis system independently.
- Future IT professionals' implementation in these assignments is part of a much more complex system. This approach will allow future IT professionals to immediately see how their implementation behaves without building a very complex system. For example, suppose the task is to analyze improper input validation by implementing such vulnerable code. In that case, future IT professionals will scan it using static analysis tools and fix it independently to make sure they understand how such vulnerabilities will arise and how they could be avoided.

### 4.4 Hands-on LAB Categories

This series of learning modules is provided in 3 different categories due to the following reasons:

- With the advent of open-source usage in the industry, it is essential to educate future IT professionals on the most common fundamental vulnerabilities of the source code. The first category bridges this gap and equips future IT professionals with proper tools to mitigate such vulnerabilities.

- OWASP serves as a critical checklist and internal Web application development standard for many of the world's largest organizations. Auditors often view an organization's failure to address the OWASP Top 10 to indicate that it may be falling short regarding compliance standards (Synopsis, 2021). Integrating the Top 10 into computer science education demonstrates an overall commitment to industry best practices for secure development. The second category covers OWASP's Top 10 and introduces vastly used industry-level applications to future IT professionals that could be utilized to identify the security risks in a Web application.
- As a development community, we include third-party libraries in our applications containing well-known published vulnerabilities (OWASP, 2020). Graduates must have the proper skills to detect publicly disclosed vulnerabilities within a project's dependencies. The third category of hands-on labs teaches future IT professionals how to integrate tools that will allow them to detect vulnerabilities in their project's dependencies. This category equips future IT professionals with automated vulnerability detection engines that could be integrated into their IDEs.

The first category includes source code vulnerability detection using static analysis. During these series, future IT professionals are introduced to common security vulnerabilities published on CVE, CWE, NIST, such as code quality issues, information leakage, insufficient input validation, etc. Each of these hands-on labs comes with a lecture that introduces the concepts covered in the lab. Figure 23 below demonstrates.

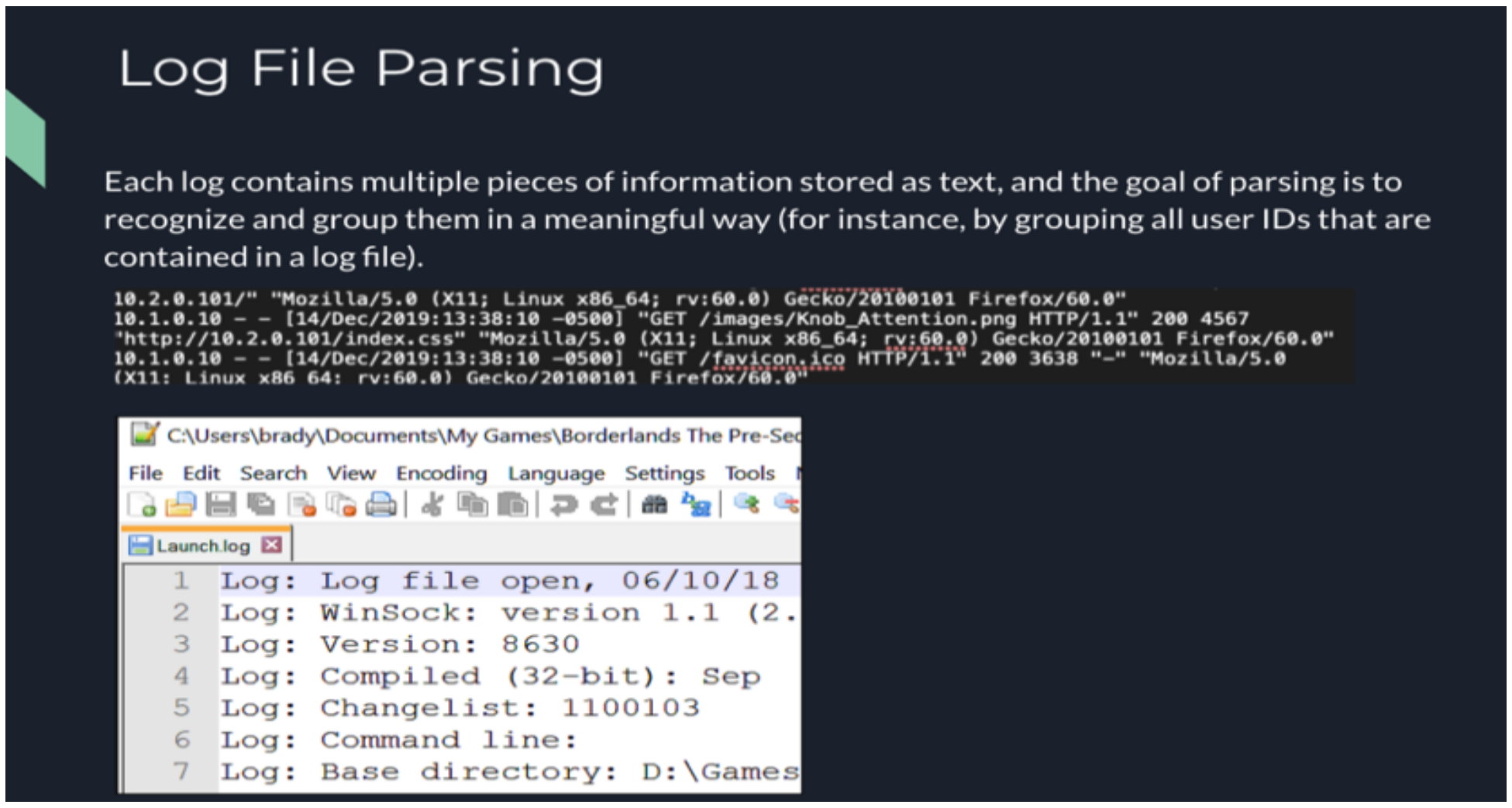

Figure 21 Sample of Lab Lecture

### 4.4.1 Type I Source Code Vulnerability Detection Using Static Analysis Tools

After being introduced to these vulnerabilities, future IT professionals were provided with open-source static analysis tools such as FlawFinder, Visual Code Grepper, and Clang Tidy to analyze noncompliant source code and identify any vulnerabilities using static analysis tools. As mentioned before, each lab includes a Description, Objective, and instructions to any special environment or Data that it requires, as demonstrated in Figure 24.

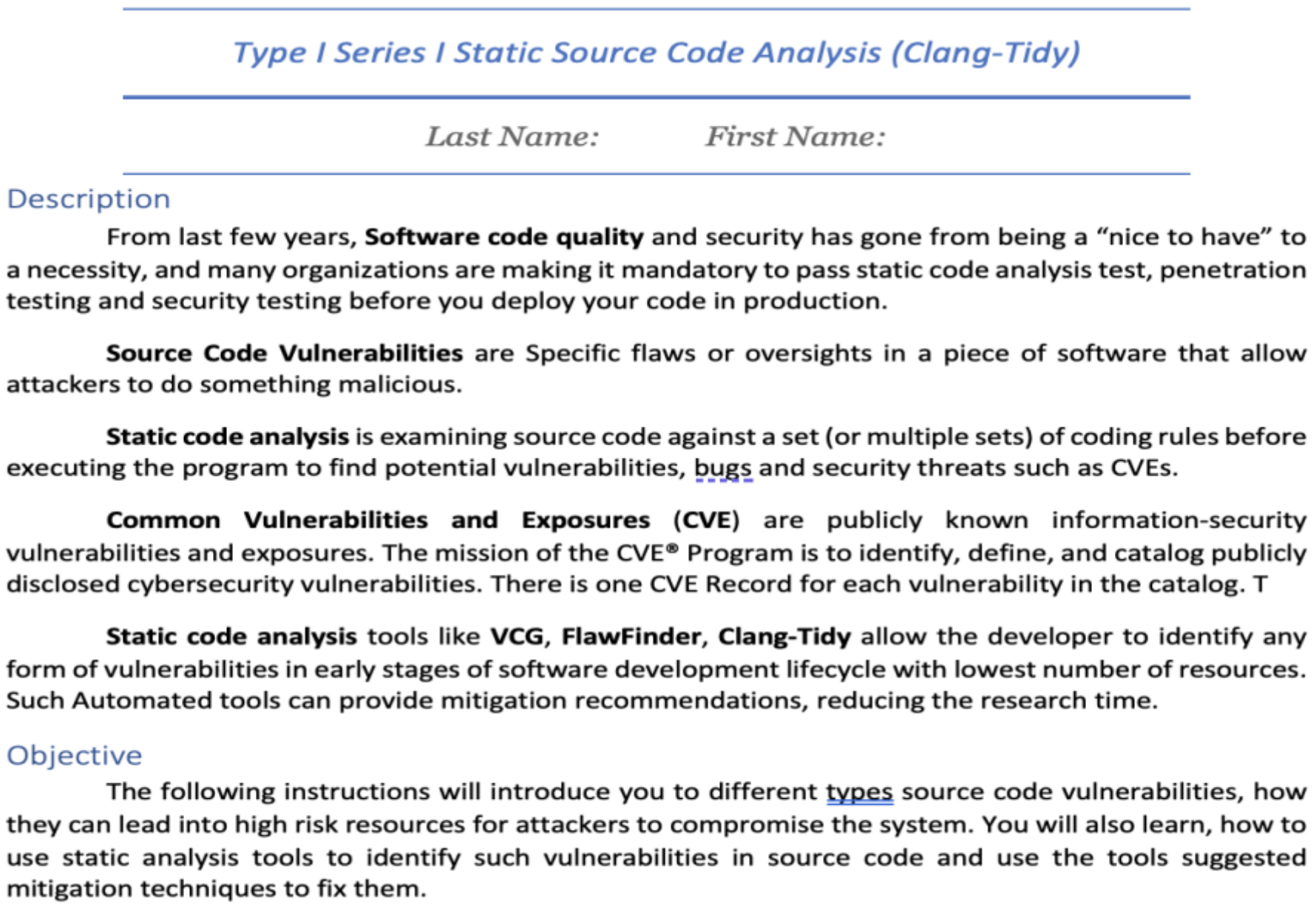

*Type I Series I Static Source Code Analysis (Clang-Tidy)*

*Last Name:* *First Name:*

Description

From last few years, **Software code quality** and security has gone from being a "nice to have" to a necessity, and many organizations are making it mandatory to pass static code analysis test, penetration testing and security testing before you deploy your code in production.

**Source Code Vulnerabilities** are Specific flaws or oversights in a piece of software that allow attackers to do something malicious.

**Static code analysis** is examining source code against a set (or multiple sets) of coding rules before executing the program to find potential vulnerabilities, bugs and security threats such as CVEs.

**Common Vulnerabilities and Exposures (CVE)** are publicly known information-security vulnerabilities and exposures. The mission of the CVE® Program is to identify, define, and catalog publicly disclosed cybersecurity vulnerabilities. There is one CVE Record for each vulnerability in the catalog. T

**Static code analysis** tools like **VCG**, **FlawFinder**, **Clang-Tidy** allow the developer to identify any form of vulnerabilities in early stages of software development lifecycle with lowest number of resources. Such Automated tools can provide mitigation recommendations, reducing the research time.

Objective

The following instructions will introduce you to different types source code vulnerabilities, how they can lead into high risk resources for attackers to compromise the system. You will also learn, how to use static analysis tools to identify such vulnerabilities in source code and use the tools suggested mitigation techniques to fix them.

Figure 22 Demo of Description, Objective and Material of Lab Type I

Once the analysis has been performed, they will be addressing (mitigating) those vulnerabilities from their source code using the instructions provided on lab and static analysis tools. Finally, to prove the efficiency of their approach, they rescan their source code to observe the effectiveness of their taken approach. This series discussed an issue reported to Apple on October 7, 2017, and it was assigned CVE-2017-13868. Apple fixed the bug in MacOS 10.13.2 and iOS 11.2. Looking at the new kern_control.c, Apple decided to fix the bug by wrapping the code after the call to sooptcopyin in an if statement that checks whether there has been an error. Future IT professionals using VCG understood where the memory leak is coming from, which built-in function it was related to, and further analyzed the Fix to this vulnerability to develop creativity.

### 4.4.2 Type II Log File Analysis

The second category of these hands-on labs included log file analysis. This series of hands-on labs introduced future IT professionals to different types of log files such as windows system log or web application log files. They were introduced to categories and clusters of information stored in various log files formats and how to identify any system or web application vulnerability through log files. Figure 25 demonstrates one of the plots from Task 3 of the second series where SYN Flood is introduced to future IT professionals, and they analyze/recognize it using LogViewPlus.

Once familiar with the analysis of log files to identify source code vulnerabilities, they use open-source tools such as SolarWinds Loggly and Microsoft log parser to perform log file parsing and identify different web application vulnerabilities such as SQL and command injection. To expand their knowledge on the task of log file parsing, they are taught how to code a program or scanner of their own to perform a fundamental task (scanning for SYN flood) and use ML and NLP techniques to perform log file parsing using CyBERT.

3- In **SYN Flood**, there Usually is a **one-way connection to exhaust resources**. A 3 way handshake won't be completed in this attack. Only the first SYN request will be sent to the server, trying to consume all the computing capacity on the networking device to render the genuine requests useless as they won't be processed due to the Web server being unavailable.

4- Using **LogViewPlus** → Actions tab, Graph Reports, get A plot of log entries per minute. attach a screen shot and write down your guess on the type of attack that is happening on the server.

| |
|---|
| Number of requests being sent in a very short period of time, as well as the number of errors and critical events with only the first handshake request indicate the possibility of SYN Flood. |

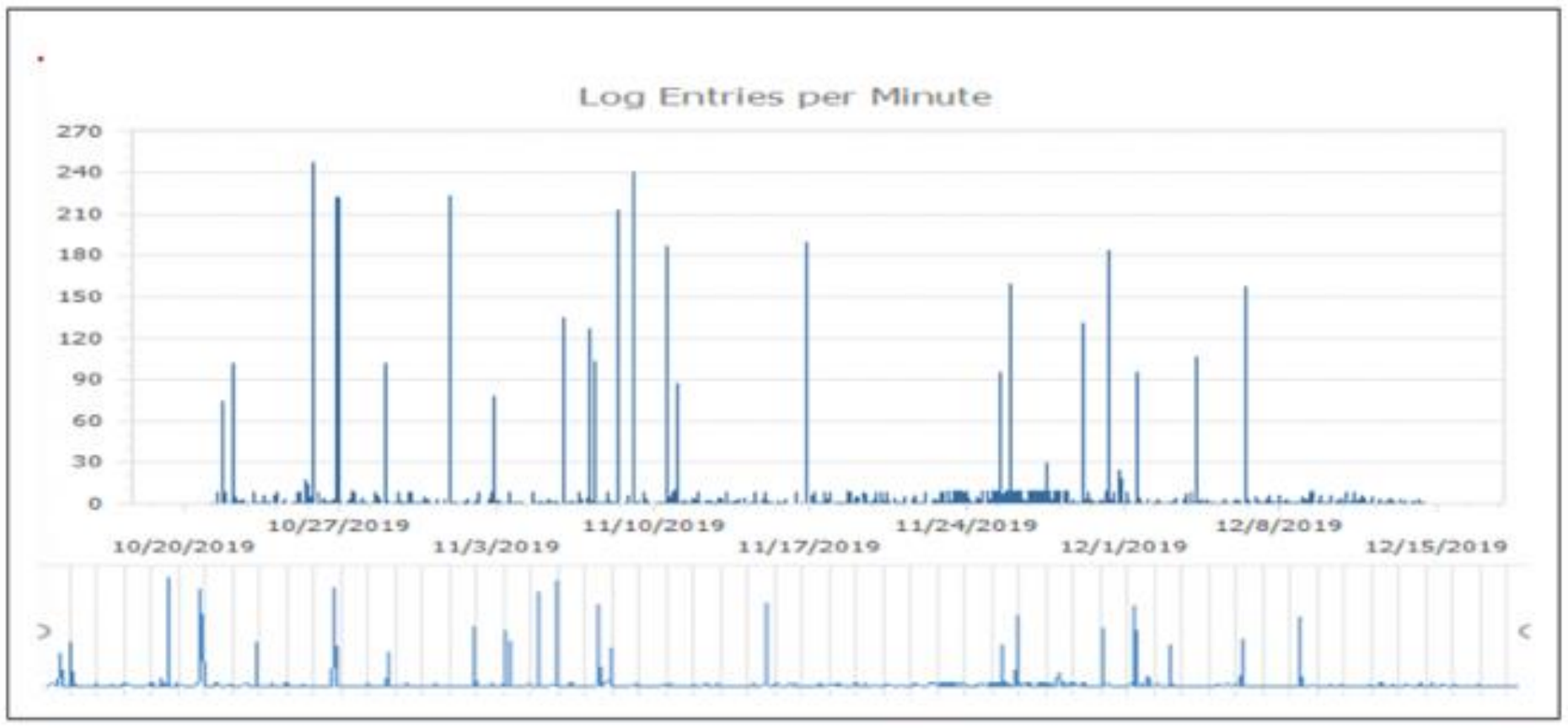

Figure 23 SYN Flood task Demo

### 4.4.3 Type III Secure Coding Practices

The third category of these hands-on lab series includes secure coding practices. Future IT professionals are introduced to developing a large-scale vulnerability detection system that interprets laxed source code using natural language processing and machine learning. This series of hands-on labs introduces AI engines currently being used to identify source code vulnerabilities when developing a software system. Future IT professionals were instructed to use Snyk and Dependabot on a development project established on GitHub to identify source code vulnerabilities and mitigate them in real-time. Using such tools, future IT professionals will always keep track of any vulnerabilities in their source code while developing software and mitigating them in the early stages of the software development lifecycle. Figure 26 demonstrates an example of a vulnerable design project scanned and fixed using Snyk in real-time.

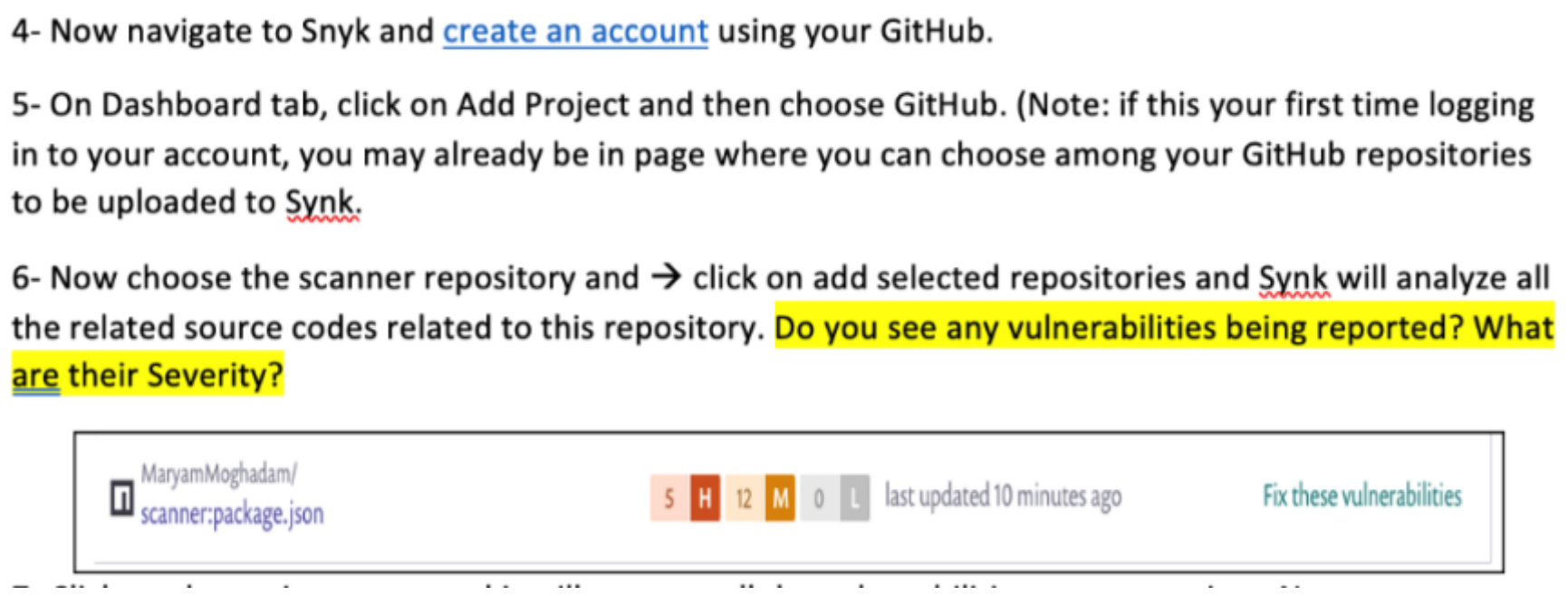
4- Now navigate to Snyk and create an account using your GitHub.

5- On Dashboard tab, click on Add Project and then choose GitHub. (Note: if this your first time logging in to your account, you may already be in page where you can choose among your GitHub repositories to be uploaded to Synk.

6- Now choose the scanner repository and → click on add selected repositories and Synk will analyze all the related source codes related to this repository. Do you see any vulnerabilities being reported? What are their Severity?

Figure 24 Snyk Application Demo

To expand their knowledge, they are also instructed to dynamic vulnerability scanners to identify an XSS vulnerability in a deliberately vulnerably designed webpage. They will then use the same source and knowledge to write their scanner, identifying cross-site scripting vulnerability.

### 4.5 Source Code Vulnerability Detection Using Static Analysis Tools

As previously discussed, each category of the hands-on lab covered different concepts with various teaching goals in mind. This section will cover the vulnerabilities that have been obscured and the tools that future IT professionals were equipped with.  Table 2 provides an overview of the vulnerabilities and tools that have been used in this category of labs.

Table 2  List of Vulnerabilities and Tools Covered in First Category

| # | Vulnerabilities | Tools |
|---|---|---|
| 1 | Format string attack (Tainted Data) | Visual Code Grepper |
| 2 | Invalid String Format | FlawFinder |
| 3 | Undefined Behavior Due to Un sequenced Modification and Access to Variables | Clang-Tidy |
| 4 | Input Validation | |
| 5 | Buffer Overflow Without User Input | |
| 6 | Insufficient Input Sanitization | |
| 7 | Memory Allocation (Errors & Leaks) | |

Figure 27 demonstrates an overview of tools and vulnerabilities covered in Type I labs

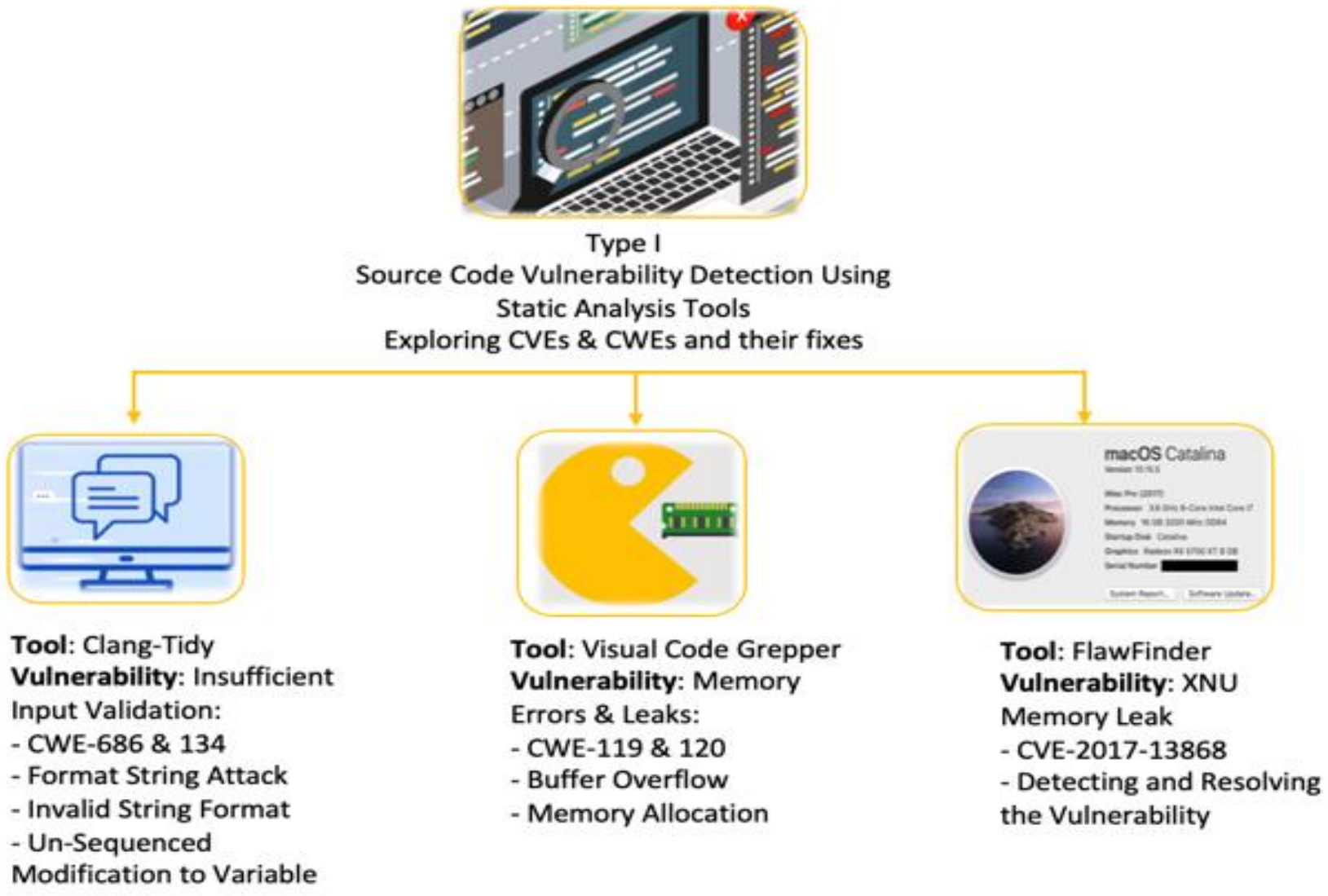

Figure 25 Type I Lab Overview

### 4.5.1 Vulnerabilities

Below is a list of all vulnerabilities covered in the first category of the hands-on lab.

**1- Format string attack (Tainted Data):** Formatted console input/output functions are used to take one or more inputs from the user at the console. It also allows the programmer to display one or multiple values in the output of the user at the console. In terms of secure programming, it’s a best practice to consider all **unchecked input valu**es as “**tainted**.” By demonstrating how such vulnerability can arise, this task instructs future IT professionals on leveraging **Clang-Tidy** to mitigate it.

**2- Invalid String Format:** Future IT professionals are familiarized with invalid variable formats and their potential risks in this task. They are instructed to read about CWE-686 and CWE134, common vulnerabilities associated with invalid string formats demonstrated in figure 28.

**3- Undefined Behavior Due to Un sequenced Modification and Access to Variables:** This task introduces modifying an object, calling a library I/O function, accessing a volatile-qualified value, or calling a function that performs one of these actions are ways to alter the state of the execution environment and instructs future IT professionals on mitigating and avoiding them using **Clang-Tidy**.

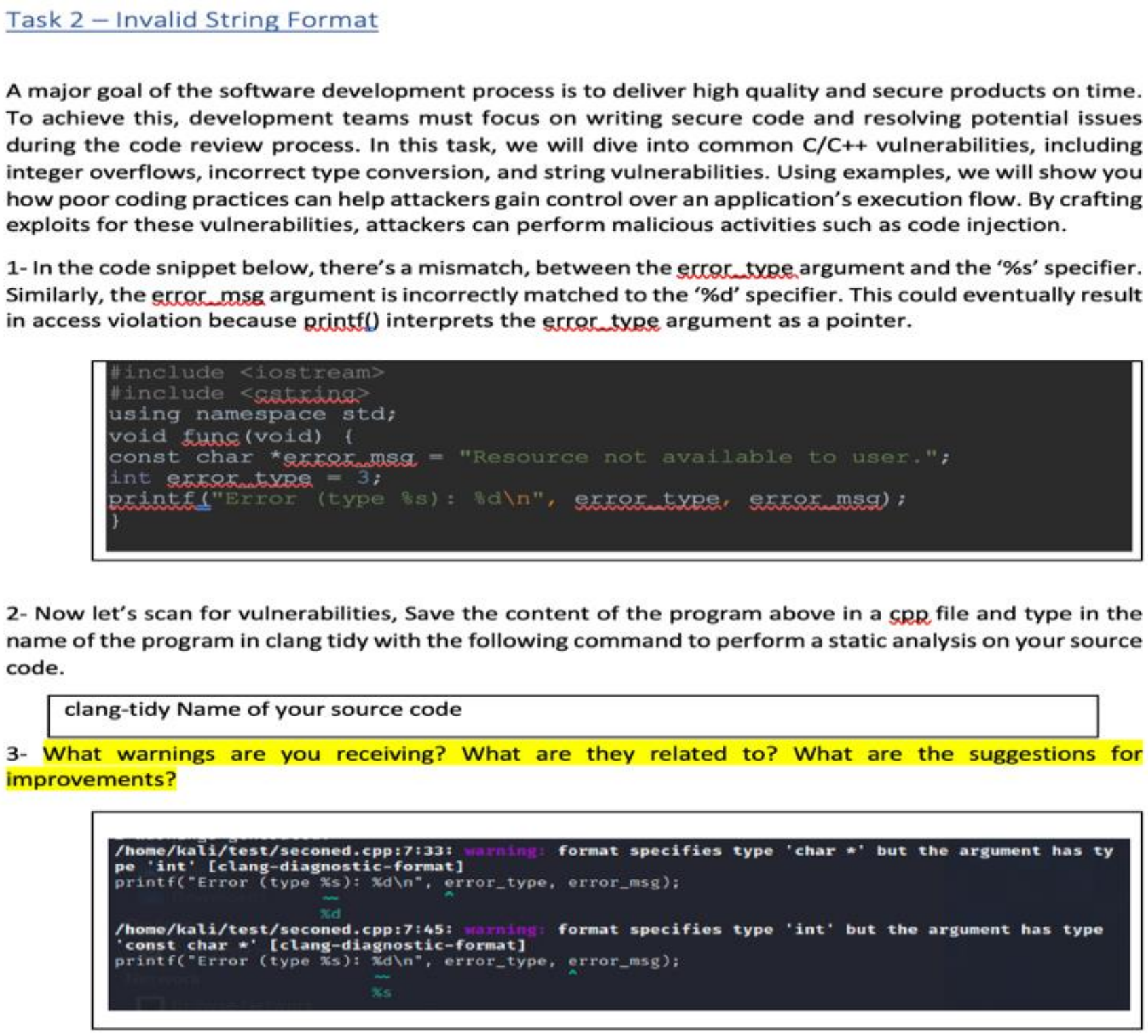

Task 2 – Invalid String Format

A major goal of the software development process is to deliver high quality and secure products on time. To achieve this, development teams must focus on writing secure code and resolving potential issues during the code review process. In this task, we will dive into common C/C++ vulnerabilities, including integer overflows, incorrect type conversion, and string vulnerabilities. Using examples, we will show you how poor coding practices can help attackers gain control over an application's execution flow. By crafting exploits for these vulnerabilities, attackers can perform malicious activities such as code injection.

1- In the code snippet below, there's a mismatch, between the error_type argument and the '%s' specifier. Similarly, the error_msg argument is incorrectly matched to the '%d' specifier. This could eventually result in access violation because printf() interprets the error_type argument as a pointer.

```
#include <iostream>
#include <cstring>
using namespace std;
void func(void) {
const char *error_msg = "Resource not available to user.";
int error_type = 3;
printf("Error (type %s): %d\n", error_type, error_msg);
}
```

2- Now let's scan for vulnerabilities, Save the content of the program above in a cpp file and type in the name of the program in clang tidy with the following command to perform a static analysis on your source code.

```
clang-tidy Name of your source code
```

3- What warnings are you receiving? What are they related to? What are the suggestions for improvements?

```
/home/kali/test/seconed.cpp:7:33: warning: format specifies type 'char *' but the argument has ty
pe 'int' [clang-diagnostic-format]
printf("Error (type %s): %d\n", error_type, error_msg);
                    ~~          ^
                    %d
/home/kali/test/seconed.cpp:7:45: warning: format specifies type 'int' but the argument has type
'const char *' [clang-diagnostic-format]
printf("Error (type %s): %d\n", error_type, error_msg);
                        ~~                  ^
                        %s
```

Figure 26 Invalid String Format Demo

**4- Input Validation:** Validate input means that input from all untrusted data sources must be sanitized. Proper input validation can eliminate the vast majority of software vulnerabilities. This task teaches future IT professionals to be suspicious of most external data sources, including command-line arguments, network interfaces, environmental variables, and user-controlled files, and use **VCG** to find and mitigate such vulnerabilities by trying different built-in functions and testing their security level.

**5- Buffer Overflow Without User Input:** A buffer is a temporary area for data storage. When more data (than was initially allocated to be stored) gets placed by a program or system process, the extra data overflow. It causes some of that data to leak into other buffers, corrupting or overwriting whatever information they were holding. This lab teaches future IT professionals that in a buffer overflow attack, the extra data sometimes contains specific instructions for actions intended by a hacker or malicious user; for example, the data could trigger a response that damages files, change data, or unveils private information. future IT professionals execute some vulnerable code to see the buffer overflow in action and leverage **FlawFinder** to mitigate them using more certain built-in functions of the C language.

**6- Insufficient Input Sanitization:** In the previous task, we analyzed the functions used by the programmer to copy the content of a buffer. This tasks instruction teaches future IT professionals how a buffer overflow can happen from user input which a cause of bad programming practice. Future IT professionals in this task are encouraged to compare the improvements and suggestions of **VCG** and **FlawFinder** together, as shown in figure 29.

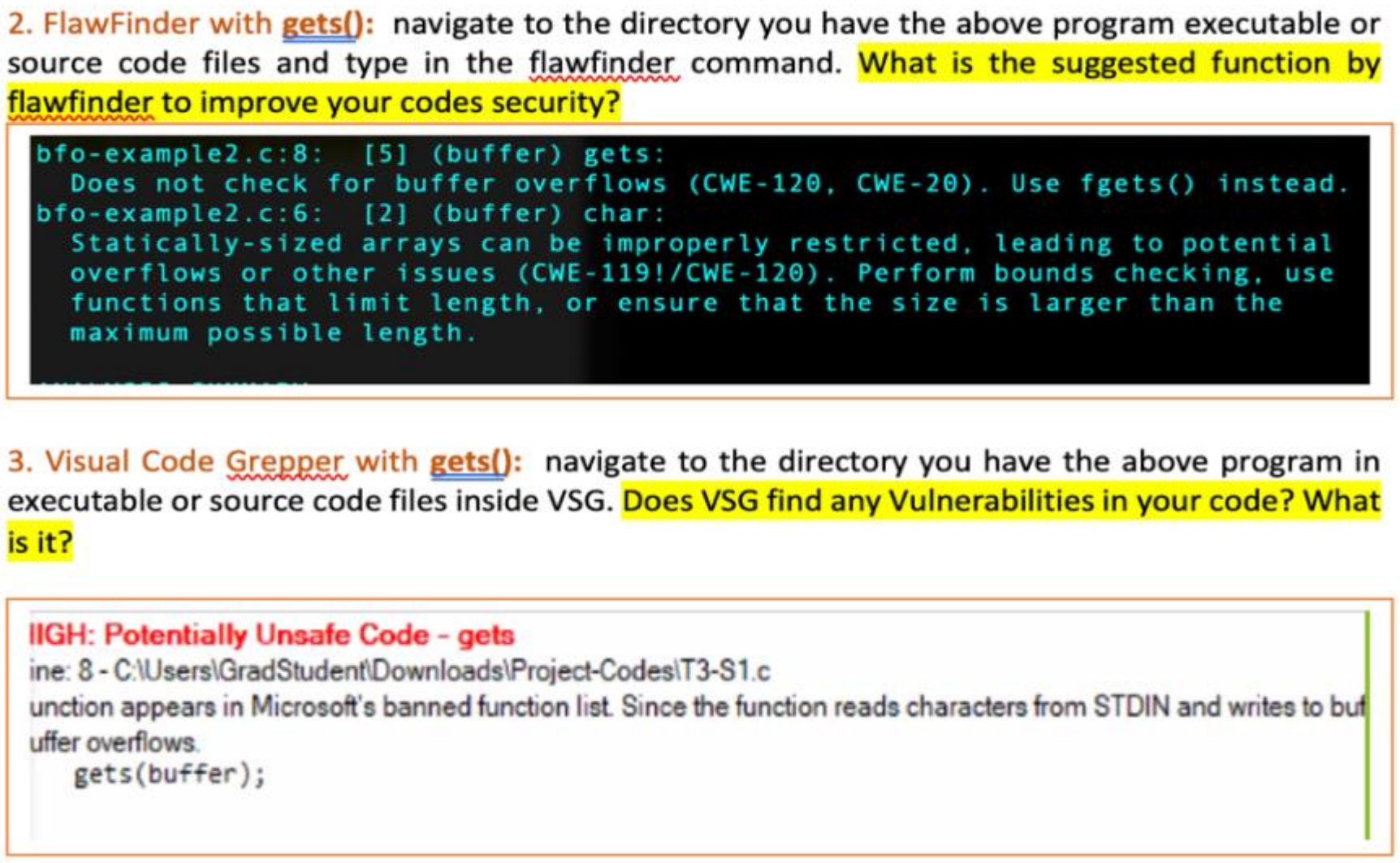

2. FlawFinder with gets(): navigate to the directory you have the above program executable or source code files and type in the flawfinder command. What is the suggested function by flawfinder to improve your codes security?

```
bfo-example2.c:8:  [5] (buffer) gets:
  Does not check for buffer overflows (CWE-120, CWE-20). Use fgets() instead.
bfo-example2.c:6:  [2] (buffer) char:
  Statically-sized arrays can be improperly restricted, leading to potential
  overflows or other issues (CWE-119!/CWE-120). Perform bounds checking, use
  functions that limit length, or ensure that the size is larger than the
  maximum possible length.
```

3. Visual Code Grepper with gets(): navigate to the directory you have the above program in executable or source code files inside VSG. Does VSG find any Vulnerabilities in your code? What is it?

IIGH: Potentially Unsafe Code - gets
ine: 8 - C:\Users\GradStudent\Downloads\Project-Codes\T3-S1.c
unction appears in Microsoft's banned function list. Since the function reads characters from STDIN and writes to buf
uffer overflows.

```
gets(buffer);
```

Figure 27 Comparison of FlawFinder & VCG Demo

**7- Memory Allocation (Errors & Leaks):** An array is a collection of a fixed number of values. Once the size of an array is declared, you cannot change it. Sometimes the size of the array you declared may be insufficient. Future IT professionals are instructed to allocate memory manually during run-time using **VCG's** warnings on a sample program to solve this issue.

### 4.5.2 Tools

As previously mentioned, Clang-Tidy, VCG, and FlawFinder were the three different tools that future IT professionals have been equipped with to identify vulnerabilities and mitigate them. All of these three tools will be described in the sections below.

**1- Clang-Tidy:** by Team (2021) is a static source code analyzer that scans your code for any patterns or usage of vulnerable built-in functions listed in CVEs. Clang-tidy is a clang-based C++ "linter" tool. Its purpose is to provide an extensible framework for diagnosing and fixing typical programming errors, like style violations, interface misuse, or bugs deduced via static analysis. **Clang-tidy** is modular and provides a convenient interface for writing new checks.

**2- FlawFinder:** by Wheeler (2020) is a simple program that scans C/C++ source code and reports potential security flaws. It can be a valuable tool for examining software for vulnerabilities, and it can also serve as a simple introduction to static source code analysis tools more generally. It is designed to be easy to install and use. FlawFinder supports the Common Weakness Enumeration (CWE) and is officially CWE-Compatible.

**3- Visual Code Grepper:** by Zhou (2018) is an automated code security review tool that handles C/C++, Java, C#, VB, and PL/SQL. It has a few features that should hopefully make it worthwhile to anyone conducting code security reviews, mainly where time is at a premium:

1. In addition to performing some more complex checks, it also has a config file for each language that allows you to add any bad functions (or other text) that you want to search for.

2. It attempts to find a range of around 20 phrases within comments that can indicate broken code ("ToDo", "FixMe", "Kludge", etc.)
3. It provides a nice pie chart (for the entire codebase and individual files) showing relative proportions of code, whitespace, comments, "ToDo" style comments, and destructive code.

It also searches intelligently to identify buffer overflows and signed/unsigned comparisons.

### 4.6 Log File Analysis

Currently, the U.S. system of cybersecurity training and education is failing to prepare future IT professionals for these roles. Employers find graduates from many programs lacking fundamental knowledge, practical experience, and soft critical skills. This series of hands-on labs prepare future IT professionals for industry-level server/system log file analysis. Table 3 provides an overview of the vulnerabilities and tools that have been used in this category of labs.

Table 3 List of Vulnerabilities and Tools Covered in Second Category

| # | Vulnerabilities | Tools |
|---|---|---|
| 1 | SQL injection | Loggly by SolarWinds |
| 2 | Brute Force Attack | LogViewPlus |
| 3 | Cross-Site Scripting Attack | Microsoft Log Parser |
| 4 | HTTP Flood | CyBERT |
| 5 | SYN Flood | |
| 6 | Insufficient Input Sanitization | |
| 7 | Memory Allocation (Errors & Leaks) | |

Figure 30 demonstrates an overview of tools and vulnerabilities covered in Type II labs.

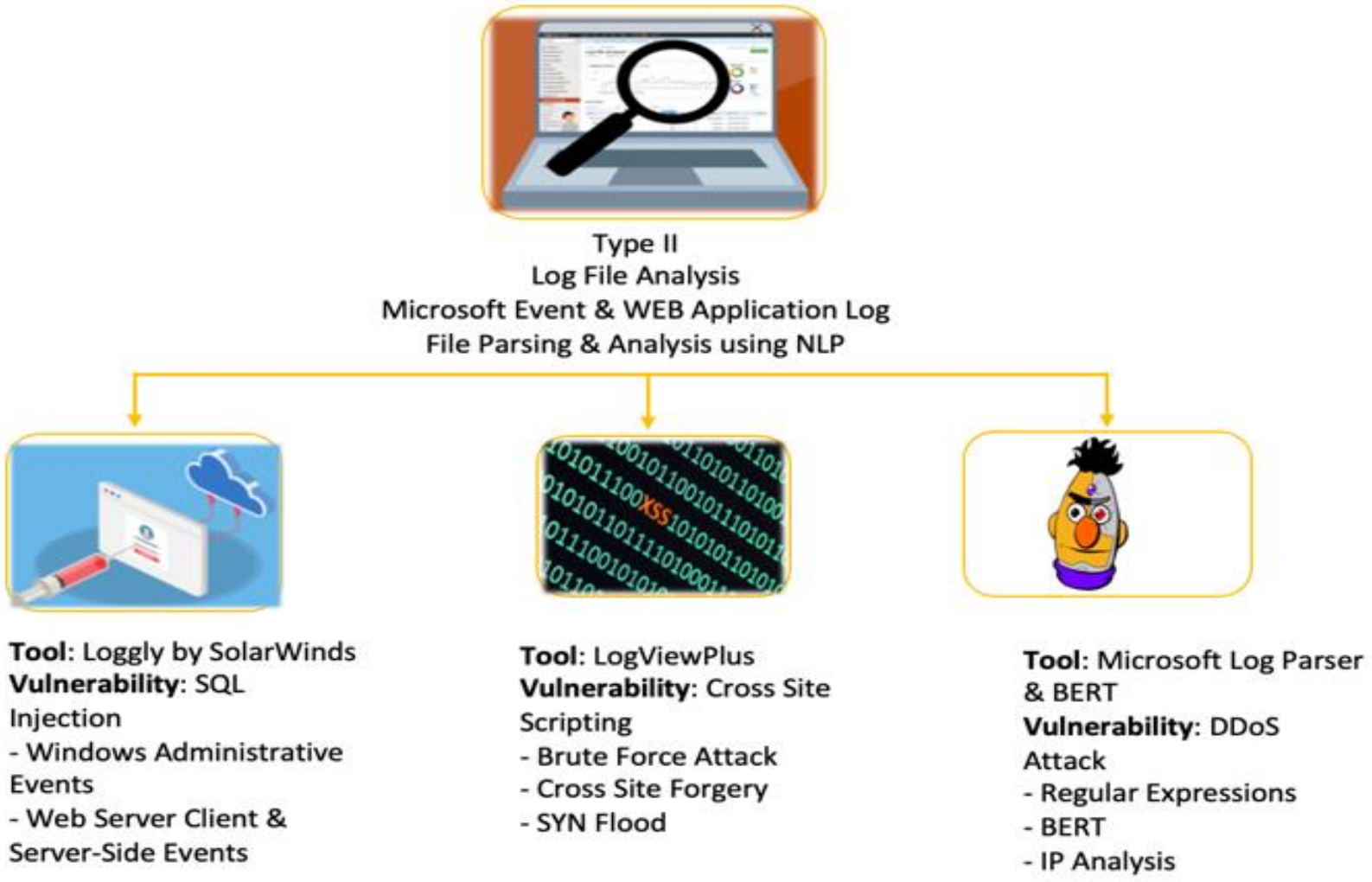

Figure 28 Type II Lab Overview

### 4.6.1 Vulnerabilities

Below is a list of all vulnerabilities covered in the second category of the hands-on lab.

**1- SQL injection:** attack directly impacts the Database server. The database can execute some malicious commands affecting the integrity of the database server and, ultimately, the Web application. SQL injection is considered a critical vulnerability.

**2- Brute force attack:** When someone tries to access our Web page by trying multiple logins. Brute force attacks use a dictionary or leaked information. **Dictionary** is a list of commonly used passwords. An attacker might use the leaked information to crack a password and gain illegitimate access. So, it is essential to change the password immediately for the account that has been informed about leaked passwords and be safe.

**3- Cross-Site Scripting Attack:** This is a type of Client-Side Injection Attack. They are primarily performed on Pages with Forms, and Forums are the most common examples. There are 2 Types of Cross-Site Scripting Attacks:

1. Stored: It occurs when a malicious script is injected directly into a vulnerable web application.

2. Reflected: The script is embedded into a link and is only activated once that link is clicked on.

Cross-Site Forgery Exploit the trust between the Web server and user's browser. User accesses malicious Web server, and the browser executes malicious commands.

**4- SYN Flood:** In SYN Flood, there Usually is a one-way connection to exhaust resources. A 3-way handshake won't be completed in this attack. Only the first SYN request will be sent to the server, trying to consume all the computing capacity on the networking device to render the genuine requests useless they won't be processed due to the Web server being unavailable.

**5- HTTP Flood:** This is a type of Distributed Denial of Service (DDoS) attack in which the attacker manipulates HTTP and POST unwanted requests to attack a web server or application. When an HTTP client, like a web browser, "talks" to an application or server, it sends an HTTP request – generally one of two types of requests: GET or POST. A GET request is used to retrieve standard, static content like images, while POST requests are used to access dynamically generated resources.

### 4.6.2 Tools

To improve cybersecurity education in the United States, we should look to the most successful cybersecurity workforce initiatives to identify best practices adopted by other programs to help prepare future IT professionals for cybersecurity careers. This series of hands-on labs leveraged a set up most common and heavily used tools in the industry to get future IT professionals started with leveraging such tools in vulnerability detection, which will be further discussed in the section below. future IT professionals can perform all the tasks for the hands-on lab on a cloud station without any changes or modifications to their system.

**1- Loggly by SolarWinds:** Loggly by Worldwide (2020) Visual with infrastructure and application environments spanning on-premises, hybrid, and public cloud environments, IT operations, application, and SRE teams are inundated with unrelated events, issues, and logs. Every outage or slowdown directly impacts the business, either in lost productivity or lost revenue. Cases must be diagnosed rapidly and resolved across all the dynamically changing components underpinning your heterogeneous web applications, services, and infrastructure. SolarWinds® Loggly® is a cost-effective, hosted, and scalable full-stack, multi-source log management solution combining powerful search and analytics with comprehensive alerting, dashboarding, and reporting to proactively identify problems and significantly reduce Mean Time Repair (MTTR).

**2- LogViewPlus:** LogViewPlus by Clearcove (2020) is a professional log viewer that can parse, read and analyze log files in a variety of different formats. LogViewPlus includes built-in support for technologies like SFTP, FTP, SCP, SSL, mapped drives, and Samba shares. Filtering log files is better than searching because filters can be chained. For example, you can narrow your log file down to a particular thread and then search just within that thread. Filters can also be updated automatically while tailing the log file. This is similar to tail and grep but completely redesigned for Windows

**3- Microsoft Log Parser:** LogParser is a flexible command-line utility initially written by Gabriele Giuseppini, a Microsoft employee, to automate tests for IIS logging (Wikipedia, 2020). It was intended for use with the Windows operating system and was included with the IIS 6.0 Resource Kit Tools. The default behavior of LogParser works like a "data processing pipeline", by taking an SQL expression on the command line and outputting the lines containing matches for the SQL expression.

Microsoft describes LogParser as a powerful, versatile tool that provides universal query access to text-based data such as log files, XML files, and CSV files, as well as critically data sources on the Windows operating system such as the Event Log, the Registry, the file system, and Active Directory. The results of the input query can be custom formatted in text-based output, or they can be persisted to more specialty targets like SQL, SYSLOG, or a chart.

**4- CyBERT:** CyBERT by Alarcon (2020) is an ongoing experiment to train and optimize transformer networks for the task of flexibly and robustly parsing logs of heterogeneous cybersecurity data. It’s part of CLX (read our overview blog about CLX), a set of cyber-specific applications built using RAPIDS. Since BERT was designed for natural human language and more traditional NLP tasks like question answering, we have overcome several challenges in our implementation. Unlike the flexible sentence organization of human speech, the rigid order of some cyber logs can cause our model to learn the absolute positions of the fields rather than their relative positions.

### 4.7 Secure Coding Practices

A critical first step to developing a secure application is an effective training plan that allows developers to learn important secure coding principles and apply them. After future IT professionals are introduced to vulnerabilities, this set of hands-on labs equip them with Automated vulnerability detection techniques. Table 4 provides an overview of the vulnerabilities and tools that have been used in this category of labs.

Table 4 List of Vulnerabilities and Tools Covered in Third Category

| # | Vulnerabilities | Tools |
|---|---|---|
| 1 | Buffer Overflow | Snyk |
| 2 | Improper Restriction of Operations within the Buffer | Dependabot |
| 3 | NULL Pointer Dereference | PwnXSS |
| 4 | Use of Pointer Subtraction to Determine Size | CyBERT |
| 5 | Improper Input Validation, Use of Uninitialized Variable, Buffer Access with Incorrect Length Value, etc. | Code Ocean |
| 6 | XSS | (Rebecca L. Russell) |

Figure 31 demonstrates an overview of tools and vulnerabilities covered in Type III labs.

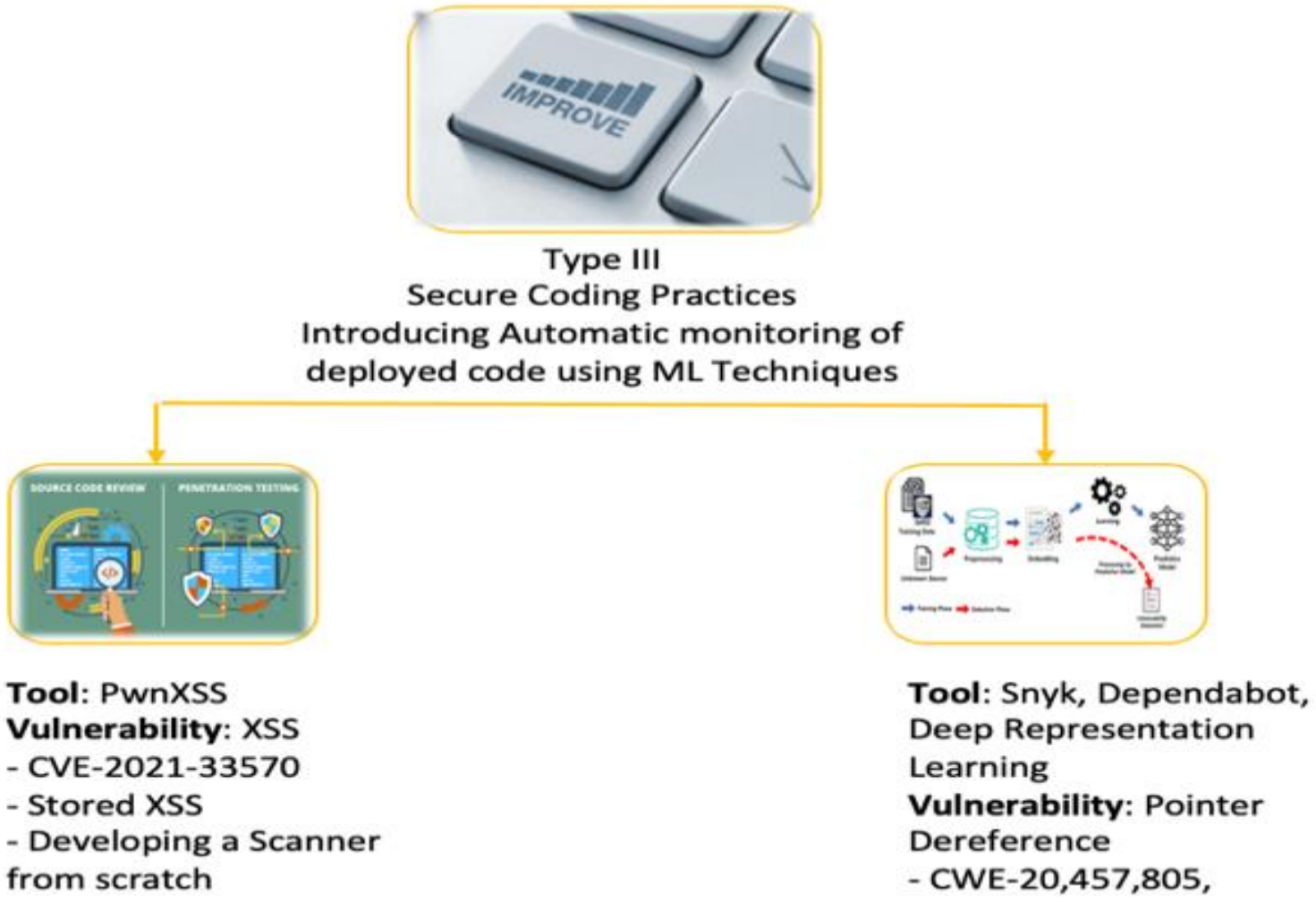

Figure 29 Type III Lab Overview

### 4.7.1 Vulnerabilities

In this set of hands-on labs, we have covered the vulnerabilities that have been covered in Russell et al. (2018) shown in the table below.

Table 5 Vulnerabilities Covered in the First set of the hands-on lab of the Third Category

| Vulnerability | Description |
|---|---|
| CWE-120,121,122 | Buffer Overflow |
| CWE-119 | Improper Restriction of Operations within the Buffer |
| CWE-476 | NULL Pointer Dereference |
| CWW-469 | Use of Pointer Subtraction to Determine Size |
| CWE-20,457,805 | Improper Input Validation, Use of Uninitialized Variable, Buffer Access with Incorrect Length Value, etc. |

Cross-site scripting, a type of security vulnerability typically found in web applications, has also been covered in these labs. XSS attacks enable attackers to inject client-side scripts into web pages viewed by other users.

XSS flaws can be challenging to identify and remove from a web application. The best way to find flaws is to perform a security review of the code and search for all places where input from an HTTP request could make its way into the HTML output. Future IT professionals have been also instructed on developing their XSS Scanner while developing software to scan their implementation and avoid vulnerabilities.

### 4.7.2 Tools

**1- Snyk:** by Snyk (2021) is a cybersecurity company that develops security analysis tools to identify open-source vulnerabilities.

The company is a developer-first security company that helps organizations use open-source code and stay secure. It helps software-driven organizations find and fix vulnerabilities in open-source dependencies and container images and provides a tool used by developers to scan their code for vulnerable open-source components.

**2- Dependabot:** Dependabot by Busoli (2020) is a tool for automatic dependency management created initially as an external service before being acquired and integrated natively into GitHub. We have been using it extensively since its early versions to automatically upgrade versions of the packages used by our repositories. Continuous dependency upgrades allow our applications and packages to stay up to date with the latest features, bugs, and security fixes by spreading the effort of doing so over a more extended period. Conversely, delaying package updates postpones the effort to a time when upgrading may be too costly or even impossible.

**3- PwnXSS:** PwnXSS by Mohdshariq (2020) is a free and open-source tool available on GitHub. This tool is specially designed to find cross-site scripting. This tool is written in python. You must have python 3.7 installed in your Kali Linux. There are lots of websites on the internet that are vulnerable to cross-site scripting (XSS). This tool makes finding cross-site scripting easy. This tool works as a scanner. The Internet has millions of websites and web apps a question comes into mind whether your website is safe or not. The security of our websites plays an important role. Cross-site scripting or XSS is a vulnerability that can be used to hack websites. This tool helps to find such exposure easily.

**4- Code Ocean:** Code Ocean by EBSCO (2020) is a centralized platform for creating, sharing, publication, preservation, and reuse of executable code and data. With Code Ocean, researchers can quickly analyze, organize, and execute work and publish in repositories and journals. Using this platform, we were able to share code and content with the educators and execute python scripts for implementation of NLP into Vulnerability detection on the cloud without any resources on their browser.

**5- Automated Vulnerability Detection in Source Code Using Deep Representation Learning:** Increasing numbers of software vulnerabilities are discovered every year, whether they are reported publicly or found internally in proprietary code. These vulnerabilities can pose a severe risk of exploitation and result in system compromise, information leaks, or denial of service. Russell et al. (2018) have leveraged the wealth of C and C++ open-source code available to develop a largescale Vulnerability detection system that interprets lexed source code using Natural Language Processing & Machine Learning.

They evaluated a tool on code from both fundamental software packages and the NIST SATE IV benchmark dataset. Their results demonstrate that deep feature representation learning on source code is a promising approach for automated software vulnerability detection**.**

### 4.8 Summary

The three of these hands-on lab categories will introduce future IT professionals to the most common CVEs, CWEs, and NIST vulnerabilities and OWASP top 10 web application security risks that can exist in source code and can be identified using log files. With the first category of labs, we assure that future IT professionals are aware of vulnerabilities that may arise from programming languages' built-in function/API used in their source code. Using the second category of labs, we emphasize the importance of log files in both operating system and web application security risk detection. We introduce log file analysis tools that could help source code vulnerability detection on software or web application. Finally, with the third category of the hands-on labs, we present the means that leverage machine learning and NLP that are more potent than the static analysis tools. We teach future IT professionals how the integration of such tools could improve the security assurance of their development at the very early stages of the software development life cycle. We also try to expand their knowledge and be more creative by teaching them to design and implement their own scanner by considering the purpose and goal of their development.

# Chapter 5 Implementation

Pedagogical research has shown that practical laboratory exercises are critically important to successfully enhance future IT professionals' understanding of new technologies such as source code vulnerability detection in the classroom Du and Wang (2008). However, education and training institutions in the United States have found it challenging to keep pace with the growing need for cyber talent. Rowe et al. (2011) demonstrated the need for security programs in the curriculum. They have suggested that curriculums should be organized to include latest cybersecurity standards, report, and analysis techniques which helps future IT professionals for real cyber world vulnerabilities. To fill this gap, this research has developed an instructional framework called SeCodEd to address the need to fill this gap systematically. This chapter will examine the SeCodEd Framework Architecture and build laboratory exercises on top of the framework; each activity challenges future IT professionals to understands the various methods, concepts, applications, and uses cases for Source code and log file analysis technologies. The instructional system is designed in such a way so our undergraduate computer science future IT professionals can further reinforce their IT curriculum.

## 5.1 SeCodEd Hands-on Laboratories

SeCodEd Framework Architecture, as seen in Figure 32, has developed a series of hands-on labs and introduction lectures. These hands-on labs are built upon the first categories of labs, "Source Code Vulnerability Detection Using Static Analysis Tools," from fundamental to intermediate and advanced. Following this, future IT professionals will either continue learning more about other vulnerabilities and static analysis tools or learning about log file parsing and analysis or secure coding practices. Future IT professionals are advised to finish all three levels of hands-on labs in each category and then move on to the rest. Still, it entirely depends on their background knowledge, level of familiarity with the topis, and interest.

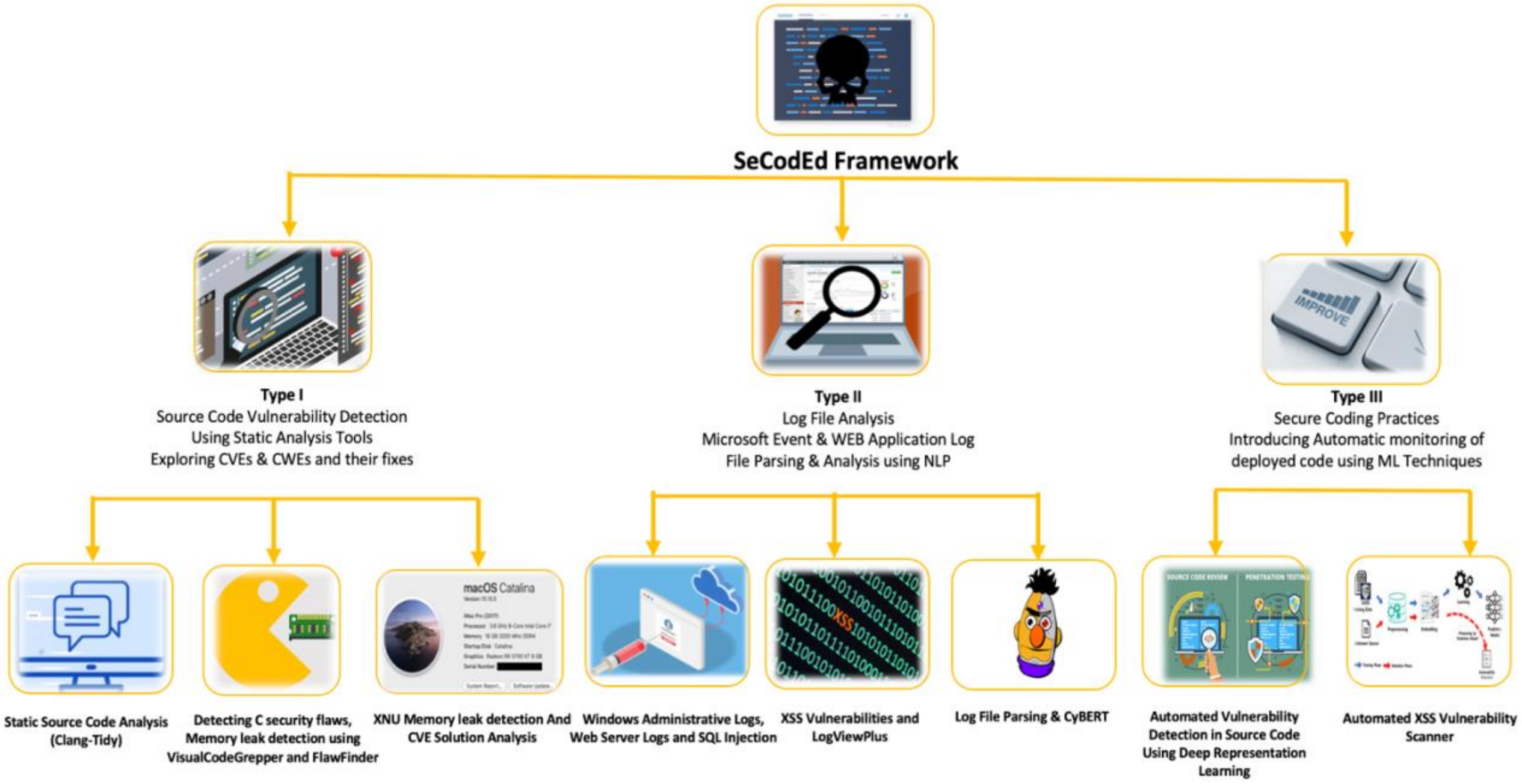

Figure 30 SeCodEd Framework Architecture

## 5.2 Data-Driven Personalized Learning

Data analytics is the process of examining raw data to determine a conclusion about the evaluated information. Several techniques and strategies of data analytics have been automated into mechanical methods and algorithms that work on raw data to make firm predictions. Nowadays, more than 6.3 million students in the U.S. chose at least one online course in the recent past Friedman and Moody (2021). Implementation of data-driven education and a personalized learning approach simulates the highest possible quality of the learning. It allows educators to provide students with precisely what is required on an individual level. Focusing on students, this microlearning-content strategy is making lessons smaller and providing them with flexibility. The process for designing personalized content comprises four primary metrics, namely, accuracy, metacognition, time, and engagement. Metacognition means the student's awareness regarding what he knows or doesn’t know, along with accuracy and time. It can be measured and analyzed through data. In this work, using Pre and Post Survey, all of these matrices are counted for and analyzed. Figure 33 demonstrates future IT professionals ' understanding of the concepts learned through the lab to provide them with proper material for the next learning module through the post-survey.

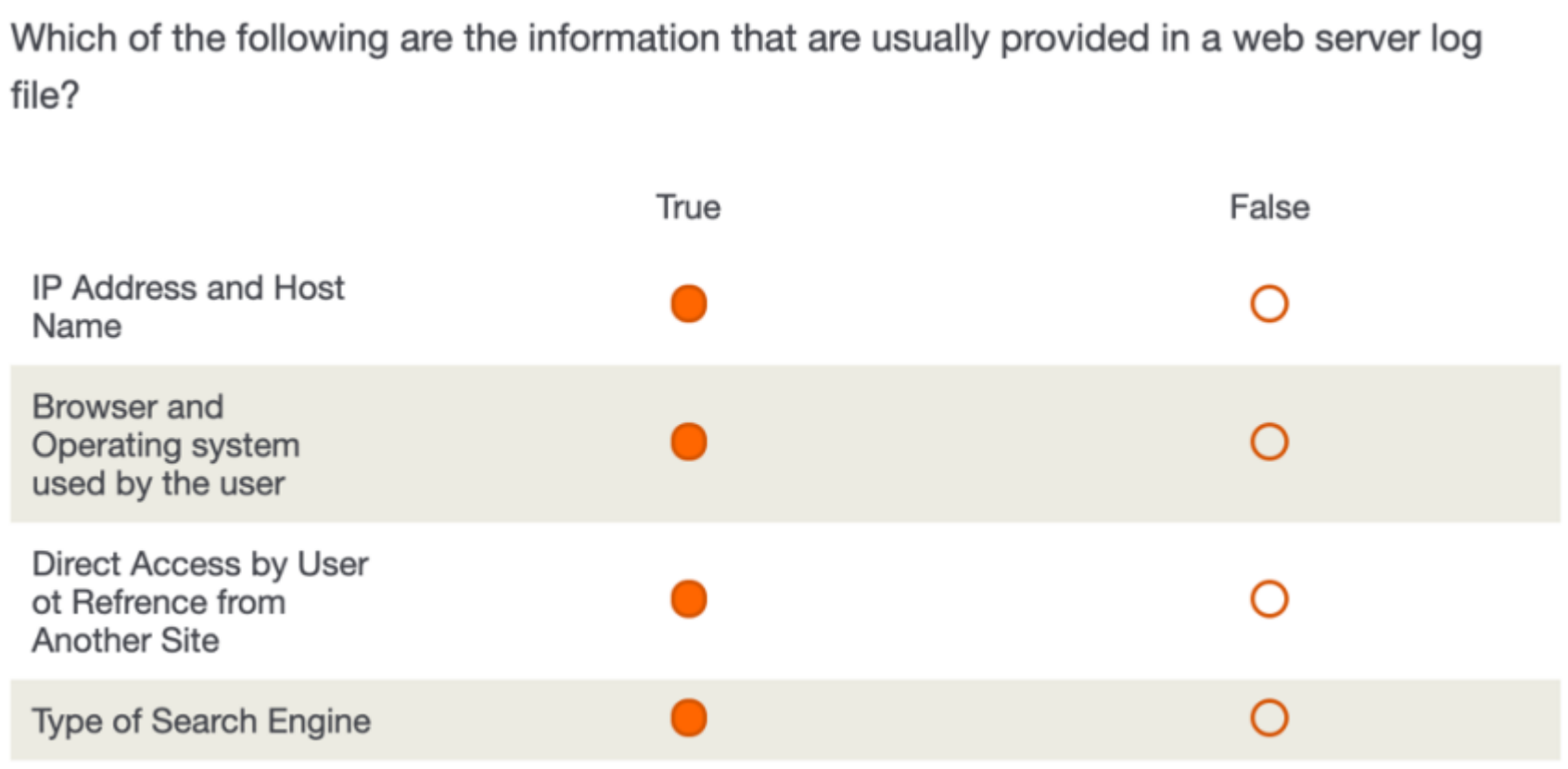

Figure 31 Post Survey question example measuring future IT professionals’ knowledge on the concepts learned

Personalized learning is gaining more attention in computer science education due to pacing learning progress and adapting instructional material for everyone. Among various instructional methods in computer science education, hands-on labs have unique requirements of understanding learner's behavior and assessing learner's performance for personalization (Deng, 2018). In this work, each hands-on lab has its goals and objectives that will allow students to have a free-style approach to do each module at their leisure at home, with all the tools accessible through the web application.

### 5.3 Principal of Designing Source Code Vulnerability hands-on labs

In the first category of labs developed, future IT professionals should understand data structures and some programming skills. Thus, targeting key concepts is ideal for reinforcing the current IT curriculum, improving future IT professionals ' programming skills, and showing them to leverage static analysis tools to form a secure programming habit. Type I of the first category of hands-on labs' purpose is to introduce future IT professionals to the most basic and common type of vulnerabilities, the reason they happen, and what they can lead to. The uncontrolled format string is a type of software vulnerability discovered around 1989 that can be used in security exploits. Originally thought harmless, format string exploits can crash a program or execute harmful code. The problem stems from unchecked user input as the format string parameter in certain C functions that perform formatting, such as printf(). During this hands-on lab, future IT professionals learn how an attack could be executed using three different methods: (1) Format Function of printf, fprintf, scanf, fscanf, (2) Format String, and (3) Format String Parameter. Using a simple noncompliant code example (Figure 34), they will learn how improper usage of Format functions can lead to acceptance of untrusted data that originates from an unauthenticated user.

```
// A simple C program with format
// string vulnerability
# include<stdio.h>

int main (int argc, char ** argv)
{
    char
    buffer[100];
    strncpy (buffer, argv[1], 100);
// We are passing command line argument to printf
    printf(buffer);

return 0;
}
```

Figure 32 Noncompliant Code Example for Format String Attack

By executing this program and using Format string parameters, they have realized how such implementation can have unsecure consequences such as accessing to memory locations of the Heap (Figure 35).

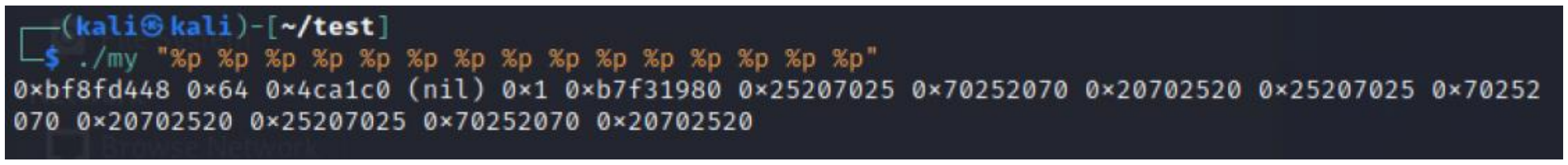

Figure 33 Access to memory location using format string parameters

Finally, students will learn how to leverage static analysis tools such as Clang-Tidy to mitigate and avoid such vulnerabilities from happening and fix the program (Figure 36).

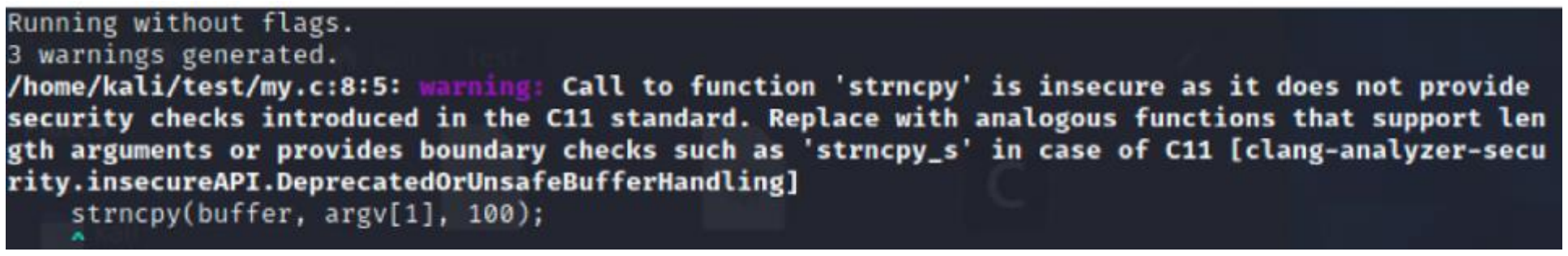

Figure 34 Clang-Tidy Demo of Vulnerability Detection

After fixing the issues, they will use the static analysis tool to rescan their new implementation to confirm the effectiveness of their work. Overall, all the hands-on labs follow the same theme and have the same goal embedded in them. This paper highlights the gaps that exist in the nation's current cybersecurity education and training landscape. It provides a successful framework that holds promise as a model for addressing the skills gap in software vulnerability detection. Targeting key concepts is ideal for reinforcing the current CIS curriculum, improving future IT professionals' programming skills, and equipping them with analysis tools to form a secure programming habit. Figure 37 demonstrates the gaps between curriculum and industry, which are address in this work.

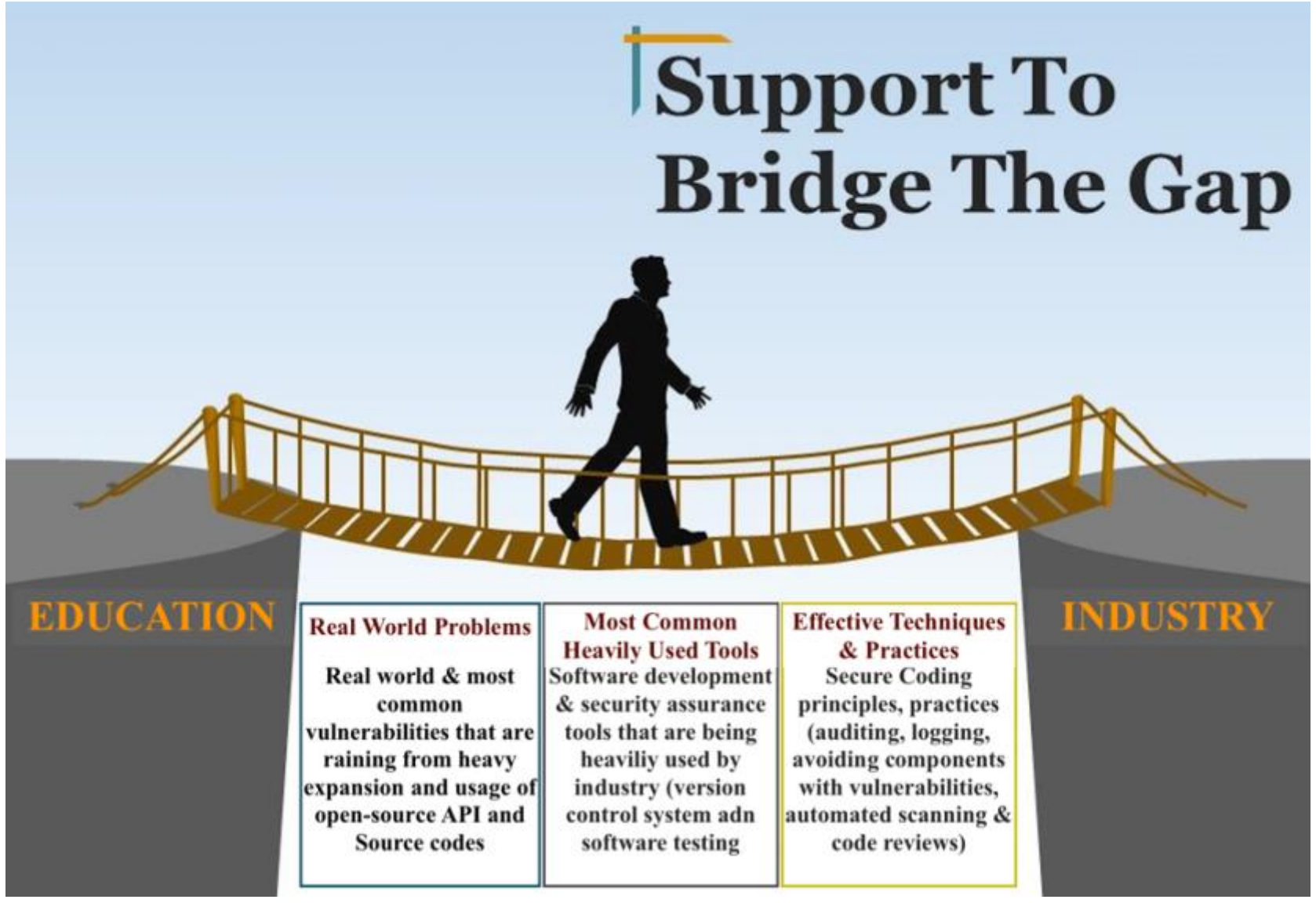

Figure 35 SeCodEd Bridging the Gap between Industry and Education

## 5.4 Platform Architecture

This framework was developed on a web application (piazza) that can be accessed through desktops, laptops, and mobile devices. This free, open-source platform provides several e-learning tools to provide a comprehensive e-learning environment to develop hands-on laboratories. Figure 38 further describes the SeCodEd Website Architecture.

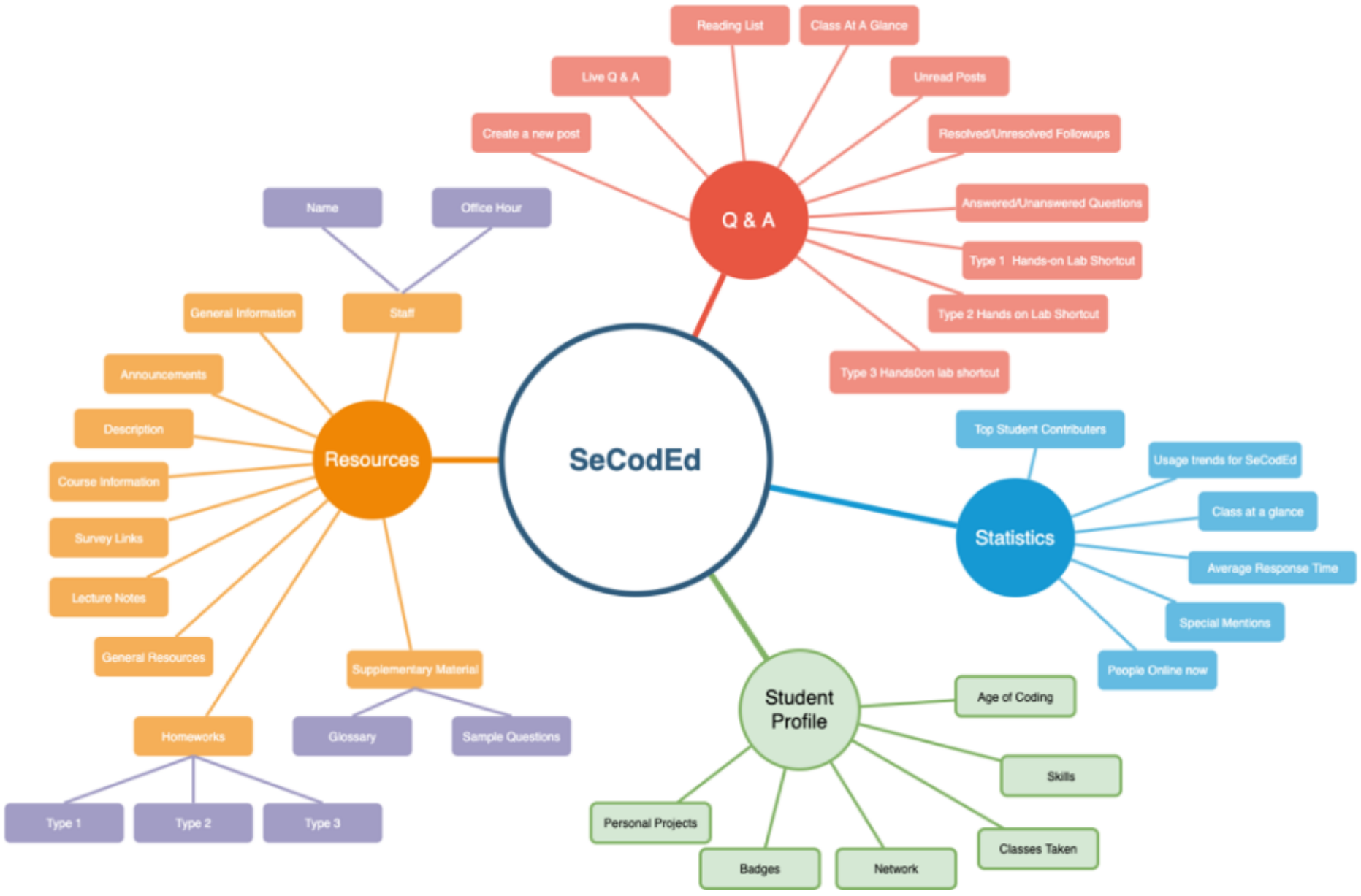

Figure 36 SeCodEd Web Application Architecture

Each module has a personal homepage with course goals and objectives that will allow future IT professionals to have a free-style approach to do each module at their leisure at home, with all the tools accessible through the web application. The hands-on labs are accessible using the following URL: piazza.com/famu/summer2021/secoded with the Access Code of 123. These hands-on labs are easily accessible for all future IT professionals to carry out lab assignments without any exceptional computing environment or special user privileges. Once they have created their username and password, they will access the web portal. A forgotten hyperlink on the login page will allow students to access their credentials if they have forgotten their username or password. To make sure future IT professionals are ready to start working on the lab and evaluate their improvements, we first conduct the pre-survey. If they mark the Pre-Survey link, they can proceed to the hands-on lab itself. Figure 39 demonstrates the assurance on students filling out the pre-survey.

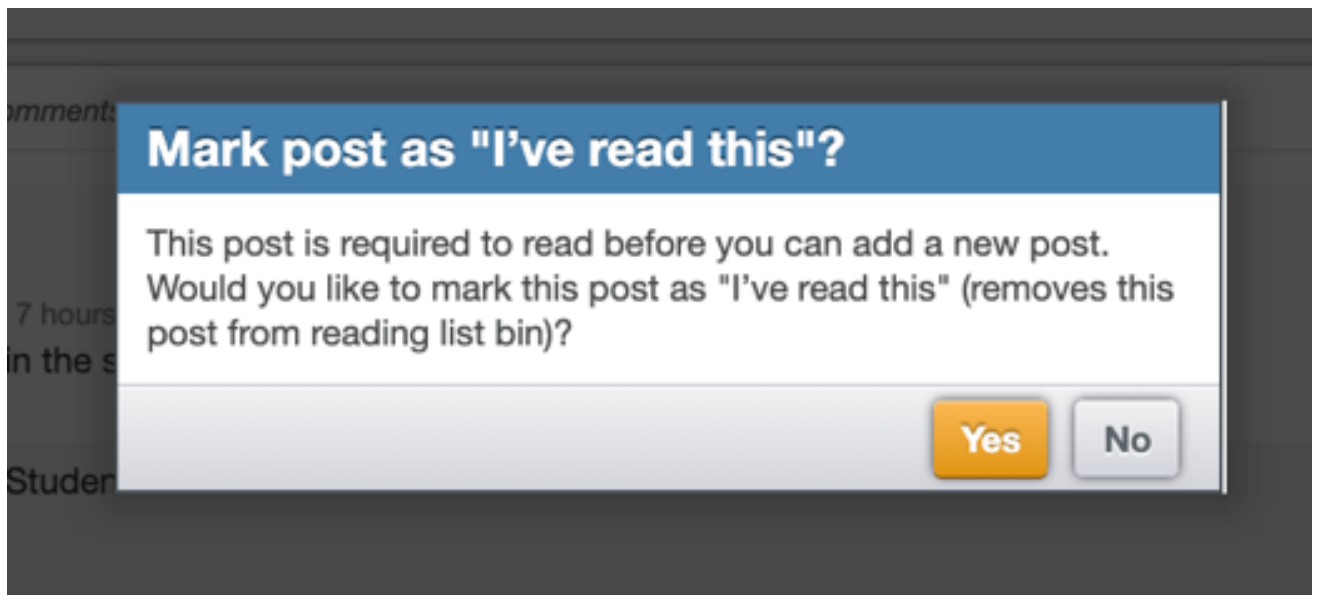

Figure 37 Assurance of doing pre survey before the hands-on lab

Future IT professionals can always see all the tasks they have to deliver on their reading list, as shown in Figure 40. They will complete a lesson in each hands-on lab that gives them an overview of the subject and topic related to Source code vulnerability, Log File Analysis, and Secure Programming Habits.

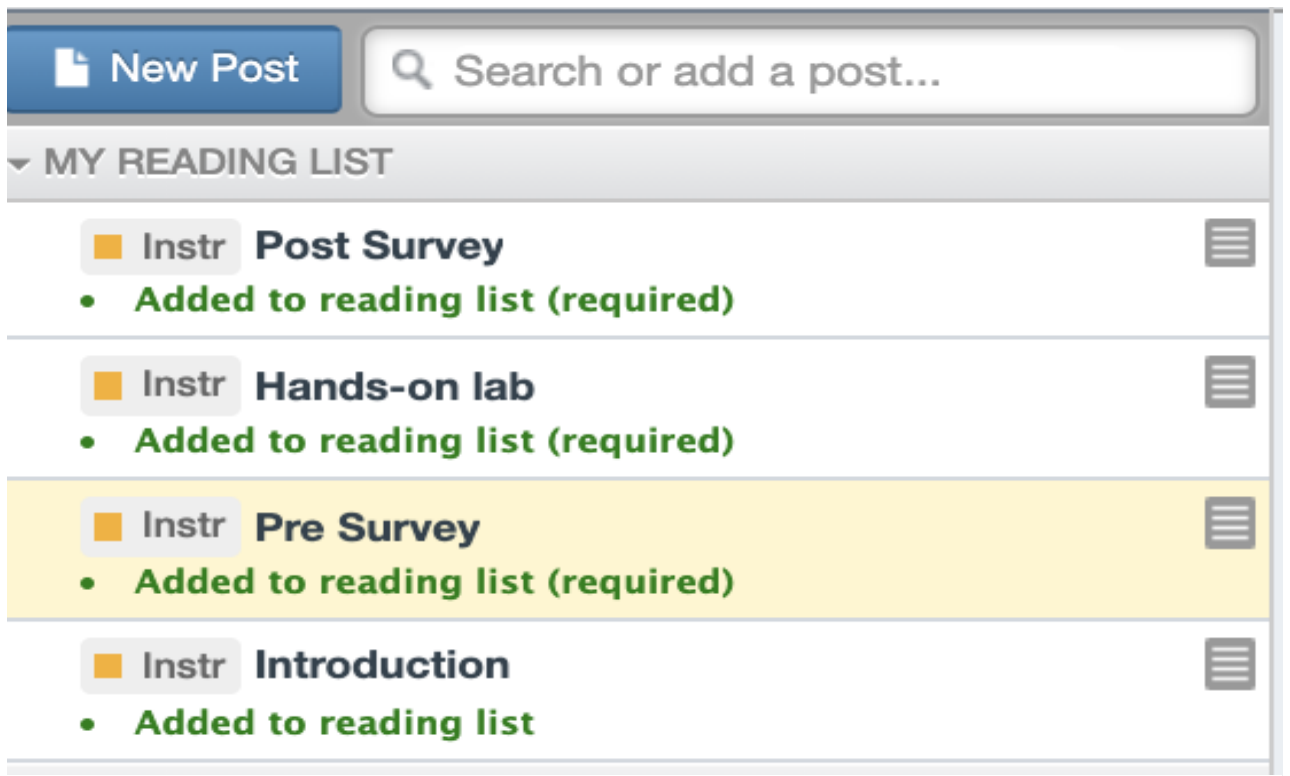

Figure 38 Reading list Demo

### 5.5 Dissemination

Educational materials are constantly created, disseminated, and implemented across academic settings, with the ultimate aim of improving curricular training. Dissemination involves planned efforts to spread experimentation with or adopt the tool (Figure 41), while implementation refers to efforts to integrate it into usual care (Held et al., 2016). Source code vulnerability detection and secure programming education tools come in many formats and can be disseminated using different strategies.

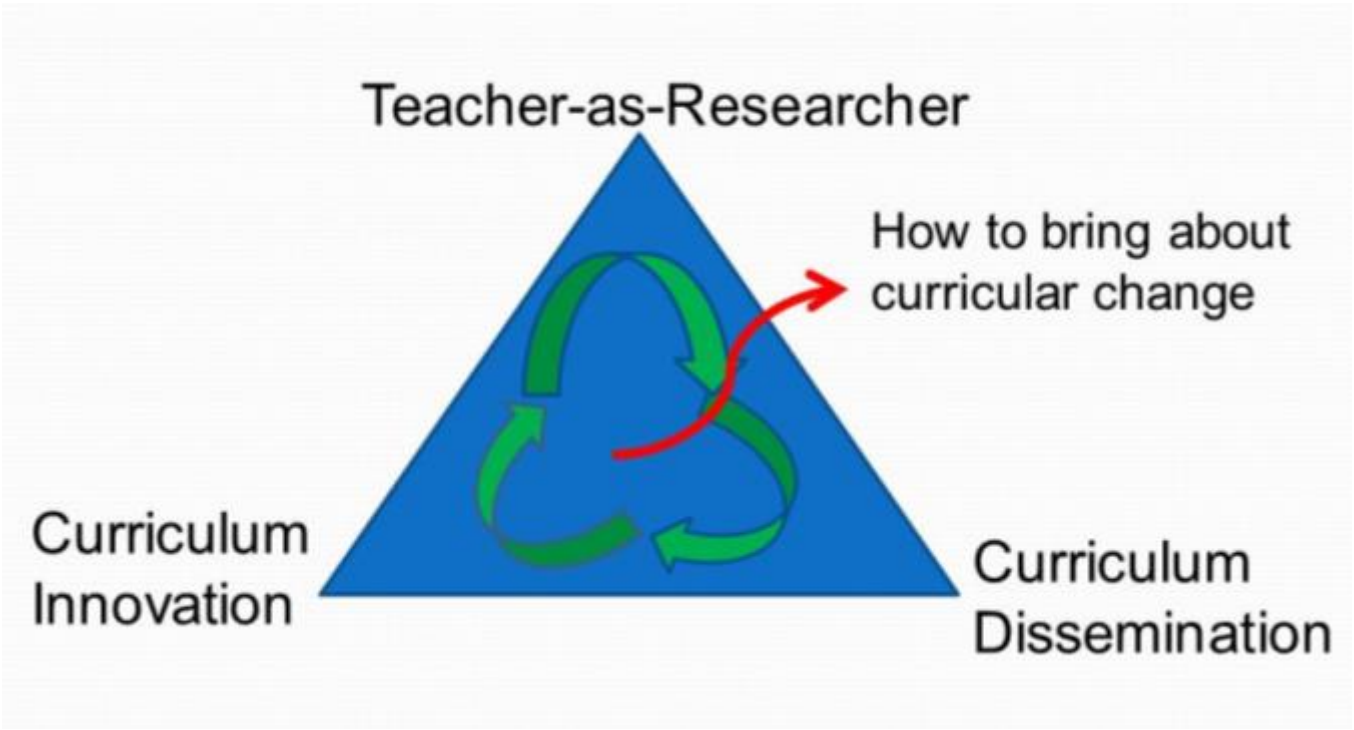

Figure 39 Planned Curriculum Architecture

Studies have shown that passive dissemination, such as handing out or mailing printed materials, is less effective than active dissemination strategies (Grimshaw et al., 2001). In this work, we have detailed the dissemination and implementation of an educational tool, SeCodEd, designed to introduce future IT professionals to Secure coding practices, most common vulnerabilities, and attacks, source code vulnerability detection using both static and log file analysis. Given the workforce gap and the missing required talent in graduates, SeCodEd focused on modalities that can be practiced independently. The goals of SeCodEd were to raise awareness on vulnerabilities that are rising with the expanded usage of open-source and prepare students to join the industry workforce with proper skills.

### 5.6 Student Feedback

This section will present the data analysis for the pre and post-survey. The results have been collected over the summer semester 2021 for a total of three months through the following courses: CIS 4360 (Introduction to Computer Security), CAP 5626 (Artificial Intelligence), COP 3014 (Fundamentals of programming), COP 3710 Database Management) and PHY 2054 (College Physics II). Table 6 represents a summary of the distribution details. Assessment feedback is extremely helpful in improving learner's decision-making and learning outcomes. However, despite the importance of feedback, it is often underutilized in higher education. Therefore, the SeCodEd framework heavily utilizes students' feedback 2 provide learning modules that will address students' needs comparative to comprehending industrial technology and the skills that are missing right now in the workforce gap. This socioecological approach which provides an exciting interpretation of human experience is considered an interplay of individual and environmental factors. This approach can be used in the adoption of technology in education. In each case, the explanation of the phenomena was enriched through the recognition of how social systems interact with each other and the individuals, and in doing so, provided a valuable framework for understanding complex human social issues (Bronfenbrenner, 1977; Ivankova & Plano Clark, 2018; Zhao & Frank, 2003).

Table 6 Hands-on Lab Distribution summary

| Lab Category | Student Background | Course | Total Participants |
|---|---|---|---|
| Type I | Data Structures, Algorithms, and Generic C, C++ Programming | CIS 4360, COP 3014, PHY 2054 | 35 |
| Type II | File Organization and SQL Database Concepts | CAP 5626, COP 3710 | 39 |
| Type III | Advanced Programming, Machine Learning | CAP 5626 | 15 |

### 5.6.1 Response Rate

Students enrolled in the classes that the survey was distributed was 74, and 45 people participated in the study in total. Figure 42 demonstrates the total number of participants in each category of the hands-on labs. It has been observed that, even though the survey is not for grading purposes, only so few people were willing to respond to them. The survey questions have been designed to take 3 minutes only. Thee-Brenan (2014) has reported that many factors contribute to low response rates, including poll fatigue and developments in technology that allow people to avoid answering the phone. Maintaining the credibility of survey research is essential. Without surveys, we would not have a lot of crucial data. We tried to encourage students to participate in surveys by giving them bonus points.

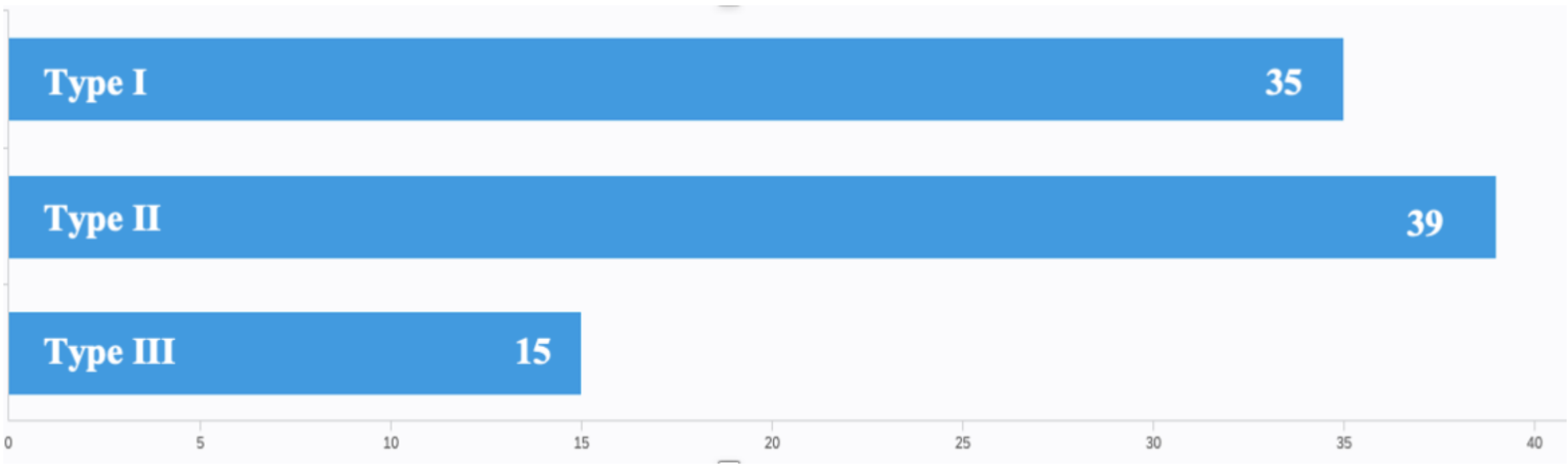

Figure 40 Total Number of Participants in each category of the hands-on lab

### 5.6.2 Pre-Survey

During the Pre-Survey, students were asked about their level of familiarity with the concepts that were going to be presented and a few quiz questions to test their knowledge and understanding after the hands-on lab. The study subjects were 39 junior students of computer science and engineering majors and 7 graduate students of the computer science major. The findings of the questionnaire are presented as follows. Students were first asked to rate their level of familiarity with specific concepts related to each lab category that would be covered in the lab.

**Type I:**

The first category of these hands-on labs was focused on teaching students the following concepts: (1) Source Code Vulnerabilities, (2) Common Vulnerabilities and exposures, and Common Weaknesses Enumeration. (3) Argument Type and Input Validation, (4) Memory Errors and Leak, (5) Static Code Analysis. Figure 43 Demonstrates that 75% of the participants did not know Source Code Vulnerability. Only 19% were somewhat familiar with the concepts, which indicates the vital need for raising awareness on source code vulnerability and detection techniques in the curriculum. Figure 44 demonstrates that almost 80% of students did not know static analysis tools before starting to work on the LAB, and 20% were somewhat familiar with the tools.

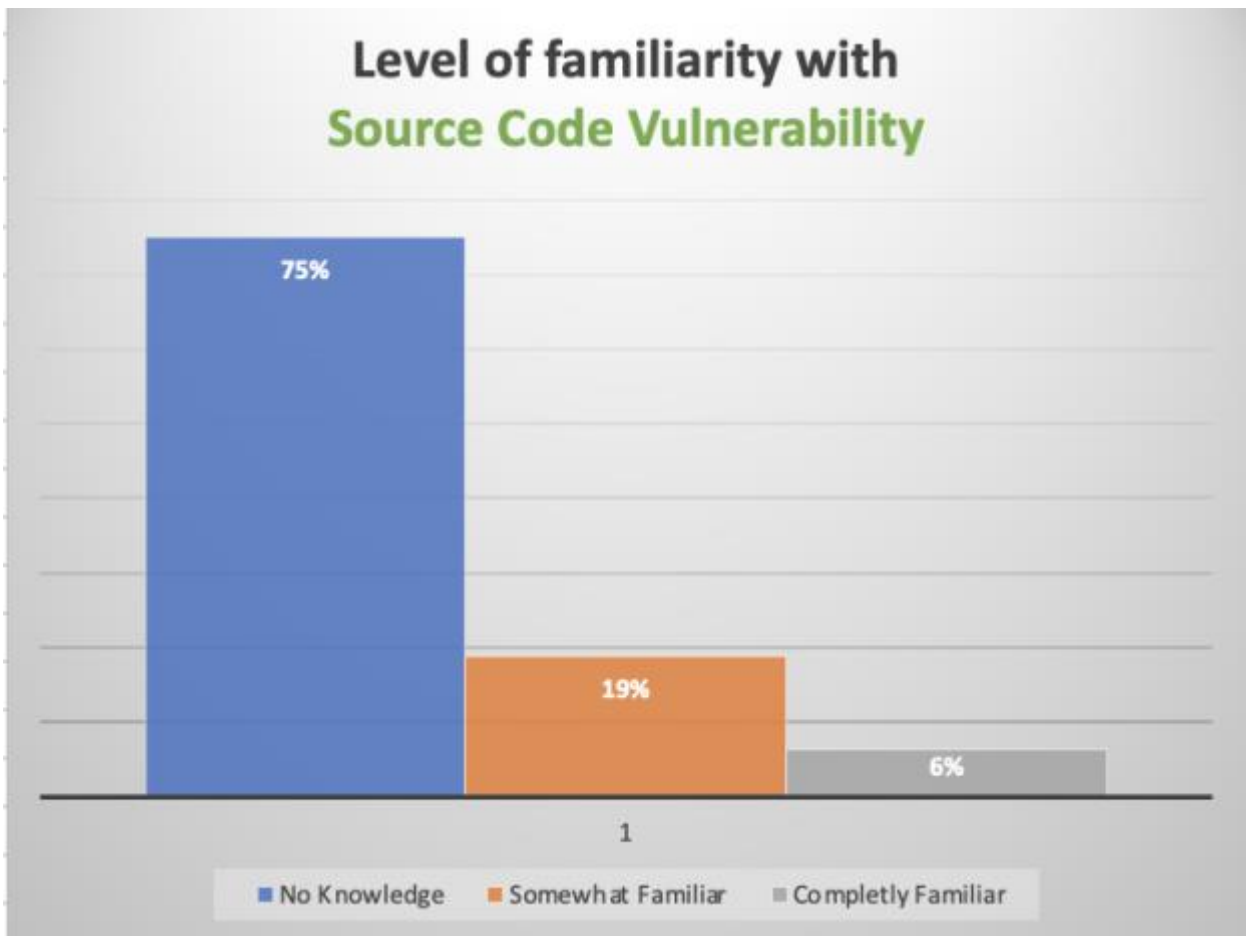

Figure 42 Students' level of familiarity with Static Code Analysis (Pre-Survey)

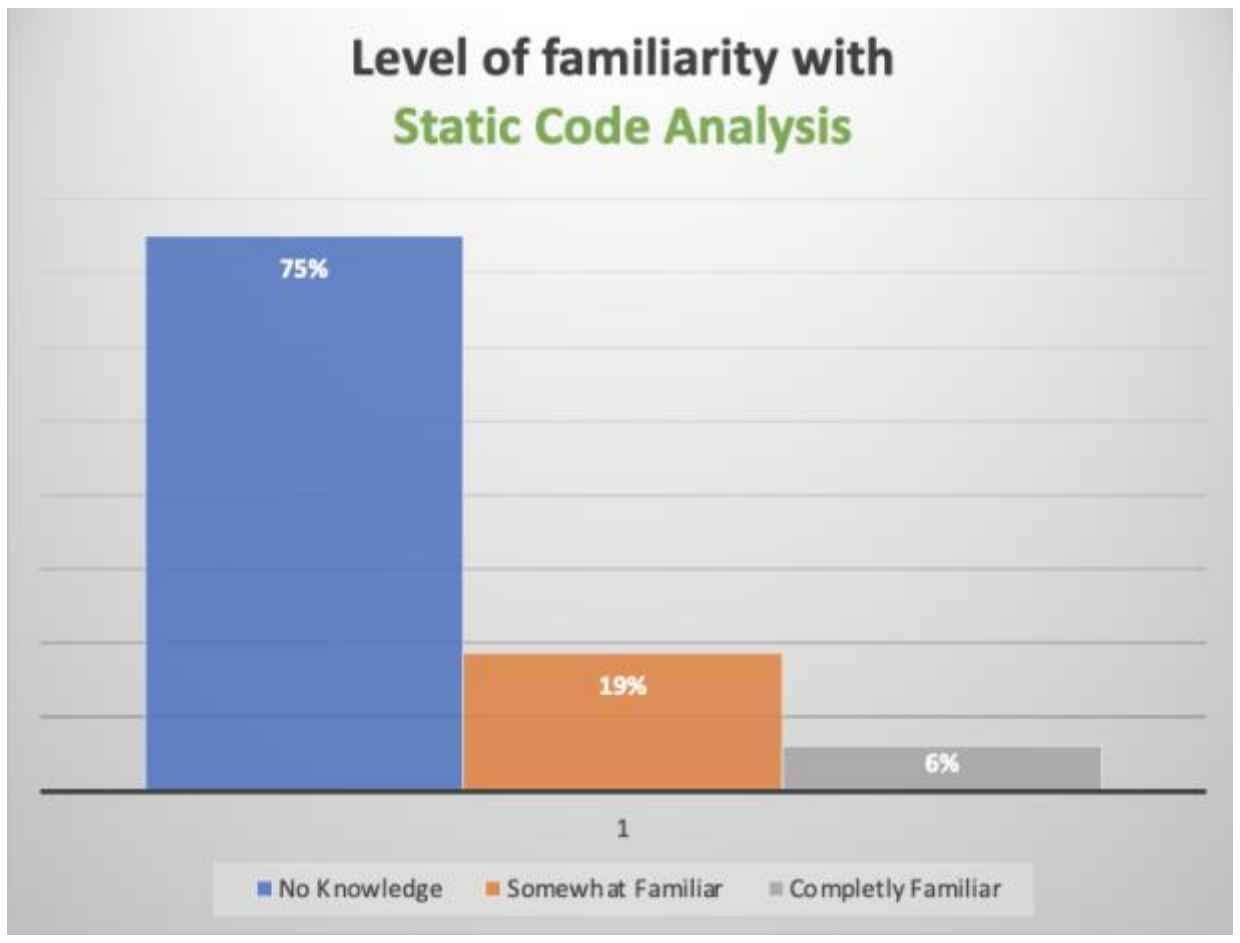

Figure 41 Students' level of familiarity with Source Code Vulnerability (Pre-Survey)

The student's level of familiarity with CVEs & CQEs, as well as Memory Errors & leak, Argument Input Validation before working on the lab, is presented in Table 7.

Table 7 Student's level of familiarity with other concepts covered in the First category (Pre-Survey)

| **Concept** | **No knowledge** | **Intermediate** | **Completely Familiar** |
|---|---|---|---|
| CVEs & CWEs | 78% | 16% | 6% |
| Memory Errors & Leaks | 59% | 34% | 6% |
| Argument Input Validation | 72% | 25% | 3% |

Figure 45 represents student's familiarity level with the tools presented in this category of the lab (0 represents not familiar at all, 5 represents somewhat familiar and 10 represents completely familiar).

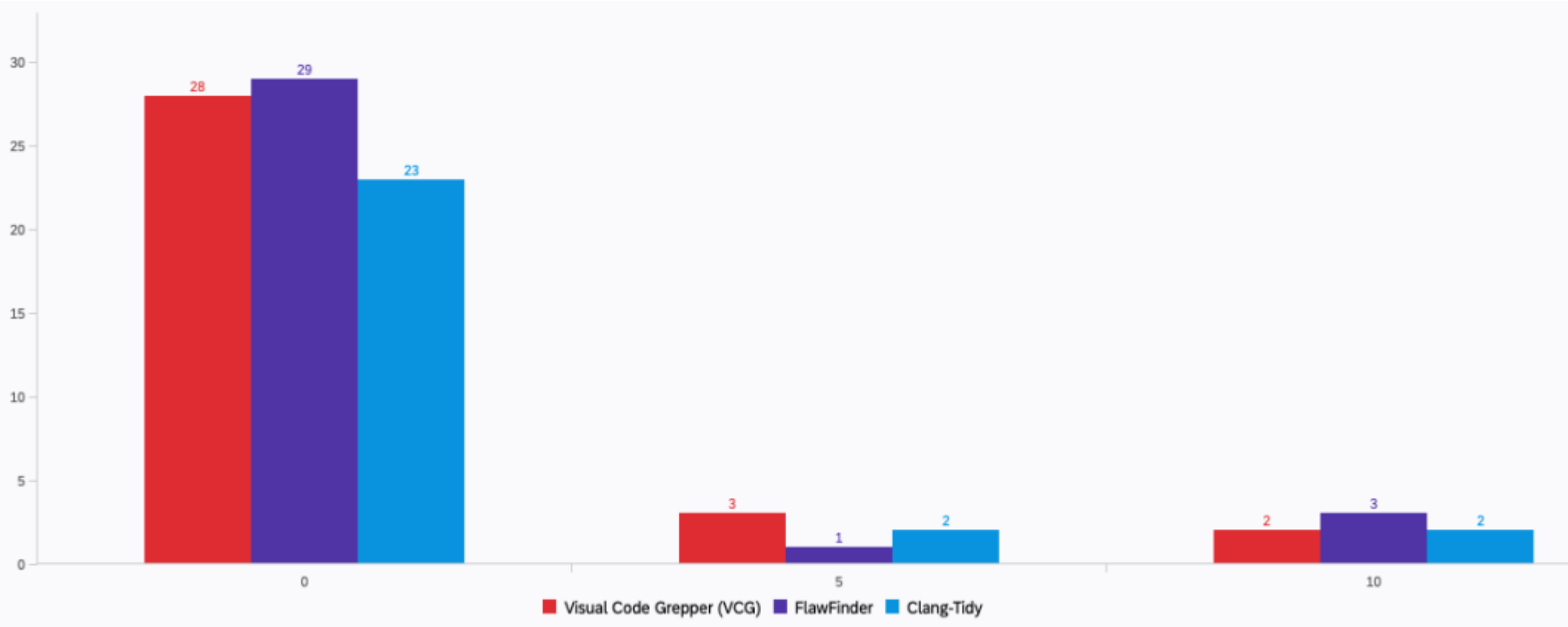

Figure 43 Students' level of familiarity with tools covered in first category of hands-on labs

**Type II:**

The second category of these hands-on labs was focused on teaching students with the following concepts: (1) Log File Analysis, (2) Log File Parsing (3) Common Web Attacks, (4) System/Web server event log files, (5) OWASP Top 10 Security Risks. Figure 46 Demonstrates that 72% of the participants had no knowledge of Log File Parsing and only 25% of them were somewhat familiar with the concepts, which indicates the vital need for raising awareness on log file analysis, its benefits, and how it could be leveraged towards source code vulnerability detection in the curriculum. Figure 47 demonstrates that 56% of students had no knowledge about Log File Analysis before starting to work on the LAB, and 42% were somewhat familiar with the tools.

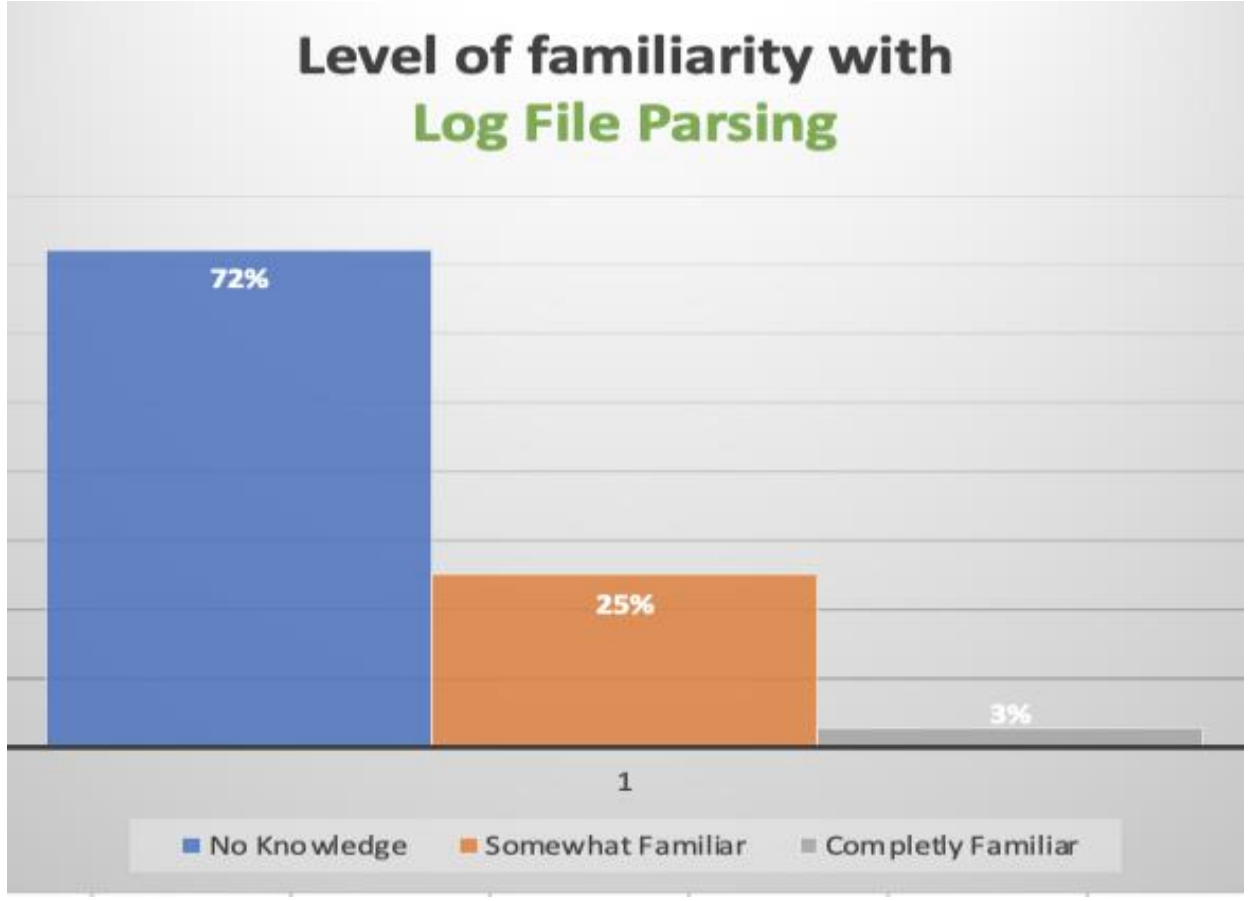

Figure 44 Students' Level of familiarity with Log File Parsing (Pre-Survey)

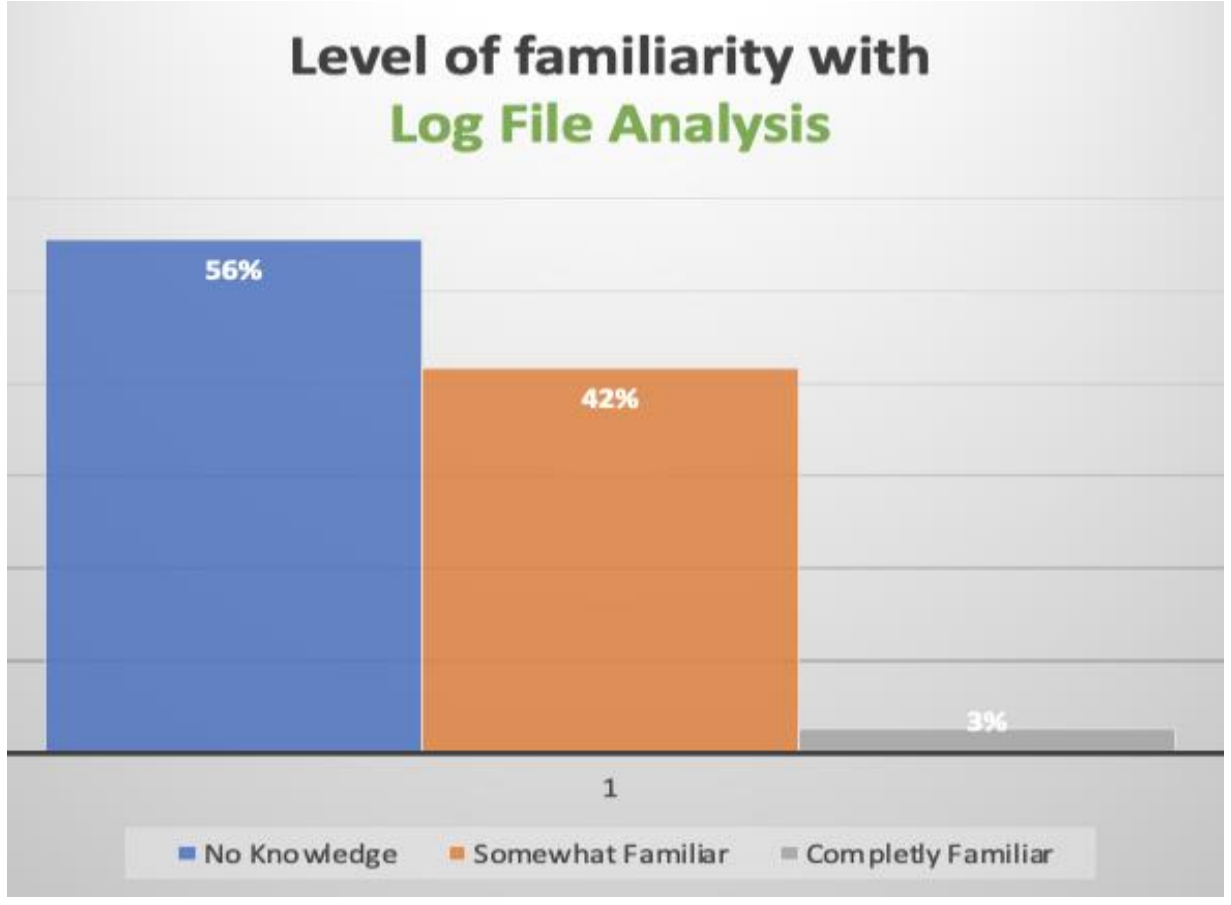

Figure 45 Students' level of familiarity with log file analysis (Pre-Survey)

Student's level of familiarity with Common Web Attacks, System/Web server event log files, OWASP Top 10 Security Risks before working on the lab is presented in Table 8.

Table 8 Student's level of familiarity with other concepts covered in the Second category (Pre-Survey)

| Concept | No knowledge | Intermediate | Completely Familiar |
|---|---|---|---|
| Common Web Attacks | 78% | 19% | 3% |
| System/Web server event log files | 81% | 19% | 0% |
| OWASP Top 10 Security Risks | 81% | 19% | 0% |

Figure 48 represents student's familiarity level with the tools presented in this category of the lab (0 represents not familiar at all, 5 represents somewhat familiar and 10 represents completely familiar).

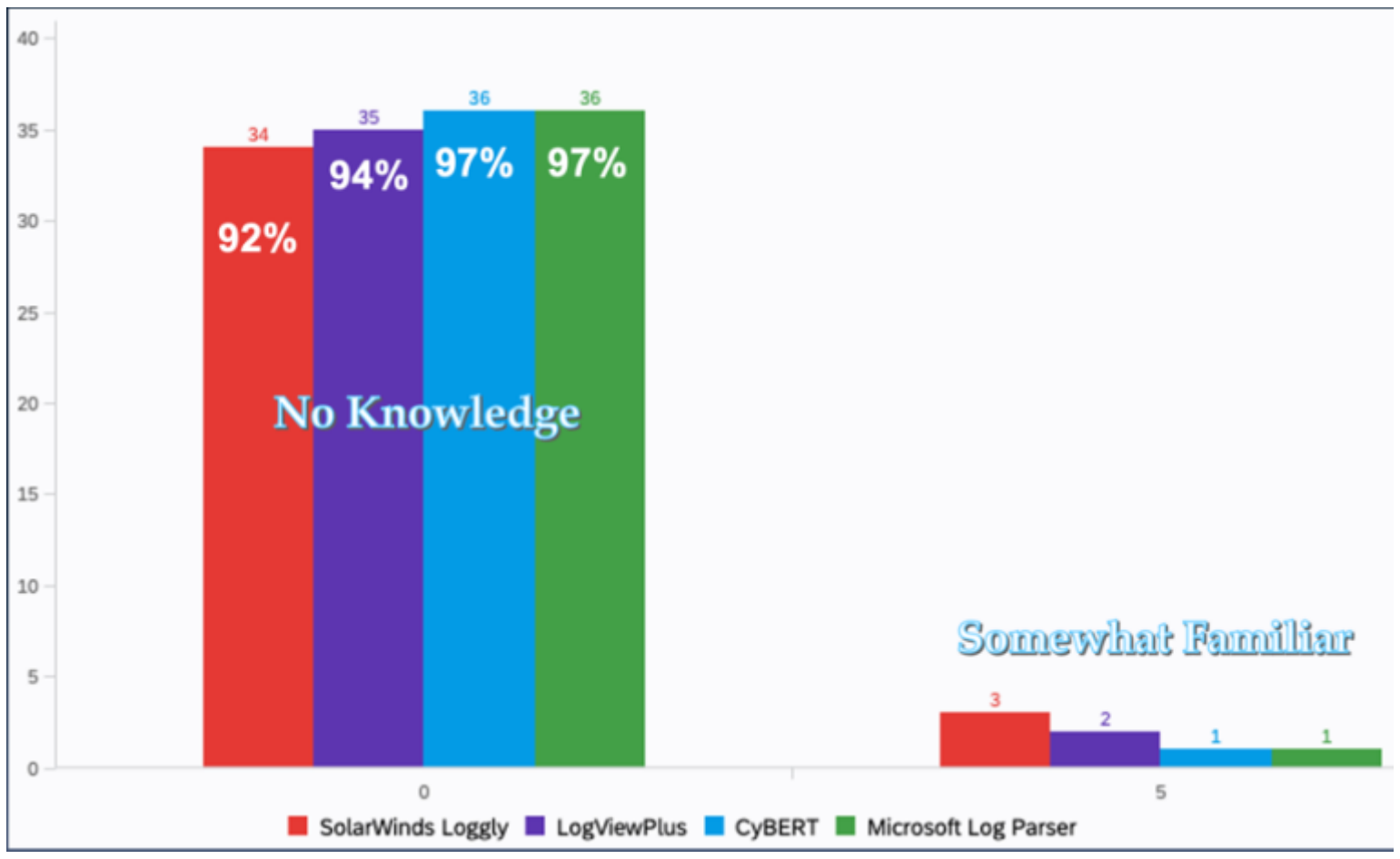

Figure 46 Students' level of familiarity with tools covered in second category of hands-on labs

**Type III:**

The second category of these hands-on labs was focused on teaching students the following concepts: (1) Application of Machine Learning & Natural Language Processing in vulnerability detection, (2) Automated Vulnerability Detection Engines, (3) Secure Programming Habits. Figure

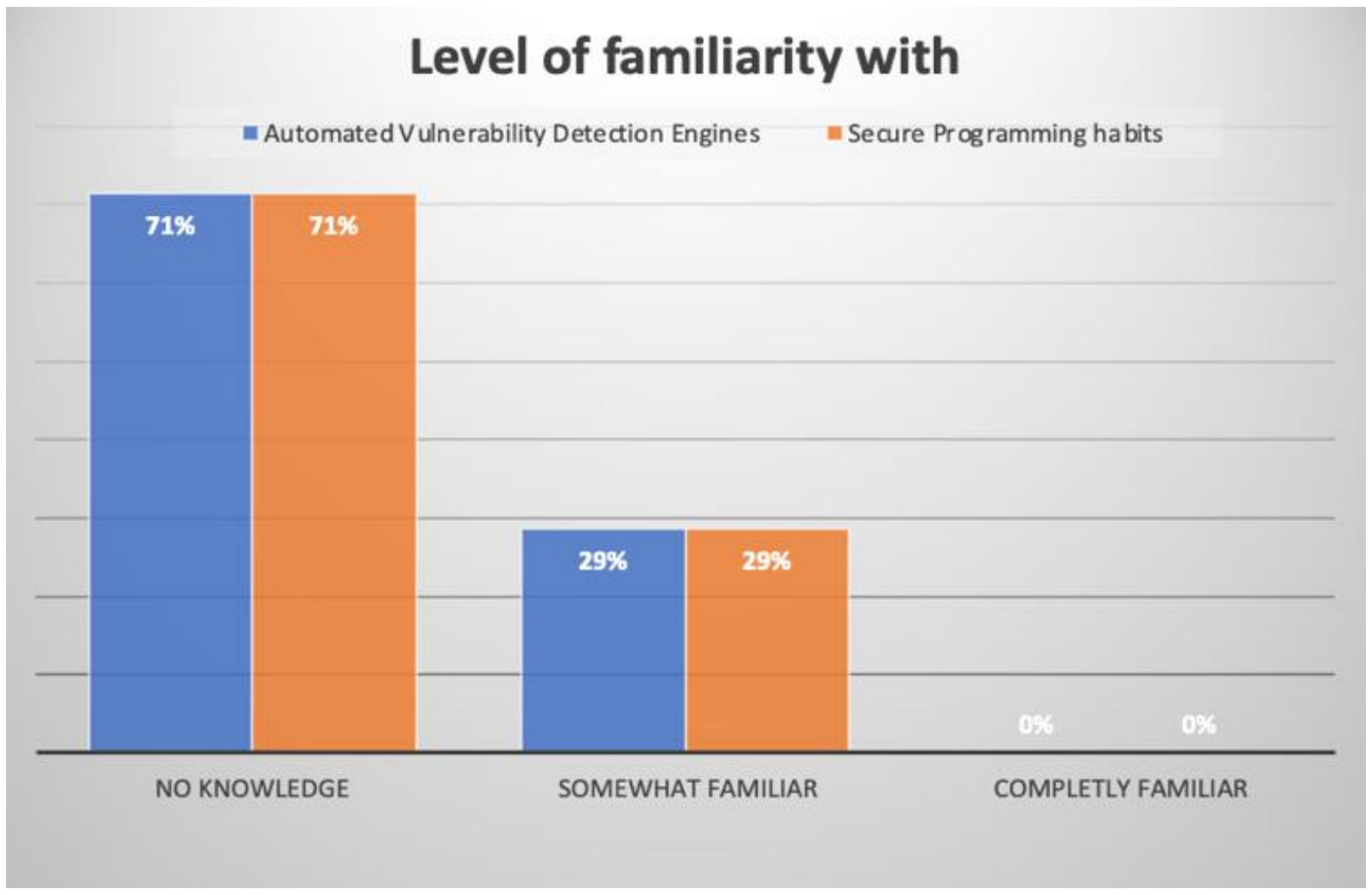

Figure 47 Students' level of familiarity with secure programming habits & Automated vulnerability detection engines (Pre-Survey)

49 Demonstrates that 71% of the participants had no knowledge of Automated Vulnerability Detection Engines and Secure Programming Habits.

Only 29% of them were somewhat familiar with the concepts, which indicates the vital need for raising awareness on secure programming habits and equip students with the tools that they could learn how to use and leverage to improve the security level of their future developments as well as develop the skills that are currently being missed in the industry by the graduates in their curriculum. The student's level of familiarity with the Application of Machine Learning & Natural Language Processing in vulnerability detection is presented in Table 9.

Table 9 Student's level of familiarity with other concepts covered in the Third category (Pre-Survey)

| Concept | No knowledge | Intermediate | Completely Familiar |
|---|---|---|---|
| ML & NLP in Vulnerability Detection | 63% | 38% | 0% |

These findings indicate that there are opportunities to further enhance the Computer Science Curriculum to raise awareness and prepare students on source code vulnerabilities and secure programming.

### 5.6.3 Post Survey

After students complete, their hands-on students are promptly asked to complete a post-survey to assess further and improve labs for future use. This section presents student's feedback (based on post-survey) after finishing each category of the lab.

**Type I:**

Students have demonstrated an overwhelming knowledge enhancement after finishing the first category of hands-on lab on Source Code Vulnerability (Beginner Level). Figures 50 and 51 demonstrate students' knowledge enhancement on Source Code Vulnerability and Static Code

Analysis. The published results are based on 39 undergraduate students who have participated in this activity

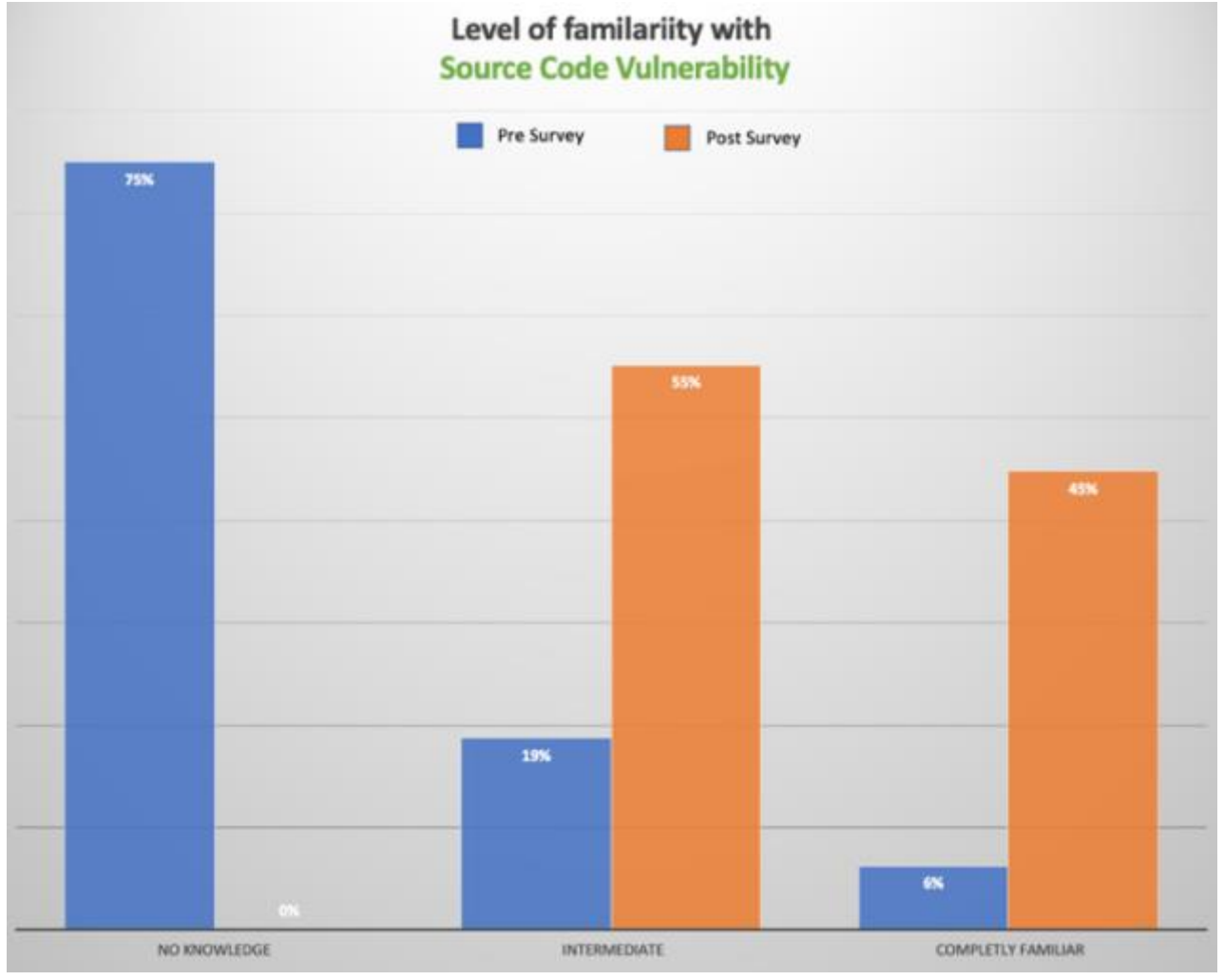

Figure 49 Familiarity level improvement on Source Code Vulnerability (Pre & Post Survey)

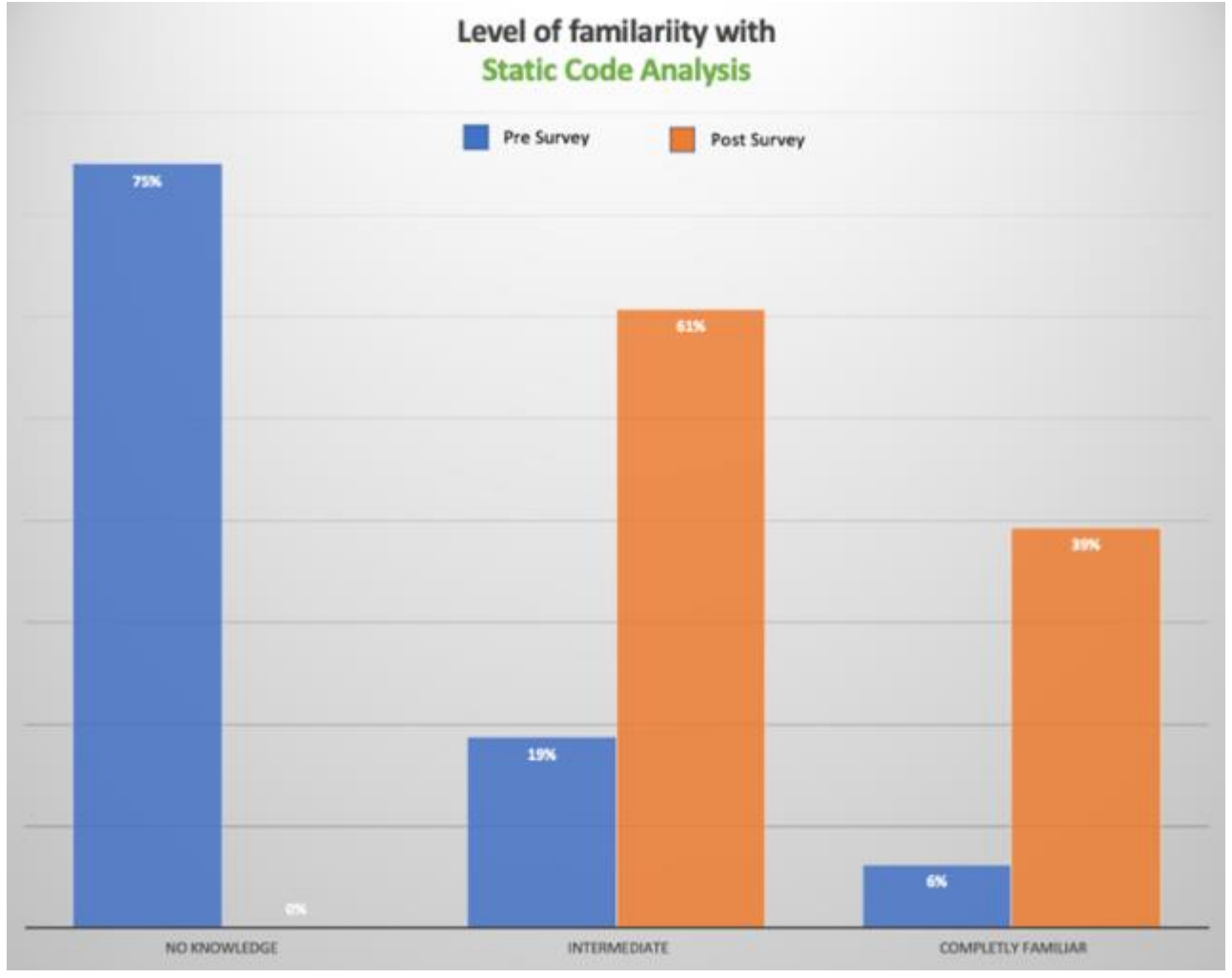

Figure 48 Familiarity level improvement on Static Code Analysis (Pre & Post Survey)

The student's level of familiarity (after finishing the hands-on experience) with the concepts covered in the first category of hands-on labs is presented in Table 10.

Table 10 Student's level of familiarity with other concepts covered in the First category (Post Survey)

| Concept | No knowledge | Intermediate | Completely Familiar |
|---|---|---|---|
| CVEs & CWEs | 0% | 52% | 48% |
| Memory Errors & Leaks | 0% | 63% | 37% |
| Argument Input Validation | 0% | 48% | 52% |

Results from the first category indicate that hands-on lab was the most successful in improving students' knowledge in the following order: 1-Argument Input Validation, 2- CVEs & CWEs, 3- Source Code Vulnerability, 4- Static analysis, 5- Memory Errors & Leaks. This conclusion is based on the percentage of thoroughly familiar students with the concepts after finishing the lab.

**Type II:**

For this category, students (30 undergraduate and 7 graduate Participants) also have demonstrated an overwhelming knowledge enhancement after finishing the second category of hands-on lab on Log File Analysis (Beginner and Advanced Level). Figures 52 and 53 demonstrate students' knowledge enhancement on Log file parsing and analysis. Undergraduate students have shown the highest amount of interest in this category of the hands-on lab.

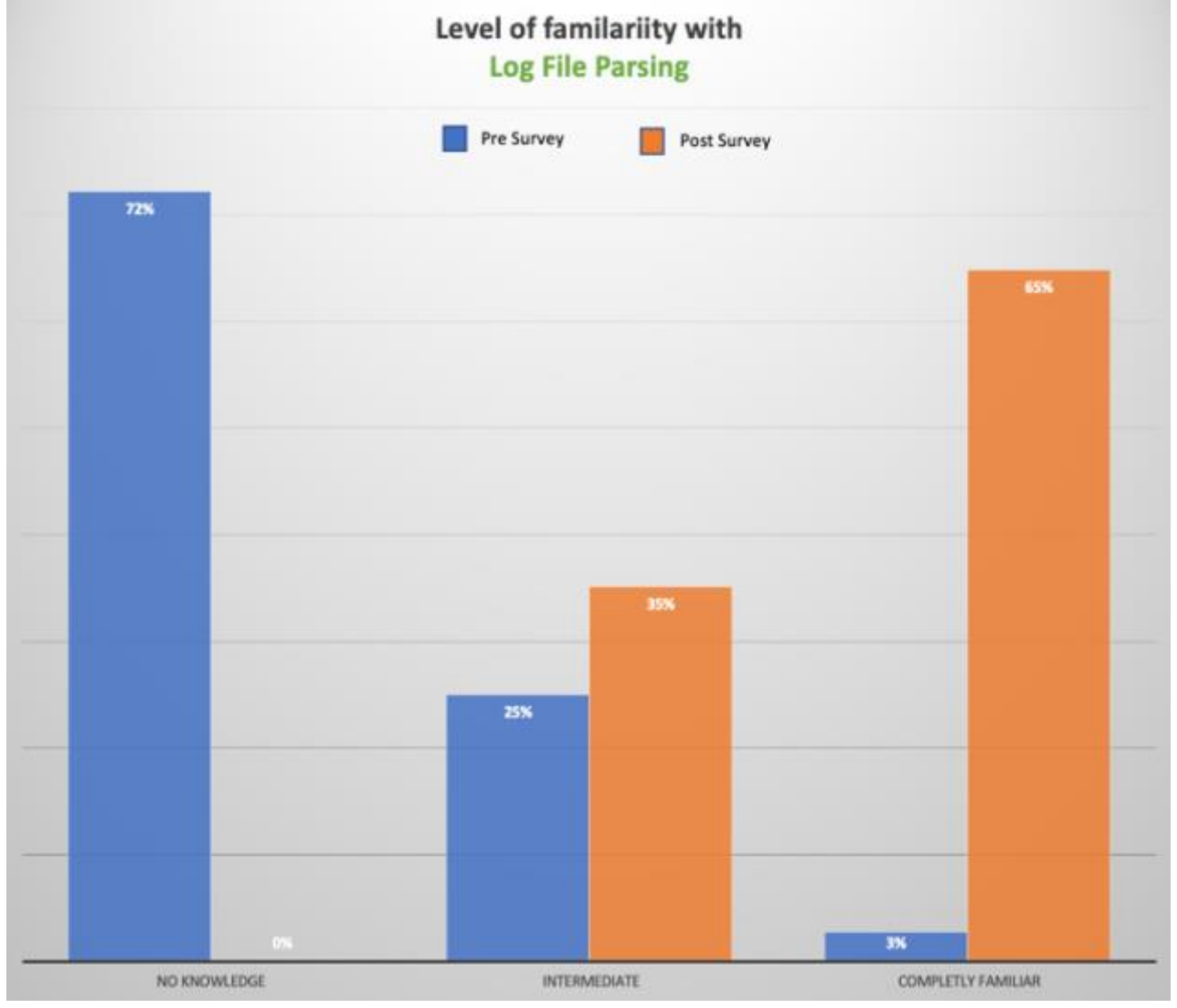

Figure 51 Familiarity level improvement on Log File Parsing (Pre & Post Survey)

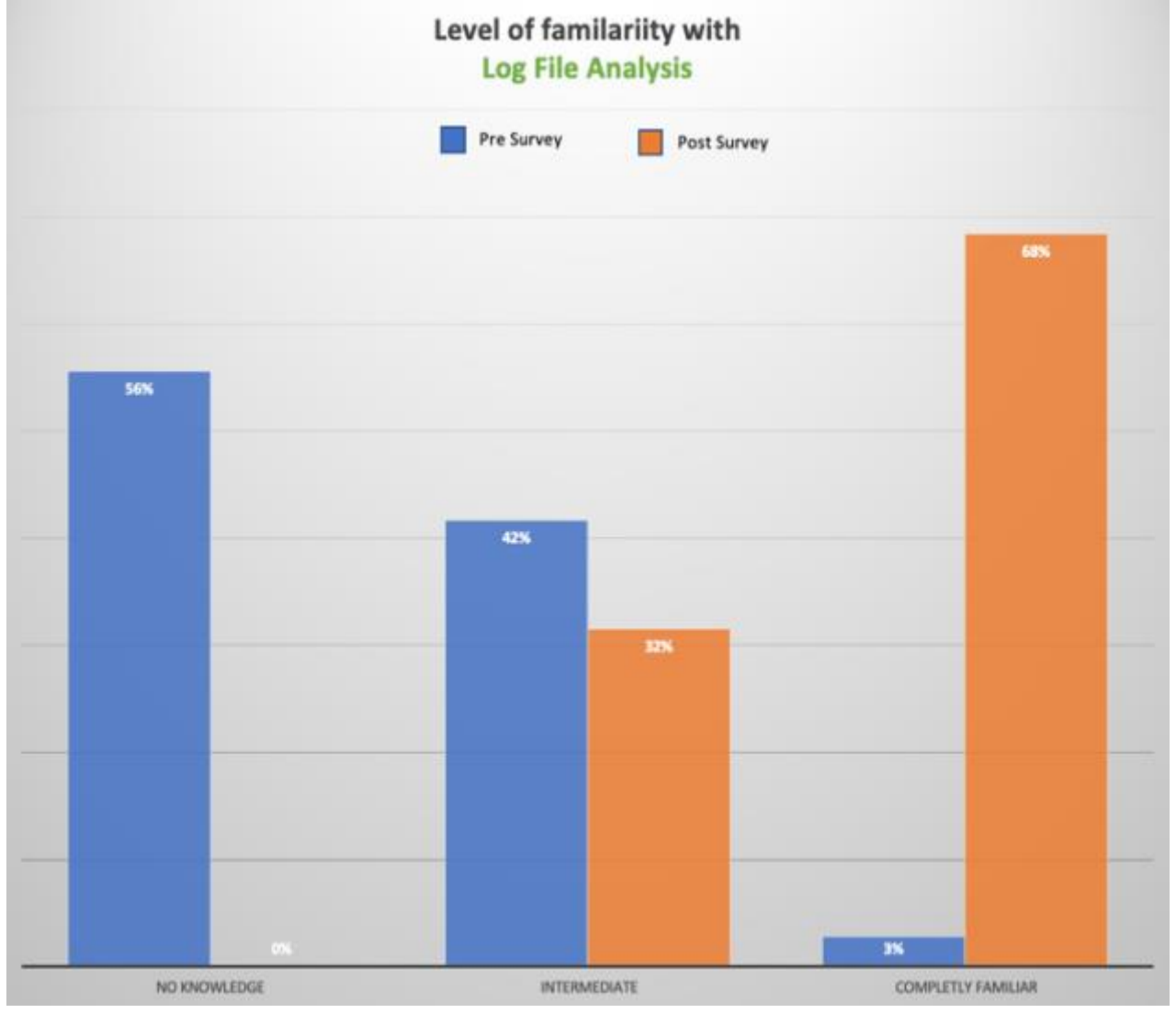

Figure 50 Familiarity level improvement on Log File Analysis (Pre & Post Survey)

The student's level of familiarity (after finishing the hands-on experience) with the concepts covered in the first category of hands-on labs is presented in Table 11.

Table 11 Student's level of familiarity with other concepts covered in the Second category (Post Survey)

| **Concept** | **No knowledge** | **Intermediate** | **Completely Familiar** |
|---|---|---|---|
| Common Web Attacks | 0% | 33% | 67% |
| System/Web server event log files | 0% | 34% | 66% |
| OWASP Top 10 Security Risks | 0% | 28% | 73% |

Results from the first category indicate that hands-on lab was the most successful in improving students' knowledge in the following order: 1-OWASP Top 10 Security Risks, 2- Log File Analysis, 3- Common Web Attacks, 4- Log File Parsing, 5- System/Web server event log files. This conclusion is based on the percentage of thoroughly familiar students with the concepts after finishing the lab.

**Type III:**

For this category, students (16 Graduate students) also have demonstrated an overwhelming knowledge enhancement after finishing the third category of hands-on lab on Secure Programming habits (Advanced Level). Figures 54 and 55 demonstrate students' knowledge enhancement on Secure Programming Habits and Automated Vulnerability Detection Engines. Graduate students have shown the highest amount of interest in this category of the hands-on lab, among all other types that they have worked on.

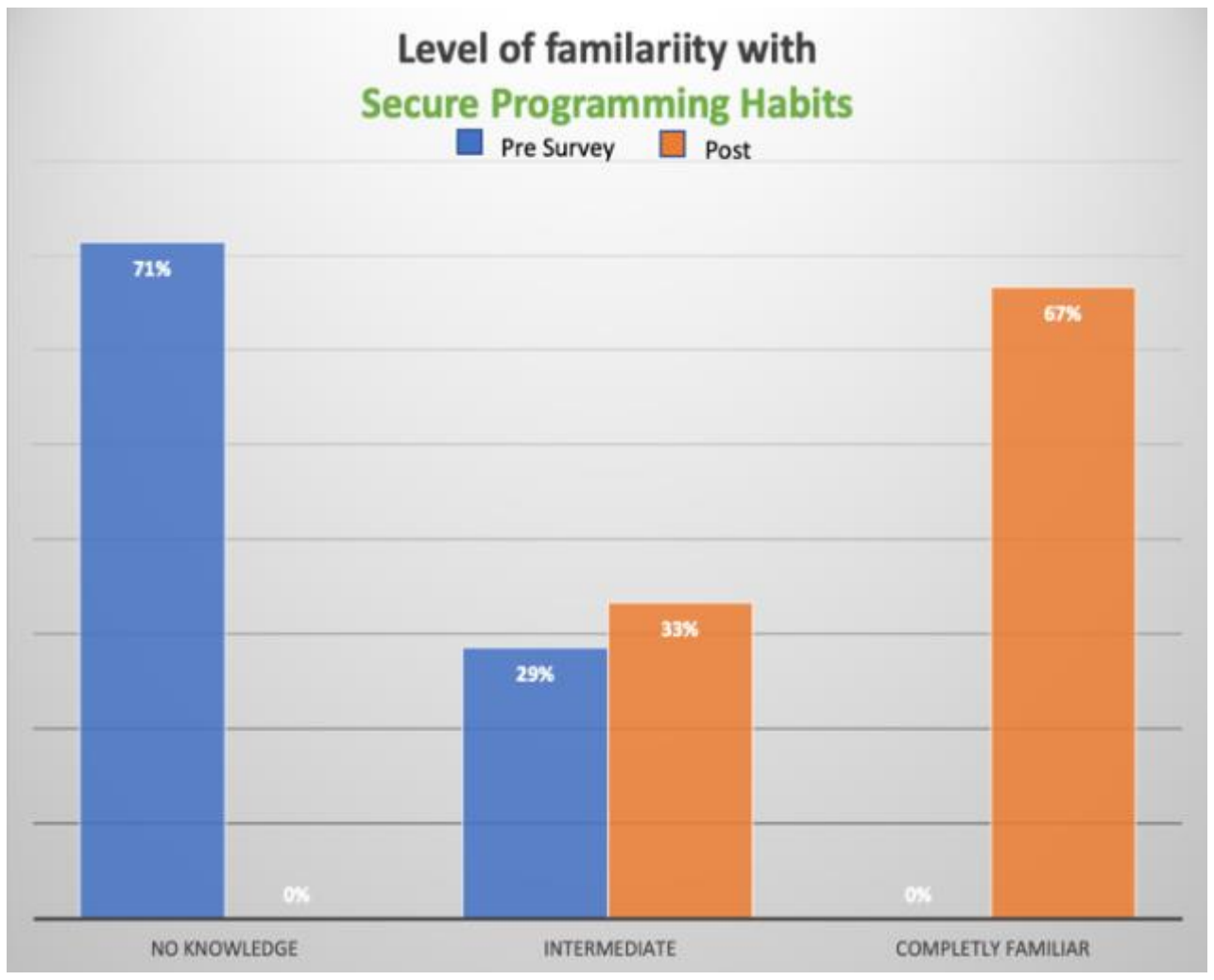

Figure 53 Familiarity level improvement on Secure Programming Habits (Pre & Post Survey)

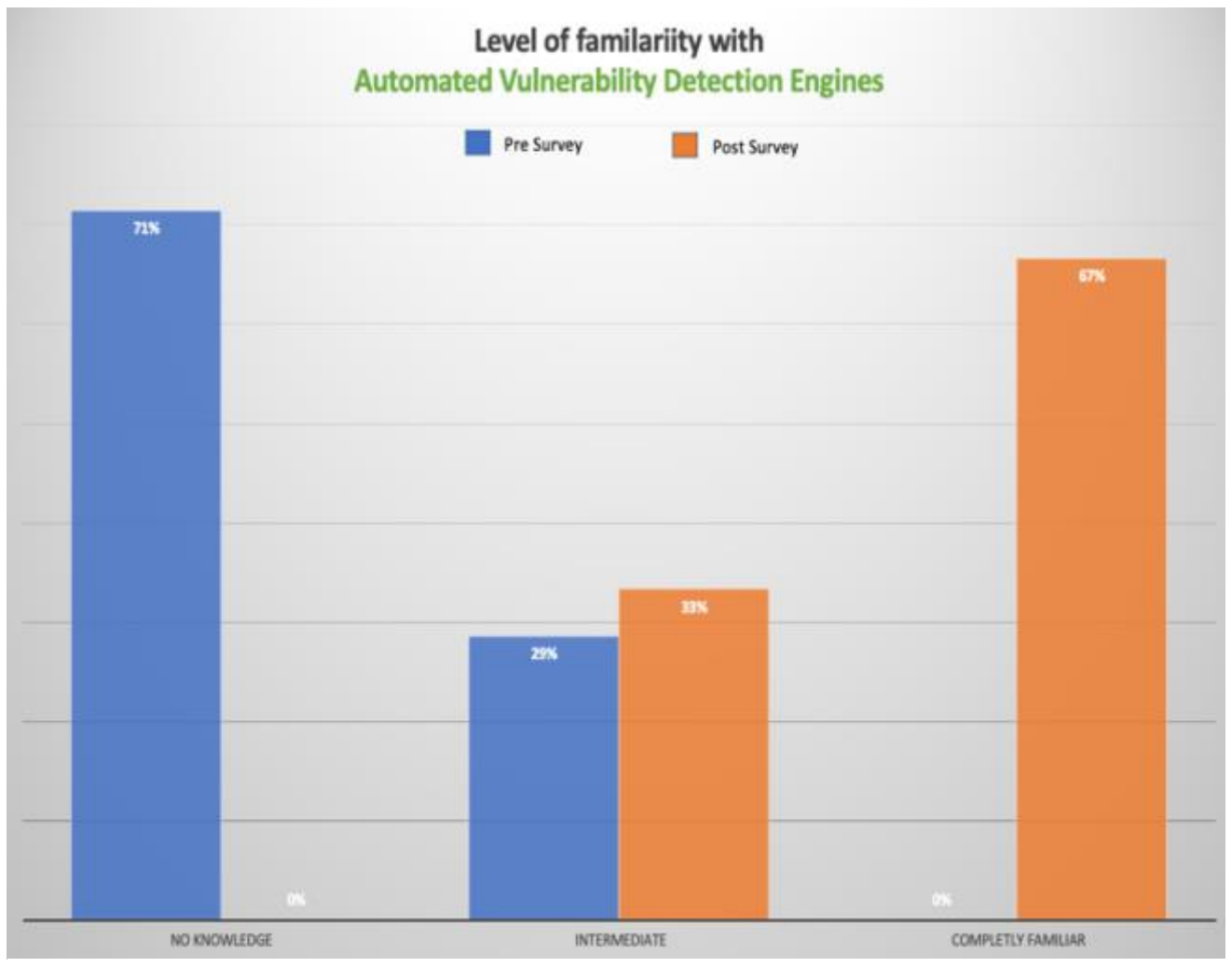

Figure 52 Familiarity level improvement on Automated Vulnerability Detection Engines (Pre & Post Survey)

The student's level of familiarity (after finishing the hands-on experience) with the concepts covered in the first category of hands-on labs is presented in Table 12.

Table 12 Student's level of familiarity with other concepts covered in the Third category (Post Survey)

| Concept | No knowledge | Intermediate | Completely Familiar |
|---|---|---|---|
| ML & NLP in Vulnerability Detection | 0% | 50% | 50% |

Results from the first category indicate that the hands-on lab was the most successful in improving students' knowledge in the following order: 1- Secure Programming Habits, 2- Automated Vulnerability Detection Engines, 3- ML & NLP in Vulnerability Detection. This conclusion is based on the percentage of thoroughly familiar students with the concepts after finishing the lab.

Overall, students have enjoyed the hands-on labs, which was part of the labs' contributions and goals. Table 13 demonstrates the contributions and objectives of this hands-on lab series.

Table 13 Objectives of Thesis

| Objective |
|---|
| Increase the awareness of Future IT professionals on Source code Vulnerability |
| Increase the awareness of Future IT professionals on Static Analysis Tools |
| Increase the awareness of Future IT professionals on Vulnerability Mitigation Techniques using ML and NLP |
| TrainFuture IT professionals how to leverage tools to detect and mitigate source code vulnerabilities |
| Introduce Future IT professionals with Common/Potential source code/log file vulnerabilities |

Based on Post Survey results, an overwhelming number of students stated that their level of familiarity with the concepts specific to each lab category has improved after finishing the hands-on lab. Students have also shown interest in having the same experience in the future to learn course material through hands-on labs. Figure 56 demonstrates that over 85% of the students have agreed that the hands-on lab introduced them to the actual life application of learning material. Figure 57 indicates that they enjoyed learning through the hands-on lab.

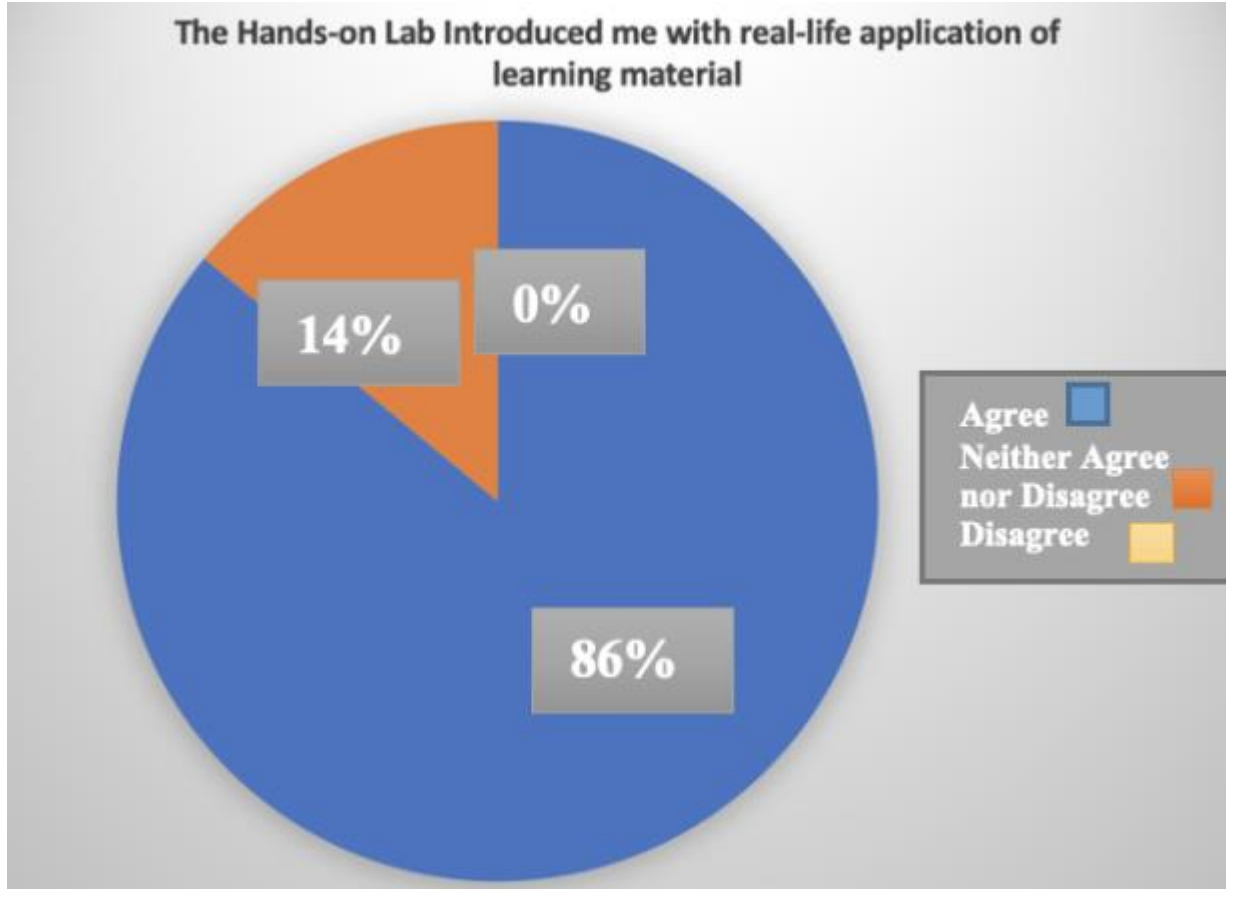

Figure 55 Lab teaching Real-world problems applied to course material

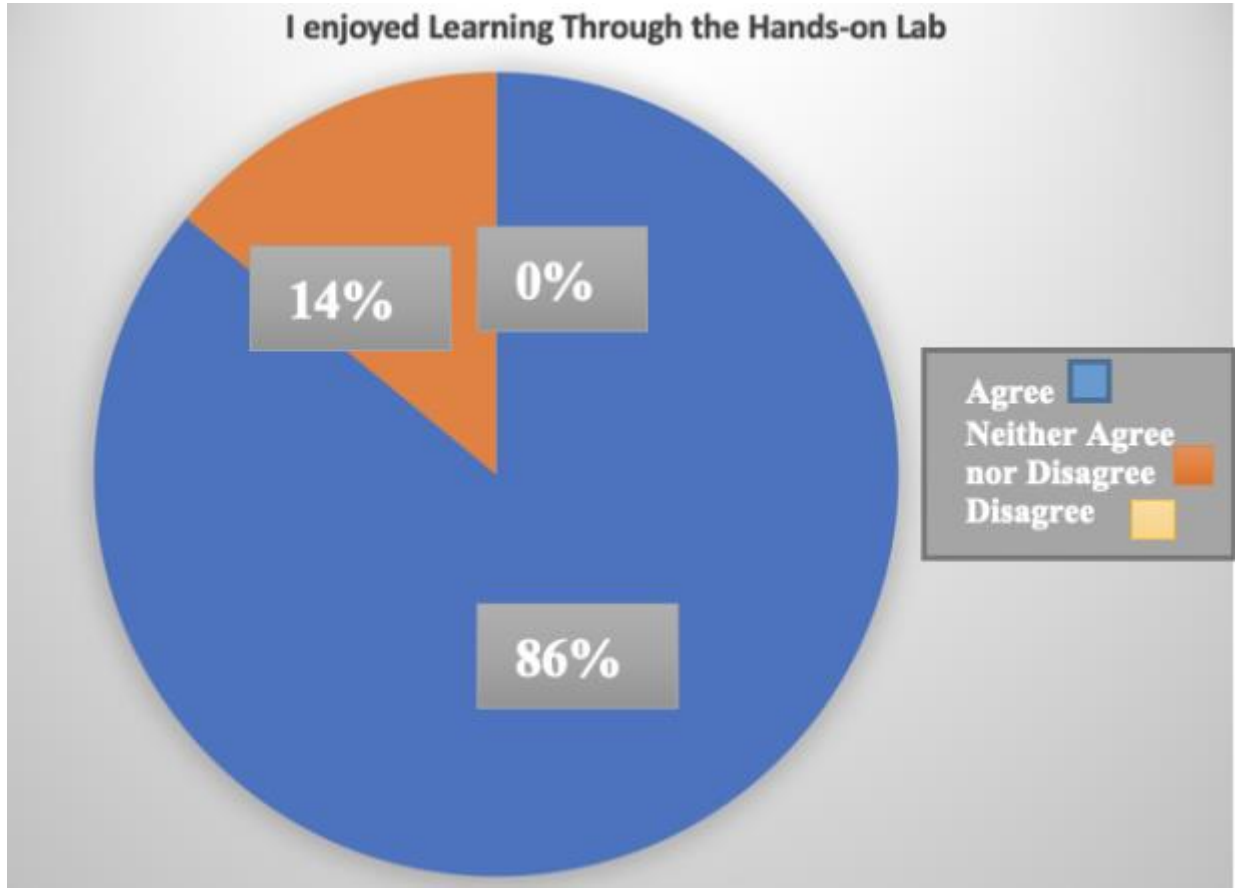

Figure 54 Lab Enjoyment Percentage

Students' reasons for interest in learning about source code vulnerabilities have shifted slightly compared to the pre-survey. Students stated that career opportunities and secure software development or programming practices are the main reason for their further interest. Students indicated a lower interest in various applications and technologies for source code/log file vulnerability detection during pre-and post-survey. However, after the lab, students' focus shifted towards Career opportunities and secure software development by a small margin (Figure 58).

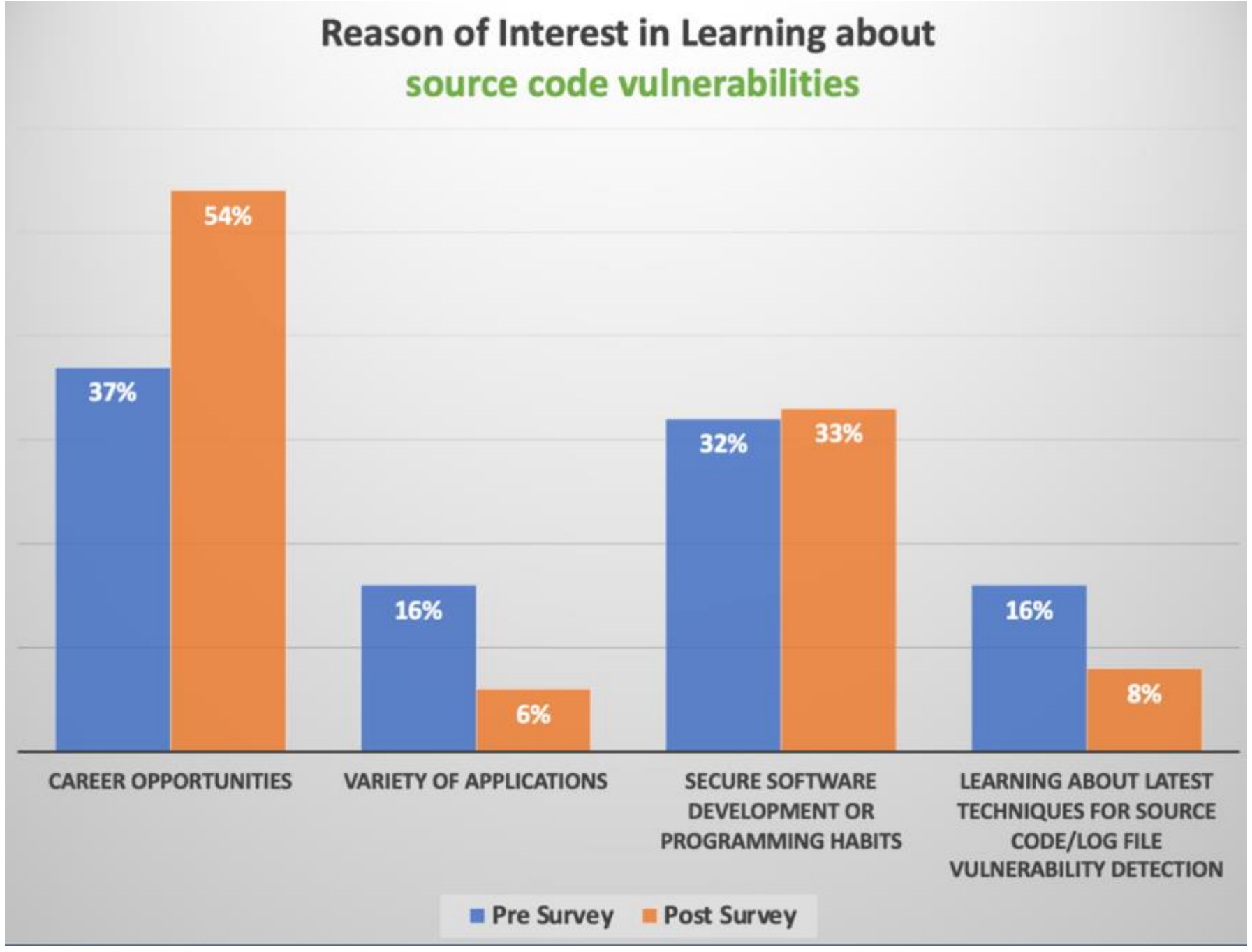

Figure 56 Reason of Interest comparison (Pre & Post) Survey

Furthermore, more than 85% of the students indicated that they learned how to use different tools to detect and mitigate source code vulnerabilities.

When students were asked to rate the lab on a scale of 1 to 5, 70% of them gave the highest level of rate to the hands-on lab, and 20% of them gave a number 4 rate to the hands-on lab, %10 gave a rate of 3 which indicates that students mostly rated the lab higher than average rate (Figure 59).

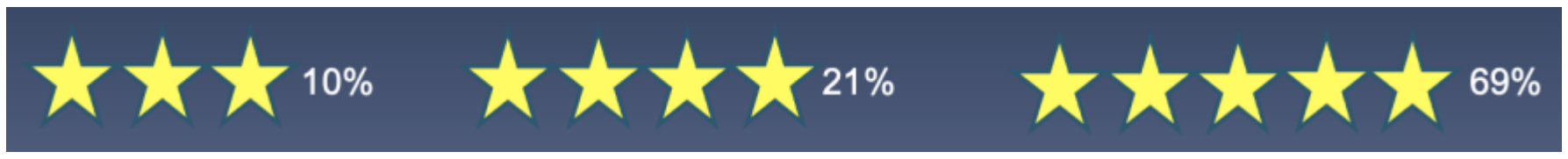

Figure 57 Students' Overall Rating to the hands-on Experience

94% of the students showed interest in having similar experiences for other concepts covered in their courses. Overall, this experience successfully found that students lack the required skills for industry and find learning through hands-on to be more effective than the lectures.

### 5.6.4 Success Measure

Statistical analysis: Unpaired Student's t-test was used to compare pre and post-survey scores on students' perception of the hands-on lab. $P < 0.05$ was considered as significant. Students' awareness and familiarity with the concepts covered in the first category of labs were significantly improved (Table 14).

Table 14 Success measure rate for the First category of hands-on labs

| Concept/Level | P-value |
|---|---|
| Source Code Vulnerability | 2.63205E-15 |
| Static Analysis | 1.6939E-09 |
| CVE & CWE | 1.61122E-11 |
| Memory Errors & Leaks | 1.58942E-10 |
| Argument Type & Input Validation | 1.46973E-08 |

The same statistical analysis using the Unpaired Student's t-test was performed to compare and evaluate the effectiveness of the hands-on lab material. The results are revealed in Table 15.

Table 15 Success measure rate for the Second category of hands-on labs

| Concept/Level | P-value |
|---|---|
| Application of ML & NLP in vulnerability detection | 0.001143378 |
| Automated Vulnerability Detection Engines | 0.000238807 |
| Secure Programming Habits | 0.00023881 |

The same statistical analysis using the Unpaired Student's t-test was performed to compare and evaluate the effectiveness of the hands-on lab material of the third category on Secure programming practices. The results are revealed in Table 16.

Table 16 Success measure rate for the Third category of hands-on labs

| Concept/Level | P-value |
|---|---|
| Log File Parsing & Analysis | 1.08806E-07 |
| Common Web Attacks | 0.000134871 |
| System/Web server event log files | 1.08806E-07 |
| OWASP Top 10 Security Risks | 2.05603E-07 |

This result demonstrates that the series of hands-on labs successfully achieved their goal and improved the learning rate (*i.e.*, raising awareness).

### 5.7 Assessment Lab (Quiz Results)

During the Pre and Post survey, students were asked a few questions about the concepts) to test their knowledge before and after the hands-on lab and see if they have learned the concepts.

During the Pre and Post survey, students were asked a few questions about the concepts (Quiz Like) to test their knowledge before and after the hands-on lab and see if they have learned the concepts.

During the Pre survey, when students were asked about the advantages of Static Program Analysis, Event Log Files, OR ML, and NLP application in Vulnerability detection, almost 80% of them did not know or have chosen the wrong answer. Surprisingly, over 50% of the students did not have a clear definition of Static Analysis. During the Post-Survey, Students were asked the same questions on their description of Static analysis and advantages of Static Program Analysis, Event Log Files, OR ML, NLP application in Vulnerability detection, and 82% of them chose the correct answer to the questions. A comparison of Pre and Post Survey Quiz Answers is demonstrated in Figure 60.

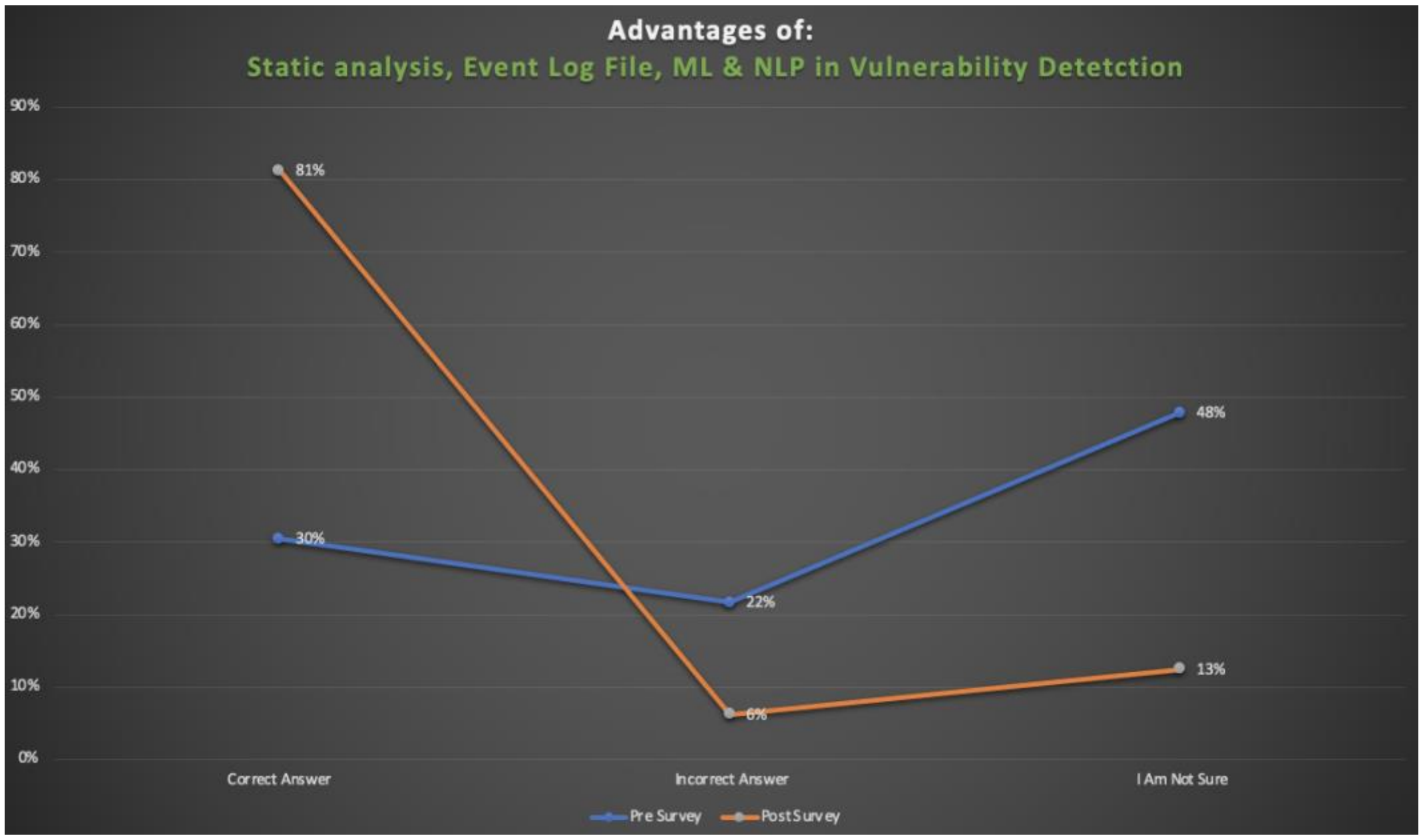

Figure 58 Comparison of Quiz Answer (Pre & Post Survey)

### 5.8 Challenges & Limitations

There have been many challenges in assisting students in reading and following the instructions of the lab. Students seem not to stay engaged or motivated enough to read every step of the instruction and follow them. They were lacking proper background to understand the root and damages of the vulnerabilities. They didn't have any knowledge on Source code vulnerability and also did not have any familiarity with the tools leveraged on the hands-on experience. None of the students were familiar with log file analysis which made it hard to explain and prepare them to understand the benefits of log files. Undergraduate students were having difficulties installing and setting up a Kali Linux environment. Therefore, I had to provide them with a Kali/Ubuntu terminal on a cloud station (Code Ocean and CoCalc) to not deal with the Linux setup. Students have indicated that the length of the lab instruction was too long, and they would like it better if it would have been shorter. Some of the students were lacking enough programming background knowledge to follow and understand the concepts. It was tough to get the students to complete both Pre and Post Survey since I could not see them every week to remind them to fill it out, and it was an optional task (not required). Table 17 demonstrates some of the feedback students provided on the lab after finishing the lab.

Table 17 Student's Feedback

| **Comments** |
|---|
| Great lab, looking forward to having more like this to widen my knowledge. |
| Excellent lab! |
| The lab was pretty straightforward and presented several new tools. |
| Good Lab! Beneficial, the hands-on lab gave me a better understanding of how Log File Analysis works. |
| It had nothing to do with my major or my field of study, but it was informative in a sense. |
| I enjoyed the lab very much, and it introduced me to something I did not know before. |
| The lab was too long. |
| The download process was the only issue for me. But after following the steps it was pretty easy going. |
| Better instructions about installing ubuntu/kali maybe? |
| More detailed in instructions step by step. System and software compatibility for students. |
| There were some minor setbacks, but it was fine for the most part. |
| Needed some more time but it was great overall. |
| Thank you for the lab and lecture it was great. |
| This lab was great, I wish we could have similar labs for other concepts in cyber security. |
| Amazing lab, I learned a lot about something I didn't know. |
| Very Straight Forward. |
| I learned a lot about this topic that I found interesting. |
| It was cool that she helped me to run Linux on my browser without virtual box installation. |
| I liked this lab better than the rest and tools introduced are very helpful. |
| I liked how I had to solve a real-life problem using new tools and techniques, but the lab was a bit long. |
| Very informative lab, learnt something new! |
| Good Lab. The lab was a good learning experience. |

# Chapter 6 Conclusion and Future Works

The objective of SeCodEd was to educate students on Source Code Vulnerabilities. Furthermore, our future IT professionals were equipped with the proper tools and skills that could be leveraged to provide a secure development in their prospective industry or academic position. A significant number of students expressed their interest in learning more about new technologies that are currently being used to identify source code vulnerabilities and analyze log files. Therefore, creating a comprehensive framework compromised of hands-on labs can bridge the gap between the rapid advancement of technology.

This thesis has demonstrated that students lack awareness and skills required to have after their graduation regarding source code vulnerability detection and secure programming habits. SeCodEd Framework has developed various hands-on labs that will help students learn and gain the missing skills to fill the gap between industry and education. Future work could work on developing an implementation method of the described framework. By leveraging ML and NLP, a knowledge graph in the source code vulnerability detection domain could be constructed and utilized to form a learner's profile (Deng, 2019). Moreover, the learner profile could be leveraged to identify the learning style from student activities and adapt learning material accordingly. Other aspects of source code vulnerabilities, such as penetration testing on a web application to demonstrate source code vulnerabilities, could also be added to this framework. SeCodEd allows for expansion into other disciplines or fields of study by analyzing students' coding habits and attitudes in real-time to better assist them in developing secure software.